\documentclass[useAMS,usenatbib]{mn2e}
\usepackage[dvips]{graphicx}
\usepackage{hyperref}
\usepackage{xcolor}

\newcommand{\bz}{$\langle B_z \rangle$}

\newcommand{\msun}{$M_{\odot}$}

\newcommand{\kms}{km\,s$^{-1}$}

\newcommand{\vsini}{$v \sin i$}
\newcommand{\teff}{$T_{\rm eff}$}

\newcommand{\mdot}{$\dot{M}$}
\newcommand{\vinf}{v$_{\infty}$}

\newcommand{\rk}{$R_{\rm K}$}
\newcommand{\ra}{$R_{\rm A}$}
\newcommand{\rark}{$\log{(R_{\rm A}/R_{\rm K})}$}
\newcommand{\mnras}{$MNRAS$}
\newcommand{\apj}{$ApJ$}
\newcommand{\apjl}{$ApJ$}
\newcommand{\aj}{$AJ$}
\newcommand{\aap}{$A\&A$}
\newcommand{\aapr}{$A\&AR$}

\newcommand{\apjs}{$ApJS$}
\newcommand{\araa}{$ARA\&A$}

\newcommand{\apss}{$Astrophys.\ Space Sci.\ $}

\newcommand{\nat}{$Nature$}

\newcommand{\an}{$Astron.\ Nachr.\ $}

\title[Evolution of the Magnetic Early B-type Stars]{The Magnetic Early B-type Stars III: A main sequence magnetic, rotational, and magnetospheric biography}
\author[M.\ Shultz et al.]{M.\ E.\ Shultz$^{1}$\thanks{E-mail:
mshultz@udel.edu}\thanks{Annie Jump Cannon Fellow},
G.\ A.\ Wade$^2$,
Th.\ Rivinius$^3$,
E.\ Alecian$^{4}$,
C.\ Neiner$^5$,
V.\ Petit$^1$, 
\newauthor{S.\ Owocki$^{1}$, A.\ ud-Doula$^{6}$, O.\ Kochukhov$^{7}$, D.\ Bohlender$^8$, Z.\ Keszthelyi$^{2,9,10}$,}
\newauthor{and the MiMeS and BinaMIcS Collaborations}\\
$^1$Department of Physics and Astronomy, University of Delaware, 217 Sharp Lab, Newark, Delaware, 19716, USA\\
$^2$Department of Physics and Space Science, Royal Military College of Canada, PO Box 17000, Station Forces, Kingston, Ontario K7K 7B4, Canada\\
$^3$ESO - European Organisation for Astronomical Research in the Southern Hemisphere, Casilla 19001, Santiago 19, Chile\\
$^4$Univ.\ Grenoble Alpes, CNRS, IPAG, 38000 Grenoble, France\\
$^5$LESIA, Paris Observatory, PSL University, CNRS, Sorbonne Universit\'e, Universit\'e de Paris, 5 place Jules Janssen, 92195 Meudon, France\\
$^{6}$Penn State Scranton, Dunmore, PA 18512, USA\\
$^{7}$Department of Physics and Astronomy, Uppsala University, Box 516, Uppsala 75120, Sweden \\
$^{8}$National Research Council of Canada, Herzberg Astronomy and Astrophysics Research Centre, 5071 West Saanich Road, Victoria, BC V9E 2E7\\
$^{9}$Department of Physics, Engineering Physics \& Astronomy, Queen's University, 64 Bader Lane, Kingston, ON Canada, K7L 3N6 \\
$^{10}$Anton Pannekoek Institute for Astronomy, University of Amsterdam, Science Park 904, 1098 XH, Amsterdam, The Netherlands \\
}
\begin{document}

\date{}

\pagerange{\pageref{firstpage}--\pageref{lastpage}} \pubyear{2019}

\maketitle

\label{firstpage}

\begin{abstract}
Magnetic confinement of stellar winds leads to the formation of magnetospheres, which can be sculpted into Centrifugal Magnetospheres (CMs) by rotational support of the corotating plasma. The conditions required for the CMs of magnetic early B-type stars to yield detectable emission in H$\alpha$ -- the principal diagnostic of these structures -- are poorly constrained. A key reason is that no detailed study of the magnetic and rotational evolution of this population has yet been performed. Using newly determined rotational periods, modern magnetic measurements, and atmospheric parameters determined via spectroscopic modelling, we have derived fundamental parameters, dipolar oblique rotator models, and magnetospheric parameters for 56 early B-type stars. Comparison to magnetic A- and O-type stars shows that the range of surface magnetic field strength is essentially constant with stellar mass, but that the unsigned surface magnetic flux increases with mass. Both the surface magnetic dipole strength and the total magnetic flux decrease with stellar age, with the rate of flux decay apparently increasing with stellar mass. We find tentative evidence that multipolar magnetic fields may decay more rapidly than dipoles. Rotational periods increase with stellar age, as expected for a magnetic braking scenario. Without exception, all stars with H$\alpha$ emission originating in a CM are 1) rapid rotators, 2) strongly magnetic, and 3) young, with the latter property consistent with the observation that magnetic fields and rotation both decrease over time. 
\end{abstract}


\section{Introduction}\label{sec:int}

Approximately 10\% of massive stars possess strong surface magnetic fields \citep{grun2012c,2017MNRAS.465.2432G}. These magnetic fields are generally believed to be fossil fields, as there are no well-established correlations between magnetic properties and rotational velocities, which are both expected and observed for magnetic fields sustained by contemporaneous dynamos in late-type stars with convective exteriors \citep[e.g.][]{dl2009}. Fossil fields are globally organized, exhibit generally simple topologies (predominantly tilted dipoles), are typically strong (of order 100s of G to 10 kG), and are stable over at least a time span of decades \citep[e.g.][]{oks2012,2018MNRAS.475.5144S}. These properties are again in contrast to those typical of the dynamo fields of stars with convective envelopes, which often exhibit tangled magnetic topologies \citep[e.g.][]{2008MNRAS.390..545D,2016MNRAS.457..580F}, are generally weak \citep[usually on the order of a few G, although fully convective M dwarfs can reach kG strengths, e.g.][]{2017ApJ...835L...4K,2019ApJ...873...69K}, and exhibit frequent topological changes consistent with magnetic activity cycles \citep[e.g.][]{2008MNRAS.385.1179D,2016MNRAS.459.4325M}.

The evolution of the magnetic fields of cool stars is relatively well understood. Young stars tend to have strong magnetic fields generated by a combination of vigorous convection and rapid rotation. Over time, magnetic braking due to torques arising from the Poynting stresses of the circumstellar magnetic field slows the star's surface rotation \citep[e.g.][]{wd1967}, leading to a less energetic dynamo and, hence, a weaker surface magnetic field \citep{2016MNRAS.457..580F}. In contrast, much remains unknown regarding the evolution of fossil magnetic fields. Since the magnetic field is not maintained by a dynamo, it is not directly related to rotation. The surface magnetic field strength is usually expected to decline due to magnetic flux conservation within the expanding stellar atmosphere \citep[e.g.][]{land2007,land2008}. on the other hand, some magnetic interior models have predicted an increase in the surface magnetic field over time \citep[][]{2004Natur.431..819B,2006A&A...450.1077B}, and such an effect has been reported by some observational studies \citep[][]{2006AA...450..763K,2015Ap.....58...29G}. Other population surveys have instead found that the surface magnetic field strength and the total magnetic flux decline over time \citep{bagn2006,land2007,land2008}. \cite{2019MNRAS.483.3127S} reported an evolutionary decline in surface magnetic field strength consistent with constant magnetic flux. \cite{2016A&A...592A..84F} compared the occurence frequency of magnetic and non-magnetic stars across the upper main sequence, and inferred that magnetic flux decay occurs and that it may accelerate with stellar mass. However, they were unable to find any trend of decreasing magnetic flux with fractional main sequence age, as would be expected in this scenario. 

Like cool stars, magnetic hot stars lose angular momentum due to magnetic spindown \citep[e.g.][]{ud2009,petit2013}, with the primary differences being that magnetic hot stars spin down much more rapidly due to their much higher mass-loss rates, and that their surface magnetic fields are not directly affected by spindown. The expected deceleration from this ``magnetic braking'' is potentially observable, and period change has indeed been directly measured in some cases \citep[e.g. HD 37776 and HD\,37479;][]{miku2008,town2010}. While HD\,37479's period is slowing down, consistent with magnetic braking, spin-up has been observed in the cases of HD\,37776 and CU\,Vir, suggesting the picture is more complicated than a simple monotonic decrease \citep{miku2011,2017ASPC..510..220M}. Nevertheless, as the projected rotation velocities of magnetic hot stars are systematically much lower than those of their non-magnetic counterparts \citep[e.g.][]{2018MNRAS.475.5144S}, it is clearly capable of creating a population of magnetic stars that is statistically more slowly rotating than non-magnetic stars, suggesting that magnetic braking dominates the rotational evolution of magnetic hot stars over the long run. \cite{2016A&A...592A..84F} noted that the spindown ages of stars between about 5 and 15 \msun~were systematically much greater than their main-sequence lifetimes, i.e.\ they are rotating even {\em more} slowly than expected. They suggested that this discrepancy could be explained if their magnetic fields were much stronger in the past. Alternatively, they could have arrived at the Zero-Age Main Sequence already slowly rotating.

The magnetospheres that act as the interfaces through which magnetic hot stars are spun down are detectable across the electromagnetic spectrum, a phenomenon that has received extensive observational attention at various wavelengths: in X-rays \citep[e.g.][]{oskinova2011,2014ApJS..215...10N}, the ultraviolet \citep[e.g.][]{brown1987,smithgroote2001,neiner2003b,2004AA...421..203S,schnerr2008,henrichs2013}, at H$\alpha$ \citep[e.g.][]{d2002,town2005b,bohl2011,wade2011,grun2012,sund2012,rivi2013,2015MNRAS.447.2551W,2015MNRAS.451.1928S,2016MNRAS.460.1811S}, in the near-infrared \citep{2015A&A...578A.112O}, and as both coherent and incoherent emission at radio wavelengths \citep{1987ApJ...322..902D, 1992ApJ...393..341L, 2000AA...362..281T, 2008MNRAS.384.1437T, leto2012, 2015MNRAS.452.1245C, 2017MNRAS.465.2160K,2018MNRAS.474L..61D, 2017MNRAS.467.2820L, 2018MNRAS.476..562L}. \citet[hereafter P13]{petit2013} showed that the presence or absence of H$\alpha$ emission is predicted by the stellar mass-loss rate, and by whether or not the star possesses a purely {\em dynamical magnetosphere} (DM), or also a {\em centrifugal magnetosphere} (CM). DMs are detectable in H$\alpha$ only around stars with high mass-loss rates (i.e.\ O-type stars), because their strong winds are able to fill their magnetospheres on similar timescales to emptying via gravitational infall \citep[e.g.][]{ud2002,ud2013}. In contrast, B-type stars, which have much weaker winds, tend not to show emission if they possess only a DM. However, if the star is rotating rapidly, corotation of the plasma with the magnetic field leads to centrifugal support capable of counterbalancing gravity, enabling the accumulation of plasma over long timescales \citep[e.g.][]{town2005c,ud2008}. 

An important limitation of previous population studies of magnetic hot stars has been that in many cases key parameters - surface magnetic field strengths, rotation periods, and atmospheric and fundamental parameters - have been poorly constrained, with only limiting values available. As a consequence, P13 were unable to determine the conditions required for the onset of H$\alpha$ emission. This may also possibly explain the inability of \cite{2016A&A...592A..84F} to unambiguously detect magnetic flux decay. 

\citet[hereafter Paper I]{2018MNRAS.475.5144S} used a large database of high-resolution spectropolarimetric time series, principally collected by the Magnetism in Massive Stars \citep[MiMeS;][]{2016MNRAS.456....2W} and Binarity and Magnetic Interactions in various classes of Stars \citep[BinaMIcS;][]{2015IAUS..307..330A} Large Programs (LPs), to obtain longitudinal magnetic field \bz~measurements and thus determine rotational periods for the stars examined by P13. Atmospheric parameters for these stars based upon spectroscopic measurements and Gaia parallaxes were presented by \citet[hereafter Paper II]{2019MNRAS.485.1508S}. In this work, we build upon the empirical foundation established in Papers I and II to derive the stellar, rotational, magnetic, and magnetospheric properties of the sample, with the aim of clarifying the evolution of these parameters. An overview of the sample stars and the dataset is provided in \S~\ref{sec:sample}. The method used to determine stellar, magnetic, and magnetospheric parameters is outlined in \S~\ref{sec:mc_method}. Stellar parameters are presented in \S~\ref{sec:stellar_par}, magnetic parameters in \S~\ref{sec:orm}, and magnetospheric parameters in \S~\ref{sec:magnetosphere}. Their evolution is examined in \S~\ref{sec:evol}, the interpretations of the results are discussed in \S~\ref{sec:discussion}, and conclusions are summarized in \S~\ref{sec:conclusion}. 

\section{Sample and Observations}\label{sec:sample}

The sample consists of all published magnetic main sequence stars with spectral types between B5 and B0 and sufficient spectropolarimetric data to confidentally assess their magnetic characteristics. The selection criteria and properties of the sample were described in Paper I, and those of 5 additional stars were described in Paper II, with the final sample consisting of 56 stars. The study is based primarily upon an extensive database of high-resolution ESPaDOnS, Narval, and HARPSpol spectropolarimetry, also described in Papers I and II. The majority of these data were acquired by the MiMeS and BinaMIcS LPs \citep{2016MNRAS.456....2W, 2015IAUS..307..330A}, with additional data from the BRITEpol LP \citep{neiner2017ppas}, as well as various PI programs (see Papers I and II). \cite{2016MNRAS.456....2W} gave an overview of the reduction and analysis of ESPaDOnS, Narval, and HARPSpol data. Five stars were primarily observed with HARPSpol and FORS2 by the B-fields in OB stars \citep[BOB;][]{2015A&A...582A..45F,2017A&A...599A..66S} LP at the European Southern Observatory. 

\section{Parameter determination}\label{sec:mc_method}

In order to ensure fully self-consistent stellar, rotational, magnetic, and magnetospheric parameters, a Monte Carlo method was developed that simultaneously accounts for all observables. A detailed description of the algorithm is given in Appendix A (online). Briefly, the algorithm takes as input all available measurements (\teff, $\log{L}$, $\log{g}$, \vsini, $P_{\rm rot}$, and \bz, as well as supplementary prior constraints on age, mass, and rotational inclination; see Appendix B, online), and combines these with evolutionary models to determine radii, masses, ages, surface rotational parameters, dipole Oblique Rotator Models (ORMs) and the resulting stellar wind and magnetospheric parameters. The algorithm takes rotational properties as a prior, rejecting test points that imply physically impossible rotational properties (i.e.\ super-critical rotation, or rotation slower than \vsini). It also rejects test points that lead to large inconsistencies between observables (e.g.\ $\log{g}$ incompatible with the observed value; age incompatible with cluster main-sequence turnoff ages). Test point rejection is probabilistic: test points are rejected if their inferred parameters ($\log{g}$, $M_*$, $t$) differ by more than 1$\sigma$ from target values drawn from Gaussian distributions. This ensures that the posterior probability distributions are approximately Gaussian. The advantage of this algorithm is that it automatically accounts for correlations in the uncertainties of derived parameters. 

At a minimum, stellar and rotational parameters were constrained by \teff, $\log{L}$, $\log{g}$, and \vsini, while the magnetic parameters were constrained via sinusoidal fits to \bz~(the mean value $B_0$ and the semi-amplitude $B_1$) and the measured value of \bz$_{\rm max}$. In most cases $P_{\rm rot}$ is also available as a key constraint. \vsini, $P_{\rm rot}$, \bz$_{\rm max}$, and sinusoidal fitting parameters to \bz~were determined in Paper I and references therein (with the exception of the five stars which have been added since the publication of Paper I, for which the parameters and references were collected in Paper II). \teff, $\log{g}$ and $\log{L}$ were given in Paper II and references therein.

Fundamental stellar and rotational parameters are given in Table C1 (online), ORM parameters in Table C2 (online), and wind and magnetospheric parameters in Table C3 (online).

\subsection{Prior distributions}

In most cases Gaussian distributions were used for the priors, with centroids and standard deviations equal to the measured values and the 1$\sigma$ error bars. The only exception is for the rotational axis inclination $i_{\rm rot}$. For most stars in the sample $i_{\rm rot}$ can be inferred from $P_{\rm rot}$, \vsini, and $R_*$ (\S~\ref{subsec:rotpars}). In the special case when $i_{\rm rot}$ cannot be rotationally constrained, $i_{\rm rot}$ was drawn from a random distribution over 4$\pi$ steradians,

\begin{equation}\label{prob_i}
P(i_{\rm rot} < i_0) = 1 - \cos{i_0},
\end{equation}

\noindent where $P(i_{\rm rot})$ is the probability that $i_{\rm rot} < i_0$. This {\em a priori} distribution was used for HD 46328 and HD 66522 (extremely slow rotators for which the measured \vsini~is basically an upper limit), and HD 52089 (for which $P_{\rm rot}$ is unknown). $P_{\rm rot}$ is also unknown for HD\,44743, HD\,58260 and HD\,136504; however alternate constraints on $i_{\rm rot}$ are available for these stars (see Appendix B3, online).

\subsection{Evolutionary models}

We adopted the solar metallicity evolutionary models calculated by \cite{ekstrom2012} and \cite{2011A&A...530A.115B}, which make different assumptions (regarding core overshooting, internal mixing, mass loss, etc.), leading to a more extended main sequence lifetime in the latter case. As demonstrated by \cite{2013A&A...560A..16M} and \cite{2014A&A...570L..13C}, within the parameter range of interest neither set of models precisely reproduces the width of the main sequence (MS), with the \citeauthor{ekstrom2012} models yielding a narrower MS and the \citeauthor{2011A&A...530A.115B} models a wider MS than observed. While \cite{2014A&A...570L..13C} found the discrepancy to be larger in the case of the \citeauthor{ekstrom2012} models, it should be kept in mind that \cite{2015A&A...584A..54P} speculated that magnetic suppression of core convection may lead to a narrower main sequence for magnetic as compared to non-magnetic stars, with non-magnetic stars presumably dominating both the \cite{2013A&A...560A..16M} and the \cite{2014A&A...570L..13C} samples. Suppression of core overshooting is supported by the asteroseismic analysis of the magnetic pulsator HD\,163472 conducted by \cite{2012MNRAS.427..483B}, who found that its core overshootting was much less than expected. If core overshooting is in fact suppressed by strong internal magnetic fields, the \citeauthor{ekstrom2012} main sequence width may indeed be more appropriate.

In the case of the \citeauthor{ekstrom2012} models, for stars with rotation periods in excess of about 6 d we adopted the non-rotating models. We also utilized non-rotating models for HD 52089 and HD 58260, which have unknown periods but sharp spectral lines. For all other stars we utilized rotating models, which assume a uniform initial rotational velocity $v_{\rm ini}$ of 0.4 times the breakup velocity $v_{\rm br}$.

In the case of the \citeauthor{2011A&A...530A.115B} models, which sample the rotational phase space more densely than the \citeauthor{ekstrom2012} models, for each star we selected the models with $v_{\rm ini}$ closest to \vsini.

%We chose the \cite{ekstrom2012} evolutionary models over the \cite{2011A&A...530A.115B} models in order to perform a self-consistent analysis, since the former assume either a uniform initial rotational velocity of $0.4$ times the breakup velocity $v_{\rm br}$, or 0~\kms, while the latter assume a diverse range of initial rotational velocities (which are not known {\em a priori} for these stars). 

In practice, neither set of models accounts for effects such as mass-loss quenching due to magnetic confinement, magnetic braking, or for the influence of fossil magnetic fields on the internal rotational profile. Preliminary models including surface magnetic braking have been presented by \cite{meynet2011}, but these assumed a constant surface magnetic field strength. While fossil fields in realistic stellar atmospheres were modelled by \cite{2010A&A...517A..58D}, calculations of stellar evolution incorporating their relaxed field structures were outside the scope of their study. \cite{2012MNRAS.424.2358P} presented models of magnetic braking for massive stars, however, they assumed a mass-dependent radiative zone dynamo, rather than fossil magnetic fields. Magnetic evolutionary models presented by \cite{2018MNRAS.477.2298Q} also assumed a dynamo mechanism, and were furthermore limited to 3 \msun~stars. Models including mass-loss quenching were developed by \cite{2017MNRAS.466.1052P} and \cite{2017A&A...599L...5G}, although these concerned stars more massive than those of the present sample. Magnetic models incorporating magnetic braking and tidal effects on stellar evolution were developed by \cite{2018A&A...609A...3S}, but they concern very close magnetic binaries that are, again, more massive than any stars in the present sample (the majority of which are furthermore single). An appropriate grid of models utilizing fossil magnetic fields and self-consistently accounting for all effects is still in development \citep[][]{2017IAUS..329..250K,2018CoSka..48..124K,2019MNRAS.485.5843K}. 

%Within the parameter range of relevance to this paper, at any given point on the HRD the \cite{2011A&A...530A.115B} models imply almost identical masses to those of the \cite{ekstrom2012} models, but predict a more extended main sequence lifetime. Thus the primary consequence of using \citeauthor{2011A&A...530A.115B} models would be to decrease the age of the oldest stars in the sample; however, general trends would remain unchanged. Below about 12 \msun, the differences between the \citeauthor{2011A&A...530A.115B} and \citeauthor{ekstrom2012} models are negligible. The differences increase above this range, but are most pronounced for rotating models, which were not used for any of the stars above 12 \msun~since these stars are uniformly slow rotators. The differences between isochrones from the two sets of models increases with time, but even for the most evolved and most massive stars in the sample (e.g.\ HD\,46328), the difference in isochrone ages is less than 0.1 dex, i.e.\ comparable to the existing uncertainty. 

Metallicity is a potentially complicating factor in determining stellar ages from the HRD \citep{bagn2006,land2007}. Due to surface abundance peculiarities of up to several dex in the photospheres of chemically peculiar (CP) stars, the bulk chemical composition of magnetic CP stars (comprising the majority of the present sample) is almost impossible to determine. However, metallicities measured in a multitude of star-forming regions and open clusters by the Gaia ESO survey, spanning about $\pm 1$~kpc of the Sun's galactocentric distance, are consistent with an essentially flat radial abundance gradient in the solar neighborhood, with a mean and standard deviation of [Fe/H] of 0.02 and 0.09, respectively \citep{2017A&A...601A..70S}. Stars within the 1 kpc range of the survey are therefore likely to have approximately solar bulk compositions. Only 8 stars in the present sample are more distant from the Sun than 1 kpc; however, given the almost flat abundance gradient reported by \cite{2017A&A...601A..70S}, their bulk compositions are unlikely to differ significantly. We therefore feel confident in adopting solar metallicity as a reasonable first approximation. 

\subsection{Rotational parameter determination}\label{subsec:rotpars}

Assuming rigid rotation, the inclination $i_{\rm rot}$ of the rotational axis from the line of sight can be determined from 

\begin{equation}\label{sini}
\sin{i_{\rm rot}} = \frac{ v\sin{i} } {v_{\rm eq} } =  v\sin{i}\frac{ 2 \pi R_{\rm eq}}{P_{\rm rot}},
\end{equation}

\noindent where $v_{\rm eq}$ is the rotational velocity at the equator and $R_{\rm eq}$ is the equatorial radius. In most cases $R_{\rm eq}$ is practically identical to $R_*$, however this sample contains some rapidly rotating stars for which this is not be the case. We therefore calculated the rotational oblateness explicitly as the ratio of the polar radius $R_{\rm p}$ to the equatorial radius $R_{\rm eq}$ \citep{1928asco.book.....J}: 

\begin{equation}\label{rpre}
\frac{R_{\rm p}}{R_{\rm eq}} = \sqrt{1 - \frac{3\Omega^2}{4\rm{G}\pi\rho}},
\end{equation}

\noindent where $\Omega$ is the angular frequency and $\rho=M_*/(4\pi R_{\rm p}^3/3)$ is the mean stellar density. In practice this correction is negligible for all but a few stars in our sample (e.g.\ HD\,142184 and HD\,182180).

$P_{\rm rot}$ and $v_{\rm eq}$ quantify the speed of rotation but do not give a sense of its dynamical importance for a given star. For this we utilized the dimensionless rotational parameter $W = v_{\rm eq}/v_{\rm orb}$, where $v_{\rm orb}$ is the velocity necessary to maintain a circular Keplerian orbit at the stellar surface \citep[][Eqn.\ 11]{ud2008}\footnote{Conversions between $W$ and more commonly used expressions for the critical rotation fraction (e.g.\ the ratio of the angular to the critical angular velocity or of the equatorial to the critical equatorial velocity) are given by \cite{2013A&ARv..21...69R}.}. Critical rotation is given by $W=1$, while a non-rotating star has $W=0$. The physical influence of rotation within the magnetosphere is measured by the Kepler corotation radius $R_{\rm K} = W^{-2/3}$, the distance from a star at which rotational and gravitational forces are in balance assuming material moving in strict corotation with the stellar surface \citep[][Eqn.\ 14]{ud2008}.

\subsection{Oblique rotator model parameter determination}

Dipolar ORMs consist of 4 parameters: $P_{\rm rot}$, $i_{\rm rot}$, the obliquity angle $\beta$ of the magnetic dipole from the rotational axis, and the surface polar strength of the magnetic dipole $B_{\rm d}$. Assuming $P_{\rm rot}$ is known and $i_{\rm rot}$ constrained, $\beta$ and $B_{\rm d}$ can be determined by solving Preston's equations \citep{preston1967} using harmonic fitting parameters to the \bz~curve and the maximum strength of \bz. Fitting parameters were defined in Paper I: $B_0$ is the mean value of the \bz~variation, and $B_1$ is the semi-amplitude of the first harmonic. $i_{\rm rot}$ and $\beta$ are then related through the Preston $r$ parameter,

\begin{equation}\label{preston_r}
r = \frac{|B_0| - B_1}{|B_0| + B_1} = \frac{\cos{(i_{\rm rot} + \beta)}}{\cos{(i_{\rm rot} - \beta)}}.
\end{equation}

The $r$ parameter is provided in Table C2. 

Obtaining $B_{\rm d}$ also requires the limb darkening coefficient, which was obtained from the tabulated values calculated by \cite{2016MNRAS.456.1294R}, where we used the B band as the closest to the mean wavelength of the line lists used to extract LSD profiles from ESPaDOnS/Narval spectra.

As noted in Paper I, 13 stars show evidence for contributions to \bz~from higher multipoles of the surface magnetic field. In most cases the amplitudes of the higher harmonic terms, reflecting multipoles, are much smaller than the amplitude of the first harmonic, describing the dipolar component; this suggests that even in these stars, the surface magnetic field is dominated by the dipole. This assumption is not always correct, as, depending on geometry, some higher spherical harmomics can yield \bz~curves dominated by the first harmonic; however, this possibility cannot be adequately explored absent Zeeman Doppler Imaging \citep[ZDI;][]{pk2002}. The contributions from higher-order multipoles fall off faster with distance from the star than those from the dipolar component, and should thus be of negligible importance to the radial extent of magnetic confinement and the resulting spindown timescale. We therefore ignore the multipolar components, and, for these stars, solve Preston's equations using only the amplitude of the first harmonic. 

We utilized the observed value of \bz$_{\rm max}$ to determine $B_{\rm d}$. For stars whose \bz~curves are well described by a first-order sinusoid, \bz$_{\rm max}$ and $B_0 + B_1$ are effectively identical. For stars whose \bz~variations require multiple harmonics to reproduce, $B_0 + B_1$ is likely to underestimate the maximum strength of the surface magnetic field. In these cases the peak surface magnetic field strength can really only be determined via ZDI. However, the `twisted dipoles' that are usually revealed by ZDI mapping \citep[e.g.][]{2019A&A...621A..47K} typically have maximum surface magnetic field strengths similar to those that would be inferred from \bz$_{\rm max}$. Using the observed value therefore avoids too badly under-estimating the surface field strength in these cases. A good example is the topologically complex star HD\,37776 \citep{koch2011}, for which $B_0 + B_1 = 0.54$~kG, but \bz$_{\rm max} = 1.7$~kG.

Since sinusoidal fits to \bz~could not be derived when $P_{\rm rot}$ is unknown, in these cases $B_0$ was approximated by the error-bar-weighted mean of \bz, and $B_1$ by the weighted standard deviation. We chose to use the weighted standard deviation for $B_1$ rather than the difference between \bz$_{\rm max}$ and $B_0$ in order to avoid outlier bias, as the stars without known $P_{\rm rot}$ generally have relatively noisy \bz~datasets. This strategy was adopted for HD\,44743, HD\,52089, HD\,58260, and HD\,136504Aa and HD136504Ab.

\subsection{Magnetospheric parameter determination}

The strength of magnetic confinement in the circumstellar environment is quantified by the dimensionless wind-magnetic confinement parameter $\eta_*$, which is the ratio of the energy density of the magnetic field to the kinetic energy density of the wind \citep[][Eqn.\ 7]{ud2002}. The physical extent of magnetic confinement, given by the distance from the star within which the magnetic field dominates the wind, is the Alfv\'en radius \ra. We determined \ra~using the approximate scaling with $\eta_*$ given by Eqn.\ 9 from \cite{ud2008}.

Magnetic confinement depends on the strength of the surface magnetic field, the radius of the star, and the momentum of the stellar wind. The net mass-loss rate is reduced within a magnetosphere \citep{ud2002,2016MNRAS.462.3672B,2017MNRAS.466.1052P}. Magnetic channeling also breaks the spherical symmetry of the wind \citep{ud2002,town2005c,ud2013}, leading to substantial changes in resultant emission line morphology that can introduce considerable inaccuracies in measurements derived from spherically symmetric models (P13). While progress is being made in developing non-spherically symmetric models designed to measure \mdot~\citep[e.g.][]{2018A&A...616A.140H,2019MNRAS.483.2814D}, such measurements are unavailable for the overwhelming majority of the sample. It is furthermore worth emphasizing that $\eta_*$ is defined by the value of \mdot~that the star would be expected to have in the absence of magnetic confinement. We therefore used theoretical mass-loss rates, i.e.\ the mass-loss rate as it is predicted to be in the absence of a magnetic field \citep{ud2002}. We emphasize that this magnetically-unconfined mass-loss rate, corresponding to the actual mass flux into the magnetosphere, is the proper quantity with which to calculate $\eta_*$ \citep{ud2002}.

We adopted two formalisms: the \cite{vink2001} mass-loss recipe, which is a standard prescription for the mass-loss rates of early-type stars, and the \cite{krticka2014} recipe, which is based upon the NLTE calculations performed by \cite{2010A&A...519A..50K}. Unlike the \citeauthor{vink2001} mass-loss rates, which are extrapolated from a relationship calibrated from measurements of the mass-loss rates of O-type stars, the \citeauthor{krticka2014} mass-loss rates were specifically developed for B-type stars. 

The Vink mass-loss rates were calculated using the {\sc idl} program {\sc cal\textunderscore tot.pro}\footnote{Jorick Vink, priv.\ comm.}, which takes as input the stellar metallicity $Z/Z_\odot$, \teff, $\log{L}$, $M_*$, and optionally \vinf. For simplicity, and as chemical abundances are available for only a few stars, $Z/Z_\odot$ was taken to be unity (i.e., solar metallicity) for all stars. \vinf~was left as a free parameter, in which case {\sc cal\textunderscore tot.pro} calculates it from the escape velocity $v_{\rm esc}$ as $v_\infty = f v_{\rm esc}$, where $f$ is a scaling factor that increases from $f=1.3$ to $f=2.6$ from the cool to the hot side of the bi-stability jump, first predicted by \cite{1990A&A...237..409P}, at \teff$\sim25$ kK \citep{vink2001}. The Vink recipe also predicts a decrease of approximately 1 dex in \mdot~from the cool to the hot side of the bi-stability jump, as a consequence of a change in the line force due to recombination of iron lines. While investigation of the UV P Cygni profiles of non-magnetic OB stars has established good evidence that \vinf~does indeed change at the bistability jump (e.g. \citealt{lamers1995}), the change in \mdot~was later found to be more gradual \citep[e.g.][]{2006A&A...446..279C,mark2008a,2017A&A...598A...4K}, and occurs at a slightly lower \teff~\citep{2016MNRAS.458.1999P,2018A&A...619A..54V}. As a consequence we fixed the \teff~for the first and second bistability jumps to 20 kK and 9 kK, respectively.

\citeauthor{krticka2014} mass-loss rates were calculated using Eqn.\ 1 in \cite{krticka2014}, which depends solely upon \teff, and then scaled according to the stellar radius\footnote{In fact, \cite{krticka2014} provided mass-loss rates for a range of surface abundance parameters, as well; since these are not known for the majority of the sample, solar abundances were assumed.}. Wind terminal velocities were determined via interpolation through their Table 2, according to the star's \teff. 

Magnetic spindown timescales $\tau_{\rm J}$ were determined from stellar, wind, and magnetospheric parameters \citep[][Eqn.\ 20]{ud2009}. $\tau_{\rm J}$ is essentially the e-folding timescale for the angular momentum $J$, which corresponds to the surface angular rotational frequency $\Omega$ given by the $J = I\Omega$ in the case of solid body rotation. $\tau_{\rm J}$ is thus sensitive to the star's moment of inertia $I = fM_* R_*^2$, where the factor$f$ is obtained from the radius of gyration $r_{\rm gyr}$ as $f=r_{\rm gyr}^2$. P13 assumed $f=0.1$ for simplicity; however, the stellar structure models calculated by \cite{claret2004} give $f=0.06$ across most of the main sequence for the mass interval of interest. Strictly speaking this quantity is age-dependent, shrinking from about 0.08 at the ZAMS to 0.02 at the TAMS as the He core grows; $f=0.06$ was chosen as the mean value for most models across the MS. Adopting $f=0.6$ rather than $f=0.1$ decreases $\tau_{\rm J}$ by a factor of almost 2. Once $\tau_{\rm J}$ is known, the maximum spindown age $t_{\rm S,max}$ is then defined as the time necessary for the star to spin down from initially critical rotation (rotation parameter $W=1$) to its present value of $W$ (P13, Eqn.\ 14). An important caveat concerning $t_{\rm S,max}$ is that it assumes $\tau_{\rm J}$ to have been constant through time, which is unlikely to be the case, as \mdot, $B_{\rm d}$, and $r_{\rm gyr}$ are all expected to change with time (Keszthelyi et al., in prep.).

\section{Parameter distributions}
\subsection{Stellar Parameters}\label{sec:stellar_par}

   \begin{figure}
   \centering
   \includegraphics[width=8.5cm, trim=50 50 30 30]{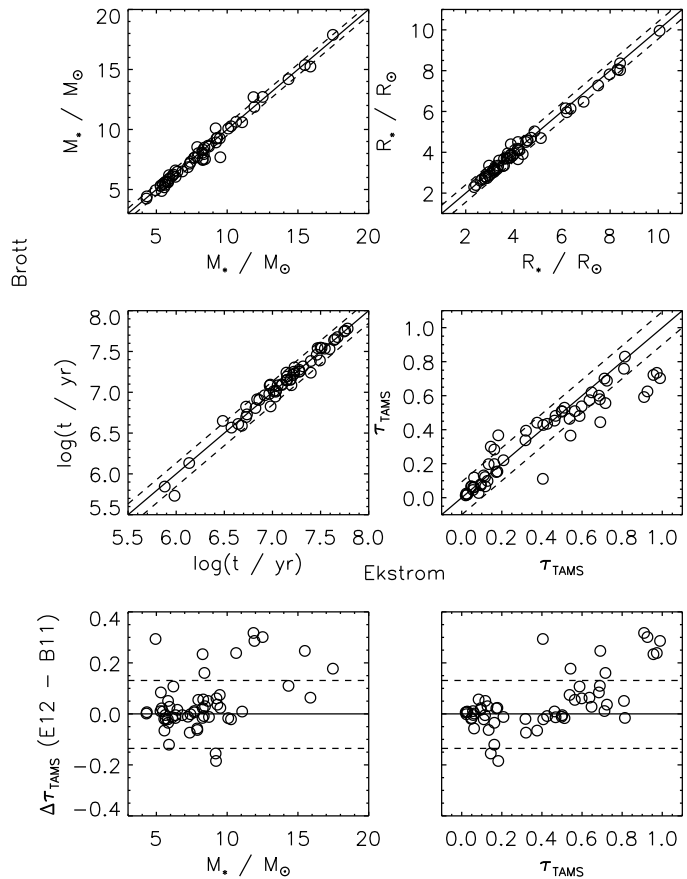}
      \caption[]{{\em Top four panels}: stellar parameters determined using \protect\cite{2011A&A...530A.115B} evolutionary models as a function of parameters determined using \protect\cite{ekstrom2012} models. Solid lines indicate $x=y$; dashed lines show the mean formal uncertainties. {\em Bottom two panels}: the difference in fractional main sequnce age $\tau_{\rm TAMS}$ determined using the two sets of models, as a function of $M_*$ and $\tau_{\rm TAMS}$ (both as determined from the \protect\citeauthor{ekstrom2012} models). Dashed lines show the mean uncertainties.}
         \label{brott_ekstrom}
   \end{figure}

To explore the degree of systematic error introduced by the choice of evolutionary models, the top four panels of Fig.\ \ref{brott_ekstrom} compare the results obtained using the \cite{2011A&A...530A.115B} models to those obtained from the \cite{ekstrom2012} models. Masses, radii, and ages are almost indistinguishable. However, there are several stars for which the \citeauthor{2011A&A...530A.115B} models yield systematically younger fractional main sequence ages $\tau_{\rm TAMS}$ than the \citeauthor{ekstrom2012} models, an expected result given that the \citeauthor{2011A&A...530A.115B} models have a more extended main sequence. Furthermore, using \citeauthor{2011A&A...530A.115B} models, there are no stars with $\tau_{\rm TAMS}$ greater than about 0.8, whereas the \citeauthor{ekstrom2012} models indicate that the entire main sequence is covered. The difference in fractional age $\Delta\tau_{\rm TAMS}$ obtained from the two sets of models is investigated in more detail in the bottom two panels of Fig.\ \ref{brott_ekstrom}. While stars with $M_* > 10$~\msun~almost all have large $\Delta\tau_{\rm TAMS}$, stars showing large discrepancies are found at all masses. $\Delta\tau_{\rm TAMS}$ tends to increase with $\tau_{\rm TAMS}$; since the most massive stars in the sample also tend to be in the second half of the main sequence, their evolutionary status is the primary reason for the discrepancies in MS age amongst this sub-sample. With the exception of these stars, there is a generally good agreement between the values of $\tau_{\rm TAMS}$ returned by the \citeauthor{2011A&A...530A.115B} and \citeauthor{ekstrom2012} models. This may at first seem surprising, since there can be large differences between the two depending on rotation, with the \citeauthor{ekstrom2012} main sequence narrowing with rotation while the \citeauthor{2011A&A...530A.115B} main sequence remains almost unchanged. However, as will be explored in \S~\ref{subsec:rot_evol}, rapid rotators tend to be very young, i.e.\ at an age at which different models return similar results. In the following, $\tau_{\rm TAMS}$ obtained from \citeauthor{ekstrom2012} models are shown; when \citeauthor{2011A&A...530A.115B} values modify the results in an important way, this is discussed in the text.

   \begin{figure}
   \centering
   \includegraphics[width=8.5cm]{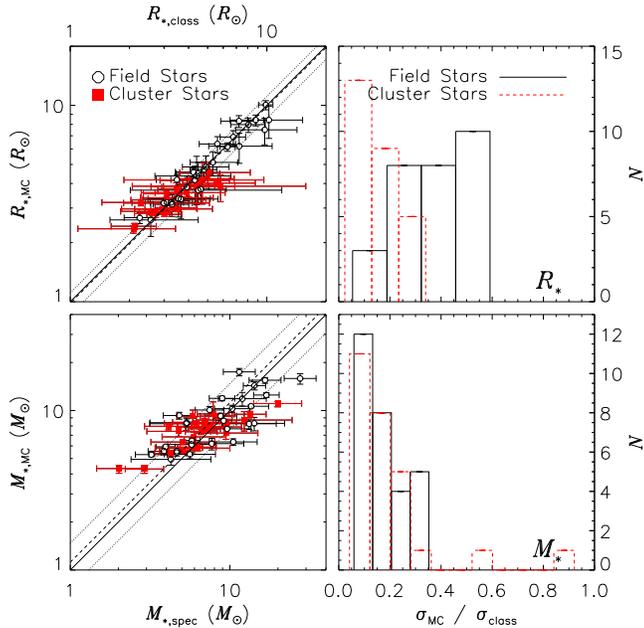}
      \caption[]{{\em Left}: Stellar radius (top) and mass (bottom) determined via the Monte Carlo method described in the text, as functions of the radius and mass calculated directly from the photometric and spectroscopic measurements. Solid lines indicate $x=y$; dashed lines show the median ratio, and dotted lines the standard deviation in the ratio. {\em Right}: histograms of the ratios of derived uncertainties for MC vs.\ classical radii (top) and MC vs.\ spectroscopic masses (bottom).}
         \label{rstar_mstar_mc_spec}
   \end{figure}

Fig.\ \ref{rstar_mstar_mc_spec} (left panels) compares the radii and masses $R_{*,{\rm MC}}$ and $M_{*,{\rm MC}}$ determined via the Monte Carlo (MC) algorithm to values calculated via standard methods. Classical radii $R_{*,{\rm class}}$ were determined using the relation $R_*/R_\odot = \sqrt{(L_*/L_\odot)/(T_{\rm eff}/T_{\rm eff,\odot})^4}$ (where $\log{(L_*/L_\odot)}$ and \teff~were given in Paper II, and $T_{\rm eff,\odot}=5.78$~kK), but without requiring consistency with $\log{g}$ or other constraints. Spectroscopic masses $M_{*,{\rm spec}}$ were determined from $R_{*,{\rm class}}$ and $\log{g}$ (from Paper II) as $M_{*,{\rm spec}}/M_\odot = g(R_{*,{\rm class}}/R_\odot)^2/G$. In both cases uncertainties were determined via error propagation. As is clear from Fig.\ \ref{rstar_mstar_mc_spec}, there is a reasonable agreement between the MC and classical values: the median of the ratio $R_{*,{\rm MC}} / R_{*,{\rm class}}$ is 0.98, and the median of $M_{*,{\rm MC}} / M_{*,{\rm spec}}$ is 1.16, both close to 1. The respective r.m.s. values of these ratios are 0.14 and 0.58, as compared to median relative uncertainties $\sigma_R / R_{*,{\rm class}} = 0.33$ and $\sigma_M / M_{*,{\rm spec}} = 0.32$. As the median difference between classical and MC values is very close to the uncertainty in the classical values, this uncertainty is probably the primary reason for the differences.

The right-hand panels of Fig.\ \ref{rstar_mstar_mc_spec} demonstrate the improvement in precision enabled by the MC method. Field stars typically see an improvement in radius error to about half of the classical uncertainty, while cluster stars are more precise by up to a factor of 10. In both cases the precision in MC mass is up to about 10 times higher than spectroscopic masses. There is no difference in mass precision between cluster and field stars as age constraints have a negligible effect on mass determination.

   \begin{figure}
   \centering
   \includegraphics[width=8.5cm]{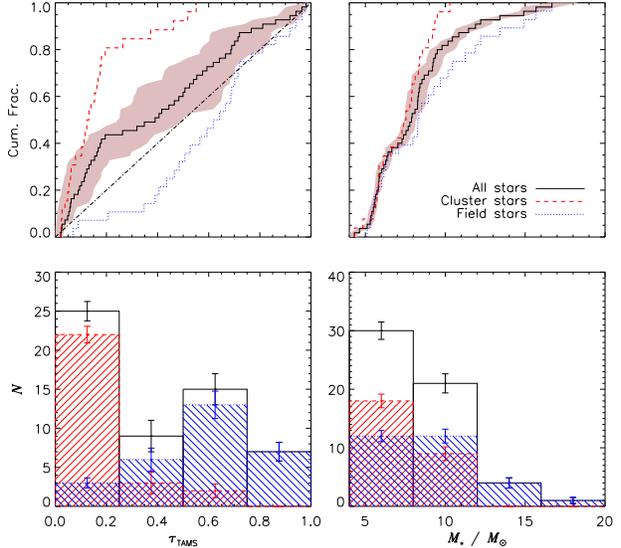}
      \caption[]{Cumulative distributions (top) and histograms (bottom) for fractional main sequence age $\tau_{\rm TAMS}$ (left) and mass $M_*$ (right), with the sample divided into cluster and field stars. Histogram uncertainties were determined from the standard deviation in each bin after generating histograms with identical bin sizes for 1000 synthetic datasets, with the values of each data point varied randomly according to the 1$\sigma$ error bars. Shaded regions around the full cumulative distribution indicate uncertainties. In the top left panel, the diagonal dot-dashed line indicates the expected cumulative distribution if $\tau_{\rm TAMS}$ is evenly distributed between the ZAMS and the TAMS.}
         \label{cluster_field_mass_fract}
   \end{figure}

   \begin{figure}
   \centering
   \includegraphics[width=8.5cm]{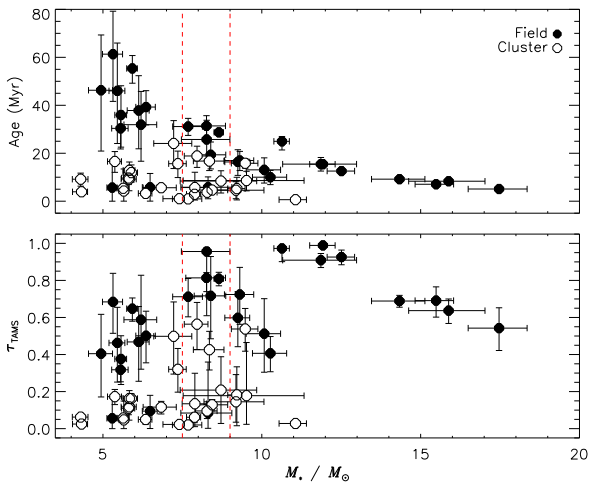}
      \caption[]{Age (top) and fractional main sequence age (bottom) as functions of stellar mass. Note that the young, low-mass sub-sample is dominated by cluster stars. The dashed lines demarcate the mass range within which coverage of the main sequence is relatively complete.}
         \label{mass_age}
   \end{figure}

Fig.\ \ref{cluster_field_mass_fract} shows cumulative distribution functions and histograms for the fractional main sequence age $\tau_{\rm TAMS}$ and the stellar mass, with the sample divided into cluster and field stars. Histogram uncertainties were determined from the standard deviation in each bin after generating histograms with identical bin sizes for 1000 synthetic datasets, with the values of each data point varied randomly according to the 1$\sigma$ error bars. The full sample is slightly biased towards less evolved stars, with a K-S test significance vs.\ an even distribution between the ZAMS and the TAMS of 0.08; however, this is not statistically significant. The age distributions of cluster and field stars are clearly different (K-S test significance of $2\times 10^{-7}$), with the cluster stars being systematically much less evolved than the field stars. This is expected, since young stars should be found close to where they were born. 

The sample is neither volume-limited nor brightness-limited. The apparent magnitude completeness of the MiMeS survey, from which our sample is primarily drawn, drops below 50\% above $V=4$, and is only about 5\% above $V=6$ \citep{2016MNRAS.456....2W}; the median $V$ magnitude of the present sample is 6.6. In spectral type, the MiMeS survey is about 30\% complete at B0 for stars with $V<8$, and drops to 10\% completeness at B5 in the same magnitude range. Given these limitations, the mass distribution of the sample is also of interest; this is shown in the right panels of Fig.\ \ref{cluster_field_mass_fract}. The sample is dominated by less massive stars, with about 50\% of the sample having $M_* < 8~M_\odot$, and a monotonic decrease in incidence with increasing mass. This is more or less as would be expected if our sample was a volume-limited survey, since less massive stars are more common than more massive stars. The distributions of cluster and field stars in mass are similar, with a K-S test significance of 0.06, although there are more lower-mass sample stars in clusters, but no very high mass (above 13 \msun) sample stars in clusters\footnote{In fact, HD\,149438 is in a cluster, but is not considered a `cluster star' for purposes of parameter determination as it is likely a blue straggler; see Appendix B1, online. If it is included as a cluster star, the K-S significance for the mass distributions of cluster vs.\ field stars rises to 0.3.}. Since there are only a handful of stars in this mass range, this is likely a result of small-number statistics (and indeed, some magnetic O-type stars are known cluster members, e.g.\ NGC 1624-2 and $\theta^1$ Ori C).

In Paper II, it was noted that coverage of the main sequence appeared to be incomplete for the less massive stars in the sample (around 5 \msun). Fig.\ \ref{mass_age} shows absolute and fractional ages as functions of stellar mass. Cluster stars dominate the low-mass, young sub-sample. The main sequence appears to be well-covered between 7.5 and 9 \msun. Below this range, there are no stars in the second half of the main sequence, as expected from the results in Paper II. Above this range the sample is dominated by older stars, likely because these stars are much larger and brighter than their younger counterparts, as well as relatively un-obscured by nebular material. The most massive stars - HD 46328, HD 63425, HD 66665, and HD 149438 - are all very close to the same fractional MS age. 

\subsection{Oblique Rotator Models}\label{sec:orm}

\subsubsection{Rotational inclinations and magnetic obliquities}\label{subsec:irot}

As an initial validation of the accuracy of the derived values of $i_{\rm rot}$ and $\beta$ we compared them to one another, and calculated the Pearson's Correlation Coefficient \citep[PCC;][]{1895RSPS...58..240P}, which is close to 0 for uncorrelated values, and close to $\pm$1 for highly correlated values. No correlation of $i_{\rm rot}$ and $\beta$ is expected, and we find PCC~$\sim 0.08$. No correlation of $i_{\rm rot}$ is expected with either $\tau_{\rm TAMS}$, $M_*$, or $W$, and indeed the PCCs of these paramaters with $i_{\rm rot}$ are below 0.1. We also find PCCs below 0.1 for $\beta$ with, $\tau_{\rm TAMS}$, $M_*$, and $W$, indicating that $\beta$ is also independent of mass, age, and rotation within our sample.

   \begin{figure}
   \centering
	\includegraphics[trim = 50 25 50 0,width=8cm]{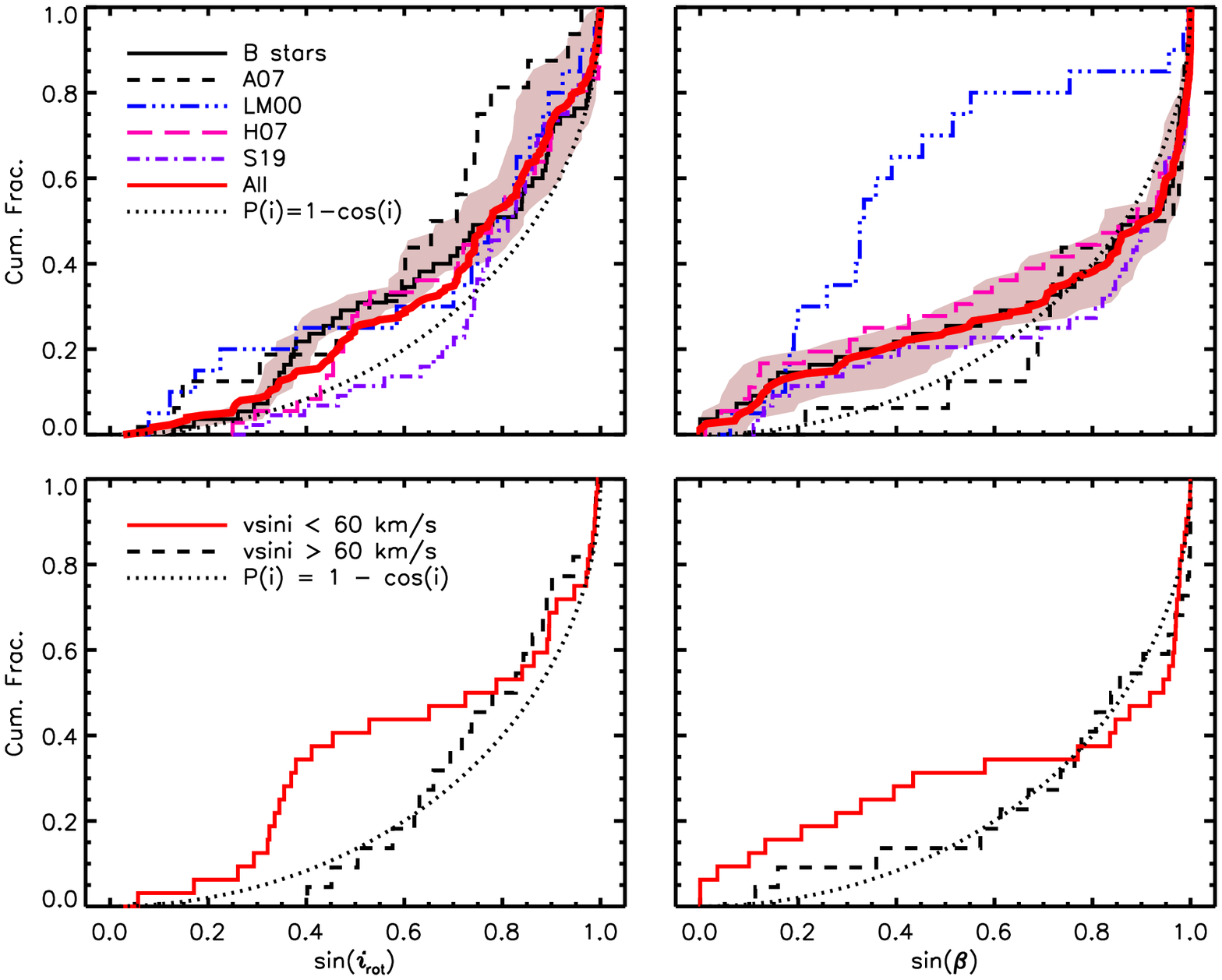}
      \caption[]{{\em Top Left}: Comparision of the cumulative distribution of early B-type star $\sin{i_{\rm rot}}$ angles, to that from previous studies of Ap/Bp stars (\protect\citet[LM00]{landstreetmathys2000}, \protect\citet[A07]{2007AA...475.1053A}, \protect\citet[H07]{2007AN....328..475H}, and \protect\citet[S19]{2019MNRAS.483.3127S}). {\em Top Right}: as before, for obliquity angles $\sin{\beta}$. {\em Bottom Left}: comparison of $\sin{i_{\rm rot}}$ distributions for \vsini~above and below the median value. {\em Bottom Right}: comparison of $\sin{\beta}$ distributions for \vsini~above and below the median value.}
         \label{inc_obl_r}
   \end{figure}

Previous investigations of the distribution of $i_{\rm rot}$ have concluded that it is compatible with a random distribution on a sphere \citep{2001AJ....122.2008A, 2010MNRAS.402.1380J}, the assumption that was used above to set the inclinations of slow rotators for which \vsini~does not provide any constraints (Eqn.\ \ref{prob_i}). The top left panel of Fig.\ \ref{inc_obl_r} compares the inferred values of $\sin{i_{\rm rot}}$ to the expected random distribution. The observed distribution approximately follows a random distribution, but is systematically offset from the expected distribution due to a larger-than-predicted number of stars with $\sin{i_{\rm rot}} \sim 0.5$; above this value, the observed and expected distributions are very similar. The K-S test probability that the predicted and observed distributions are the same is about 0.02.

The left panel of Fig.\ \ref{inc_obl_r} also compares the distribution of inclination angles found here to those reported in previous studies of Ap/Bp stars by \citet[L00]{landstreetmathys2000}, \citet[A07]{2007AA...475.1053A}, \citet[H07]{2007AN....328..475H}, and \citet[S19]{2019MNRAS.483.3127S}. The two-sample K-S test significances comparing the B star inclinations to the previous studies are, respectively, 0.6, 0.1, 0.6, and 0.1, all compatible with belonging to the same population.  Comparison of the distributions from these studies to the expected random distribution yields respective one-sample K-S test significances of 0.2, 0.003, 0.1, and 0.06, suggesting that only the A07 sample is unambiguously inconsistent with a random distribution. Combining all stars, the one-sample K-S significance is $5\times10^{-6}$. 

It should be noted that there is an unavoidable detection bias for $i \sim 0^\circ$ and $\beta \sim 90^\circ$ and vice versa, as in this case \bz~is always 0 for a dipole, and there is an exact symmetry of the line-of-sight component of the magnetic field across the stellar disk and therefore no crossover signature is expected in Stokes $V$. However, the probability of small $i_{\rm rot}$ values is intrinsically very low, and almost any orientation of $\beta$ other than exactly perpendicular (or parallel, if $i=90^\circ$) to the rotation axis would lead to a detectable signature in Stokes $V$. 

The cumulative distribution of $\sin{\beta}$, shown in the top right panel of Fig.\ \ref{inc_obl_r}, suggests an excess of small ($\le40^\circ$) angles; above this value the distribution is consistent with a random distribution. Comparing previous studies to the B-type stars with the two-sample K-S test yields significances between 0.7 and 0.9 for all samples but LM00, which yields $7\times10^{-5}$, i.e.\ only the LM00 sample is clearly drawn from a different population (although it should be kept in mind that LM00 assumed that $\beta$ is always smaller than $i_{\rm rot}$). In all four cases, the one-sample K-S test yields a higher significance level (between 0.1 and 0.2) vs.\ a random distribution than is achieved for $i_{\rm rot}$. The LM00 distribution is clearly not random (the K-S significance is 4$\times10^{-8}$) but is different from the rest, with a very strong excess below about $\beta \sim 30^\circ$. The LM00 distribution is, however, composed largely of very slowly rotating Ap stars, containing more stars with long $P_{\rm rot}$ (of order 10 days to $\sim$~years) than the remaining samples combined, and a principal conclusion of their study is that very slowly rotating Ap stars tend to have aligned magnetic and rotational axes. Combining all stars yields a one-sample K-S significance of $10^{-5}$, but this is mostly an effect of the LM00 sample as, with those stars excluded, the one-sample K-S significance increases to 0.005 i.e.\ much higher than the significance for $i_{\rm rot}$. 

These results are curious, since $i_{\rm rot}$ is expected to be random, but $\beta$ is not, and yet the results formally imply the opposite. Comparing $i_{\rm rot}$ and $\beta$ yields a two-sample K-S probability of 0.02 of belonging to the same distribution; combining all stars (but excluding the LM00 sample) yields $10^{-5}$. Therefore $i_{\rm rot}$ and $\beta$ are probably not both drawn from the same distribution, suggesting that if $i_{\rm rot}$ is in fact random $\beta$ is not (or vice versa). 

One possible explanation for the non-random distribution of $i_{\rm rot}$ is observational bias. Since Zeeman splitting is easier to detect in stars with smaller \vsini, the MiMeS survey preferred stars with narrower line profiles, which in general are likely to have smaller $i_{\rm rot}$ than the overall population of early-type stars. To test this hypothesis we divided the sample into stars with \vsini~above and below the median value of 60 \kms. The $\sin{i_{\rm rot}}$ and $\sin{\beta}$ distributions of the two sub-samples are shown in the bottom panels of Fig.\ \ref{inc_obl_r}. The slowly rotating sub-sample shows a clear excess of stars with small $i_{\rm rot}$, while the rapidly rotating sub-sample is much closer to the expected distribution. This is consistent with the hypothesis that the bias of the MiMeS survey towards more narrow-lined stars is the reason for the excess of small values of $i_{\rm rot}$. This should not have affected the S19 study, which was volume-limited. LM00 did not determine $i_{\rm rot}$ directly from $R_*$ and \vsini~(since the latter quantity was comparable to measurement uncertainties for their long-period stars), but rather assumed that $i_{\rm rot}$ was the larger of the two angles inferred from the \bz~curve properties; their study should therefore also be immune to such a bias. The H07 study was largely based on low-resolution spectropolarimetry, for which line-width is a less relevant issue, and should therefore also not be affected by this bias. Of the 4 previous population studies, only A07 used high-resolution data, for which a bias towards sharp-lined stars might be expected; as noted above, the A07 $i_{\rm rot}$ distribution is the only one of the 4 that is inconsistent with a random distribution. 

The difference in $\sin{\beta}$ distributions between early B-type stars with sharp vs.\ broad lines is not statistically significant, consistent with the expectation that $i_{\rm rot}$ and $\beta$ are independent. 

\subsubsection{Dipolar magnetic field strengths}\label{subsec:bd}

   \begin{figure}
   \centering
   \includegraphics[width=8.5cm]{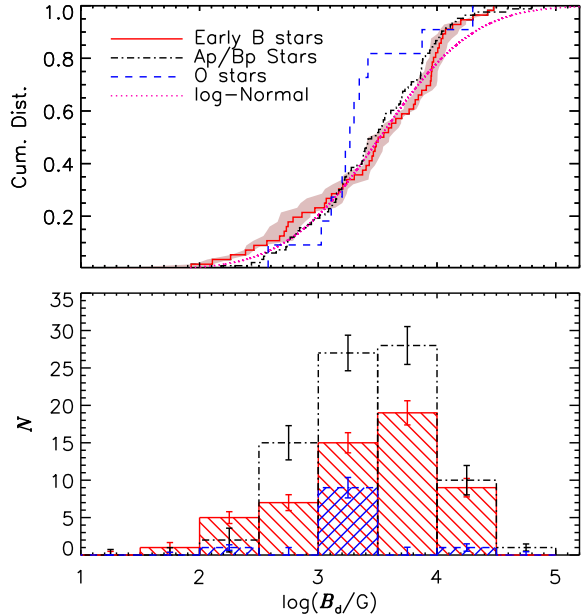}
      \caption[]{Cumulative distributions (top) and histograms (bottom) of the dipole magnetic field strengths $B_{\rm d}$ for early B-type stars, Ap/Bp stars, and magnetic O-type stars. The gray shaded region in the top panel indicates the $1\sigma$ uncertainty.}
         \label{bd_cdf_hist}
   \end{figure}

Surface magnetic field maps determined via ZDI are available for 4 stars in our sample: HD\,37479, HD\,37776, HD\,149438, and HD\,184927 \citep{2015MNRAS.451.2015O, koch2011, 2016A&A...586A..30K, 2015MNRAS.447.1418Y}. Comparing the published ZDI maps to the $B_{\rm d}$ values determined here, only HD\,37776 differs outside uncertainty, with $B_{\rm d} = 6 \pm 0.5$~kG vs.\ a peak surface strength of 30 kG via ZDI. HD\,37776 has by far the most complex magnetic topology of all the stars in this sample, and its \bz~curve is dominated by the second and third harmonics. The next most complex field belongs to HD\,149438, for which ZDI inversion determined a maximal surface magnetic field strength of $\sim 0.6$~kG, with a poloidal contribution to the total magnetic energy of $\sim$10\% \citep{2016A&A...586A..30K}. Our result ($B_{\rm d} = 0.31^{+0.11}_{-0.01}$~kG) under-estimates the surface magnetic field strength by a factor of 2, and over-estimates the poloidal strength by a factor of 5. For both stars, using \bz$_{\rm max}$ from the harmonic fits (i.e.\ $|B_0| + |B_1|$), as supposed to the measured value, would lead to even larger under-estimates of the maximal surface magnetic field strength. At any rate, $B_{\rm d}$ is not especially meaningful for either star. In consequence, when examining the evolution of magnetic field strength and flux (\S~\ref{subsec:mag_evol}), we adopted the ZDI values.

Fig.\ \ref{bd_cdf_hist} shows the cumulative distribution and histogram of the surface magnetic field strength $\log{B_{\rm d}}$ for the stars in this sample. A log-normal fit, generated using the mean and standard deviation of $\log{B_{\rm d}}$ of 3.5 and 0.6 respectively, is shown in the top panel. The K-S significance of the log-normal cumulative distribution compared to the observed distribution is 0.5, suggesting that a log-normal distribution is a reasonable fit to the data. This distribution predicts a tail extending to 100 kG, which is not observed: there is a sharp cutoff in the observed distribution at about 25 kG. This suggests an upper limit on the surface strength of fossil fields in early B-type stars.

The absence of magnetic fields above about 25 kG is likely to be real, as stronger magnetic fields are much easier to detect. The distribution in Fig.\ \ref{bd_cdf_hist} could be affected by a bias related to the intrinsic difficulty of detecting weak magnetic fields. This is particularly the case for rapidly rotating stars, as their large line widths spread Stokes $V$ over a larger number of pixels, bringing the amplitude of the magnetic Zeeman signature closer to the noise level. However, in their analysis of the MiMeS Survey, \cite{2016MNRAS.456....2W} found that the median sensitivity of the survey was around $B_{\rm d}=300$ G. This is about an order of magnitude below the median value of $B_{\rm d}$ found here for the magnetic B-type stars, and only 10\% of the present sample has $B_{\rm d} < 300$~G. This suggests that if a very large fraction of massive stars have $B_{\rm d}$ on the order of several hundred G, they should have been detected by the MiMeS Survey, and that the relatively small number of stars with magnetic fields in this range is probably not a result of observational bias. 

Fig.\ \ref{bd_cdf_hist} also shows the distributions for the magnetic O-type stars and the Ap/Bp stars. We obtained the dipole magnetic field strengths for the O-type stars from P13, with the exceptions of updated values for HD 37743 and HD 108 \citep{2015A&A...582A.110B,2017MNRAS.468.3985S}, and the more recently discovered magnetic O-type star HD\,54879 \citep{2015A&A...581A..81C}. The Ap/Bp star values were taken from LM00, A07, and S19. The K-S significance of the present sample stars vs.\ the Ap/Bp stars is 0.5, indicating they come from the same parent distribution. The early B-type sample includes marginally more stars with weak fields, and a slightly higher fraction of stars with very strong fields, than the sample of cooler Ap/Bp stars. However, this may represent an observational bias: Ap/Bp stars typically possess $\sim$ kG surface magnetic fields, while magnetic fields of the order of 0.1--1 G have been found in normal A-type stars \citep[e.g.][]{2009AA...500L..41L,2010AA...523A..41P,2011AA...532L..13P,2016MNRAS.459L..81B}. Including these stars would likely reconcile the differences between the distributions at the low end, however their inclusion may not be appropriate as it is not clear that such fields have the same origin as classical fossil fields.

The K-S significance for the O-type stars vs.\ the early B-type stars or the Ap/Bp stars is 0.01 and 0.06, respectively, consistent with belonging to the same parent distribution albeit suggestive of a difference. While the extrema of the O-type star $B_{\rm d}$ distribution extend to similar values, the most likely value ($\sim 1 - 3$~kG) is systematically lower than for either the cooler Ap/Bp stars, or the early B-type stars (between 4 and 10 kG). This may reflect an incomplete characterization of the sample of magnetic O-type stars. Many such systems have been found to be spectroscopic binaries, leading to an increase in the inferred surface magnetic field strength \citep[e.g., $\zeta$ Ori Aa and HD 148937;][]{2015A&A...582A.110B, 2019MNRAS.483.2581W}; $B_{\rm d}$ is still a lower limit for the extremely slow rotator HD\,108 \citep{2017MNRAS.468.3985S}; and while magnetic fields have still not been detected in the magellanic cloud Of?p stars \citep{2015AJ....150...99W,2017A&A...601A.136B}, preliminary models of their magnetospheric variability suggest the mean $B_{\rm d}$ of this population to be about 7 kG \citep{2018CoSka..48..149M}, consistent with that of the Galactic magnetic AB stars. 

\subsection{Magnetospheric parameters}\label{sec:magnetosphere}

   \begin{figure}
   \centering
   \includegraphics[trim=25 25 25 25, width=8.5cm]{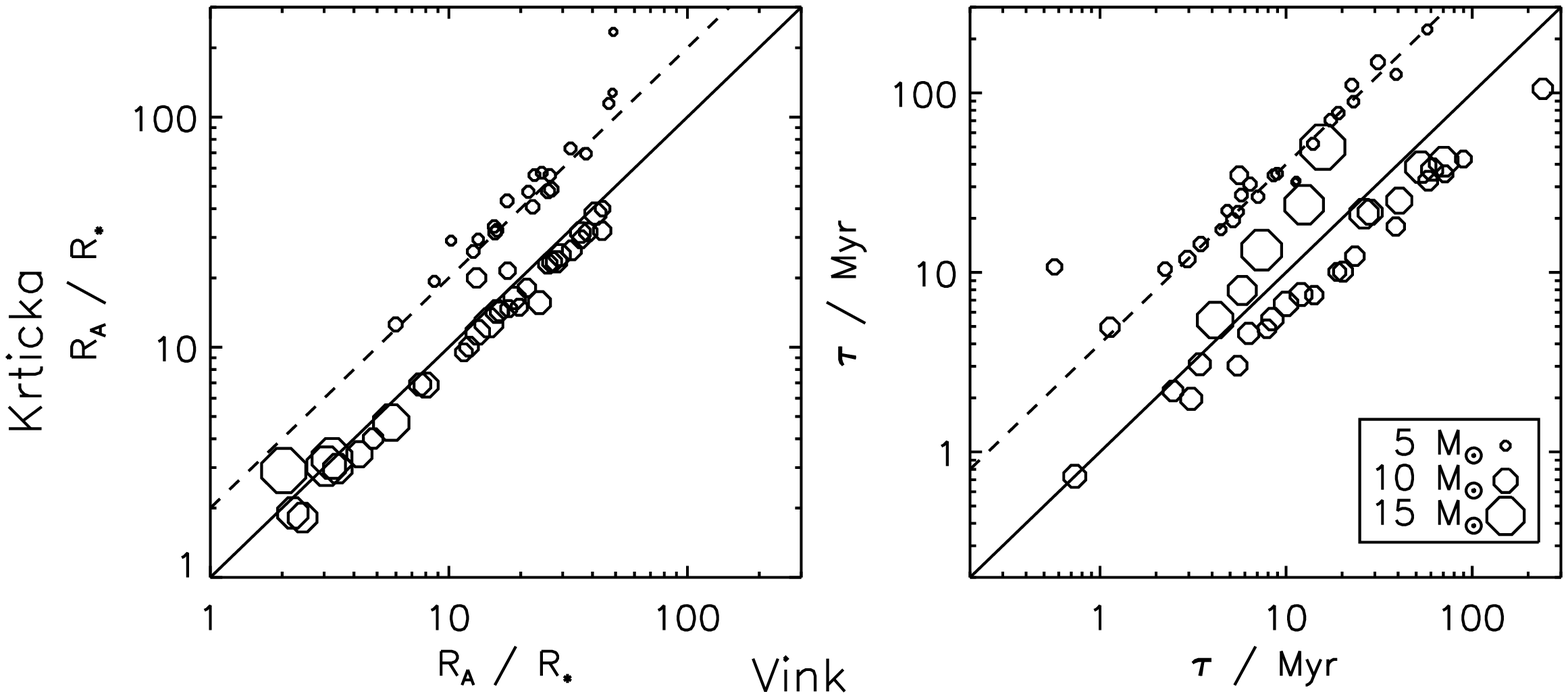}
      \caption[]{Comparison of Alfv\'en radii ({\em left}) and spindown timescales ({\rm right}) derived from \protect\cite{vink2001} and \protect\cite{krticka2014} winds. Solid lines show $x=y$; dashed lines show $x = 0.5y$ (left) and $x=0.25y$ (right).}
         \label{vink_krticka_tauj_ra}
   \end{figure}

Mass-loss rates, wind terminal velocities, the wind magnetic confinement parameter $\eta_*$, and the Alfv\'en radius \ra~are given in Table C3. The tabulated values are those determined using \cite{vink2001} mass-loss. $\eta_*$ is greater than unity for almost all stars in the sample, and ranges up to $10^6$ for some stars, therefore all are predicted to possess magnetospheres. The only possible exception is HD\,52089, which due to its extremely weak magnetic field may have $\eta_* < 1$; however, the lower \cite{krticka2014} \mdot~yields $\eta_* > 1$ for this star. 

Fig.\ \ref{vink_krticka_tauj_ra} compares \ra~and $\tau_{\rm J}$ calculated using the two wind prescriptions. At the high-mass end, there is almost no difference in \ra; towards lower masses, \citeauthor{krticka2014} \ra~diverges from the Vink predictions by a factor of up to about 2 (see dashed line) due to the rapid decline in \citeauthor{krticka2014} \mdot~towards low \teff. This is due to the much greater \teff~sensitivity of \citeauthor{krticka2014} mass-loss, which converges to Vink rates at high \teff~but drops by up to about 2 dex near 15 kK. Amongst the lowest-mass stars, $\tau_{\rm J}$ is generally longer by a factor of about 4 (dashed line) when calculated using \citeauthor{krticka2014} wind parameters, again due to the lower \mdot; intermediate mass stars tend to have similar $\tau_{\rm J}$ since the two prescriptions give similar \mdot~and $v_\infty$ in this mass and \teff~range (i.e.\ just after the bistability jump). 

For the sake of simplicity, we have generally selected parameters derived using Vink mass-loss rates for presentation, as the general distribution of stars in e.g.\ the rotation-magnetic confinement diagram (Fig.\ \ref{ipod}) is not qualitatively different between the two wind prescriptions.

\subsubsection{Revisiting the Rotation-Confinement Diagram}

   \begin{figure}
   \centering
   \includegraphics[trim = 25 25 50 50, width=8cm]{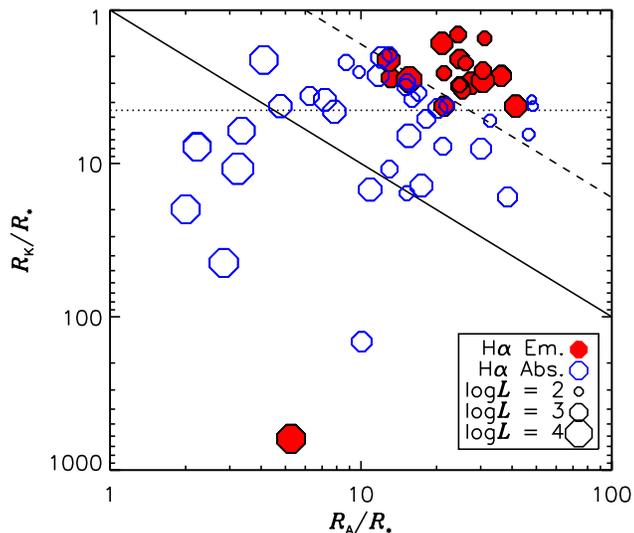}
      \caption[]{The rotation-magnetic confinement diagram (using Vink wind parameters). Symbol size is proportional to the log luminosity $\log{L}$. The solid line divides dynamical magnetospheres (below) from centrifugal magnetospheres (above). The dashed line shows \ra$=6$\rk; the dotted line indicates \rk$=4.5 R_*$: almost all stars above these lines display H$\alpha$ emission.}
         \label{ipod}
   \end{figure}

The rotation-magnetic confinement diagram (RMCD) using the Vink mass-loss recipe is shown in Fig.\ \ref{ipod}, with H$\alpha$-bright stars indicated by filled red symbols. Evaluations of the H$\alpha$ emission status of individual stars were provided by P13 and references therein; those stars for which P13 did not have H$\alpha$ observations have all been subsequently examined, and references are given in the caption of Table C3. The H$\alpha$-bright stars are concentrated in the upper right of the diagram, in the regime of rapid rotation (small \rk) and strong magnetic confinement (large \ra). The only exception to this rule is HD\,46328, which is located in the bottom left of the diagram. This star's H$\alpha$ emission properties are consistent with an origin in a dynamical magnetosphere \citep{2017MNRAS.471.2286S}, in contrast to the stars in the upper right which all display double-humped emission profiles consistent with an origin in a centrifugal magnetosphere. 

It is clear from the RMCD that merely possessing a CM is an insufficient condition for H$\alpha$ emission. The diagonal solid line indicates \ra$=$\rk: stars above this line are predicted to possess a CM, and many of these have H$\alpha$ in absorption. It is typically the stars with \rark~$>0.6$ that are H$\alpha$ bright (diagonal dashed line). The majority of the emission-line stars also have \rk~$<4.5~R_*$ (horizontal dotted line). The Kepler radius sets the minimum distance of the plasma accumulation surface from the star \citep{town2005c}. Taken together, these results suggest that there are at least two conditions necessary for H$\alpha$ emission: 1) \rk~must be close to the star; 2) \ra~must be sufficiently large to enforce strict corotation of the plasma out to a much greater distance than \rk. 

The thresholds for H$\alpha$ emission were set in order to exclude as many stars without H$\alpha$ emission as possible, however there is some overlap at the borders between stars with and without emission. The absorption-line stars in the region of overlap  are typically less luminous than the emission-line stars. This suggests that there may be a third condition for emission: that the wind feeding rate be above a certain threshold. If true, this is suggestive of a plasma leakage mechanism competing with mass loading, a possibility suggested by P13 and developed further by \cite{2018MNRAS.474.3090O}. This is because, in a breakout scenario, CMs should eventually fill regardless of \mdot~(P13). Since the wind plasma feeding the magnetosphere is spread out over a larger surface area with increasing distance from the star, the requirement that \rk~be close to the star for H$\alpha$ to be detectable may also point to the operation of a leakage mechanism. 

\section{Evolution}\label{sec:evol}

\subsection{Magnetic Field Evolution}\label{subsec:mag_evol}

   \begin{figure*}
   \centering
\includegraphics[trim = 20 0 10 20, width=18 cm]{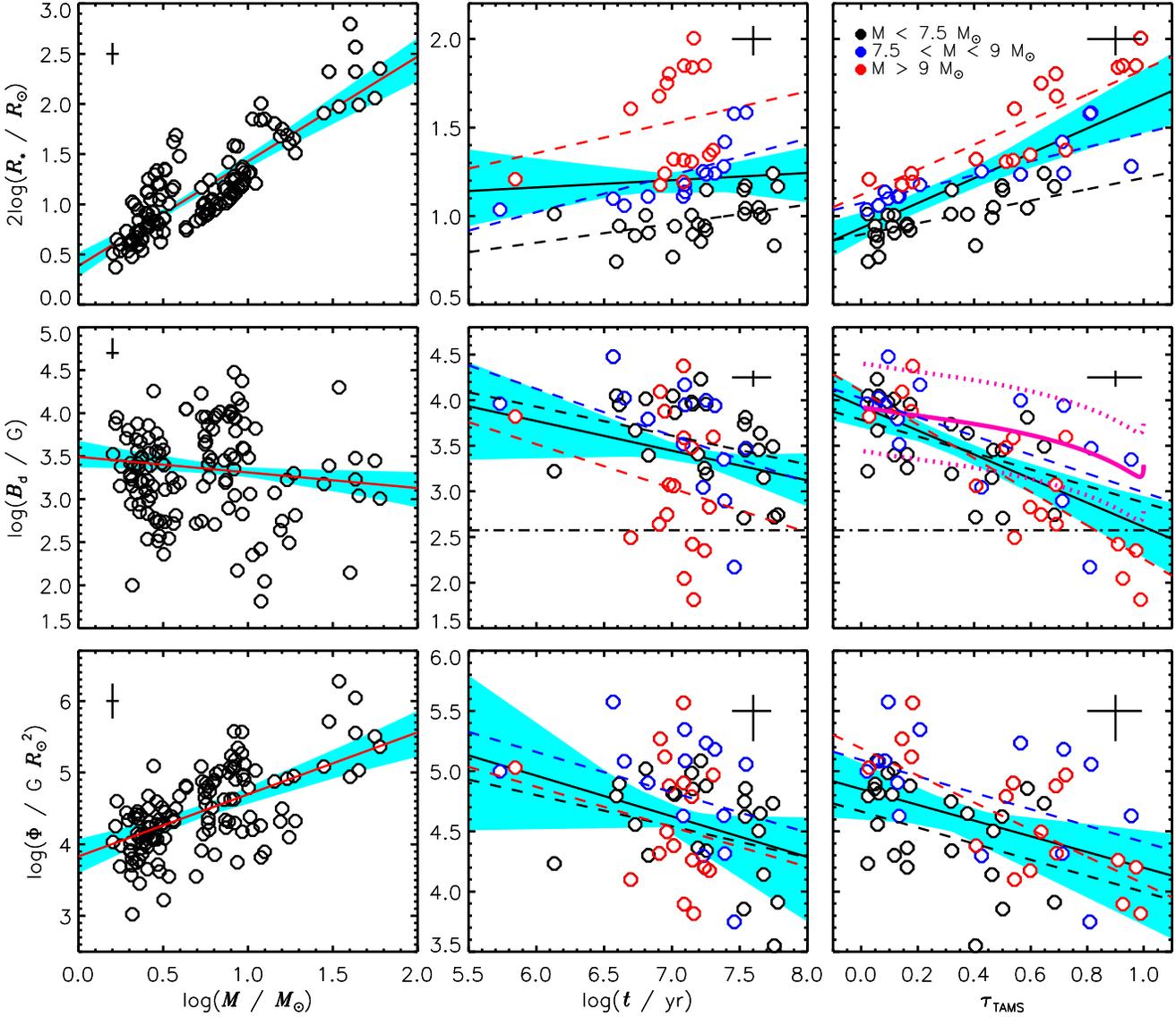}
      \caption[]{{\em Left panels}: Log radius (top), log surface dipole magnetic field strength (middle), and log unsigned magnetic flux (bottom) as functions of stellar mass. Solid lines and shaded regions indicate linear fits and the maximal (approximately $3\sigma$) uncertainties. Parameters for A and O-type stars were obtained from A07, S19, and P13. Crosses in the top left corners indicate mean error bars. {\em Middle and right panels}: log radius (top), log surface dipolar magnetic field strength (middle), and log unsigned magnetic flux (bottom) as functions of stellar age (left) and fractional main sequence age (right). Symbol colour indicates mass. Mean error bars are indicated by crosses in the top right corners. Solid lines and shaded regions indicate linear fits and uncertainties when the whole dataset is considered. Dashed lines indicate fits to mass bins of corresponding colour. Dot-dashed horizontal lines in the middle panels indicate the median MiMeS survey dipole sensitivity \protect\citep{2016MNRAS.456....2W}. In the middle right panel, the thick purple curve shows an expected evolution of $B_{\rm d}$ assuming flux conservation (see text).}
         \label{bd_fract_age}
   \end{figure*}

\begin{table}
\caption[]{Linear regression slopes (top numbers) and $p$ values (bottom numbers) for magnetic field evolution for the full sample, and for various sub-samples. Slopes in boldface indicates both a slope significance of at least $3\sigma$ and a $p$ value below 0.05.}
\label{fluxtab}
%\resizebox{8.5 cm}{!}{
\begin{tabular}{l | r r }
\hline\hline
Sample                                   & $\log{t}$ vs.\ $\log{B_{\rm d}}$ & $\log{t}$ vs.\ $\log{\Phi}$ \\
                                         & $\tau_{\rm TAMS}$ vs.\ $\log{B_{\rm d}}$ & $\tau_{\rm TAMS}$ vs.\ $\log{\Phi}$ \\
\hline
All stars                                & $\mathbf{-0.38 \pm 0.09}$   & $\mathbf{-0.38 \pm 0.12}$ \\
                                         & 0.02                      & 0.008    \\       
                                         & $\mathbf{-1.3 \pm 0.1}$          & $\mathbf{-0.66 \pm 0.16}$ \\
                                         & 0.000                              & 0.0005               \\
Cluster stars                            & $-0.13 \pm 0.11$    & $-0.12 \pm 0.12$    \\
                                         & 0.16                        & 0.29         \\
                                           & $-0.49 \pm 0.29$          & $-0.07 \pm 0.38$  \\
                                         & 0.12                                & 0.85               \\
Field stars                              & $-0.14 \pm 0.20$        & $-0.27 \pm 0.14$     \\
                                         & 0.74                       & 0.12                  \\
                                         & $\mathbf{-1.4 \pm 0.3}$            & $-0.59 \pm 0.29$  \\
                                         & 0.0001                             & 0.06            \\
$M_* < 7.5 M_\odot$                      & $\mathbf{-0.34 \pm 0.16}$   & $-0.3 \pm 0.2$  \\
                                         & 0.04                       & 0.11          \\
                                         & $\mathbf{-0.9 \pm 0.3}$          & $\mathbf{-0.7 \pm 0.2}$ \\
                                         & 0.002                              & 0.03                \\
$7.5 \le M_* \le 9 M_\odot$              & $-0.6 \pm 0.25$        & $-0.4  \pm 0.2$      \\
                                         & 0.03                       & 0.1       \\
                                         & $\mathbf{-1.0 \pm 0.3}$           & $-0.7 \pm 0.3$ \\
                                         & 0.02                               & 0.1                  \\
$M_* > 9 M_\odot$                        & $-0.5  \pm 0.4$            & $-0.4  \pm 0.3$ \\
                                         & 0.24                       & 0.3           \\
                                         & $\mathbf{-1.8 \pm 0.3}$          & $\mathbf{-1.1 \pm 0.3}$ \\
                                         & $0.00001$                   & 0.0003                    \\
\hline\hline
\end{tabular}
%}
\end{table}

\subsubsection{Magnetic fields and mass}

The middle left panel of Fig.\ \ref{bd_fract_age} shows $B_{\rm d}$ as a function of stellar mass for the Ap/Bp stars, the early B-type stars, and the magnetic O-type stars. Stellar parameters for the A-type stars were obtained from A07 and S19, while those of the O-type stars are from the same sources as were used for Fig.\ \ref{bd_cdf_hist}. The upper and lower extrema of $B_{\rm d}$ are basically independent of mass. 

We tested for correlations using linear regressions, PCCs, and $p$ values, where the latter give the probability that the PCC is not a result of random statistical variation ($p \ll 1$ indicates a high probability that there is a real variation). Uncertainties in linear regressions and PCCs were derived using a Monte Carlo methodology intended to control for both random and systematic errors. 1000 synthetic datasets were generated. In each synthetic dataset, one point selected at random was removed; this tested for statistical robustness against the influence of individual datapoints. Another randomly selected point was shifted in both the x and y direction, using a flat distribution normalized to the span of the full dataset; this tested for the influence of large systematic errors in the parameters of individual datapoints. The remaining datapoints were allowed to vary within Gaussian distributions normalized to the standard error of each datapoint. The solid lines in Fig.\ \ref{bd_fract_age} indicate the mean best fit across all synthetic datasets; the light blue shaded regions show the minimum and maximum fits, approximately corresponding to $3\sigma$ uncertainties. 

While a linear fit to $\log{B_{\rm d}}$~vs.\ $\log{M_*}$ suggests a slight downward trend, as might be expected from Fig.\ \ref{bd_cdf_hist}, the slope is much less than the range of values found at any given mass, the PCC is only $-0.09 \pm 0.06$, and the $p$ value is 0.51. There is therefore no evidence from this exercise for a change in $B_{\rm d}$ with mass.

Since stellar radius increases with mass, an essentially unchanging range of $B_{\rm d}$ means that the unsigned magnetic flux $\Phi = B_{\rm d}R_*^2$ must increase with mass. This is demonstrated in the bottom left panel of Fig.\ \ref{bd_fract_age}. The PCC of $\log{\Phi}$ vs.\ $\log{M_*}$ is $0.47 \pm 0.05$ ($p = 0$), and the slope is $0.91 \pm 0.07$, indicative of a significant positive relationship. While many stars with very low magnetic flux could easily remain undetected, the increase in flux with stellar mass is probably real: for an A-type star to have the same magnetic flux as the ultra-strongly magnetized O-type star NGC 1624-2 ($\sim 10^{6.4}~{\rm G}~R_\odot^2$; \citealt{wade2012b}), it would need to possess a surface magnetic field of $\sim 400$~kG; meanwhile the most strongly magnetic Ap star, Babcock's star, has a dipole strength of 34 kG \citep{1960ApJ...132..521B}, an order of magnitude smaller. The top left panel of Fig.\ \ref{bd_fract_age} shows $2\log{R_*}$ vs.\ $\log{M_*}$; the slope of this correlation, $1.04 \pm 0.05$, is almost identical within uncertainty to the slope of $\log{\Phi}$ vs.\ $\log{M_*}$, verifying that the correlation in the latter is purely a function of the former. 

\subsubsection{Magnetic fields and time}

Assuming that fossil fields are intrinsically static on the stellar evolutionary time-scales, their surface strength should weaken over time as a star expands, with the surface magnetic field declining at a minimum as $\sim R_*^{-2}$ due to conservation of magnetic flux. While such an effect is not expected to be detectable in individual stars, over a large enough sample it would be expected that $B_{\rm d}$ should be systematically lower amongst older stars. $\log{B_{\rm d}}$ is shown as a function of $\log{t}$ and $\tau_{\rm TAMS}$ in the middle row, middle and right panels of Fig.\ \ref{bd_fract_age}. The PCCs with respect to $\log{t}$ and $\tau_{\rm TAMS}$ are respectively $-0.27 \pm 0.05$ and $-0.66 \pm 0.05$; the slopes (given in Table \ref{fluxtab}) are also negative and significant at the 3$\sigma$ level. The $p$ values of these relationships (given in Table \ref{fluxtab}), which give the probability that the correlations are artifacts produced by random sampling of an underlying random distribution, are likewise well below the usual threshold of 0.05 for statistical significance. 

As a comparison to the decline in $B_{\rm d}$ expected due to flux conservation, the middle right panel of Fig.\ \ref{bd_fract_age} shows the evolution of a 9 \msun~\cite{ekstrom2012} evolutionary model, starting from the median $B_{\rm d}=8$~kG close to the ZAMS (i.e.\ for $\tau_{\rm TAMS} < 0.2$). The relative change in radius across the MS varies insignificantly within the mass range of interest here, and does not differ between rotating and non-rotating models. The model predicts a very slow decrease in $B_{\rm d}$ over the first half of the MS, accelerating towards the end, with a net change of about $-0.7$ dex. For comparison, the least-squares fit to the full dataset yields a net change of $-1.4$ dex. This suggests that flux is not conserved. 

The upper middle and right panels of Fig.\ \ref{bd_fract_age} show $2\log{R_*}$ vs.\ $\log{t}$ and $t_{\rm TAMS}$. Radius increases with time, although the fit to $2\log{R_*}$ for the full sample vs.\ $\log{t}$ is not particularly meaningful due to the dependence of radius on mass. 

%

%, with an apparently somewhat steeper relationship in the high-mass bin; this is a consequence of the relatively wide mass range of this sub-sample

A less model-dependent test of flux change is illustrated in the bottom middle and right panels of Fig.\ \ref{bd_fract_age}, which show $\log{\Phi}$ as a function of $\log{t}$ and $\tau_{\rm TAMS}$. The PCCs are respectively $-0.30 \pm 0.07$ and $-0.38 \pm 0.09$, with respective slopes of $-0.38 \pm 0.12$ and $-0.66 \pm 0.16$. Once again the $p$ values (Table \ref{fluxtab}) are close to 0. 

To check the robustness of these results, we divided the sample into different sub-samples. Considering cluster stars alone, the PCC of $\tau_{\rm TAMS}$ vs.\ $\log{B_{\rm d}}$ is $-0.23 \pm 0.12$; $\log{t}$ vs.\ $\log{B_{\rm d}}$ is $-0.19 \pm 0.13$; $\tau_{\rm TAMS}$ vs.\ $\log{\Phi}$ gives $-0.03 \pm 0.14$; and $\log{t}$ vs.\ $\log{\Phi}$ gives $-0.15 \pm 0.14$. Slopes and $p$ values are given in Table \ref{fluxtab}, and are of similarly low statistical significance. Within this sub-sample there is no compelling evidence of a decay of magnetic flux. The results are broadly similar for field stars, with respective PCCs of $-0.57 \pm 0.09$, $-0.17 \pm 0.15$, $-0.29 \pm 0.13$, and $-0.27 \pm 0.10$. Within the field star sub-sample, only $\tau$ vs.\ $\log{B_{\rm d}}$ yields statistically significant results (see Table \ref{fluxtab}). However, the cluster and field star sub-samples are heavily biased towards young and old stars, respectively (Fig.\ \ref{cluster_field_mass_fract}), which make them individually limited tests of evolutionary change. These sub-samples furthermore group together stars with very different masses and, hence, very different evolutionary timescales.

We also divided the sample into three different mass bins: low-mass ($M_* < 7.5$~\msun), mid-mass ($7.5 < M_* < 9$~\msun), and high-mass ($M_* > 9$~\msun). These were chosen based on the identification of the mid-mass sample as containing the most complete sampling of the main sequence (Fig.\ \ref{mass_age}). Linear fits to the three sub-samples are indicated with dashed lines in Fig.\ \ref{bd_fract_age}, where the colour corresponds to the stellar mass. All three sub-samples show uniformly negative trends in both $\log{B_{\rm d}}$ and $\log{\Phi}$. Unsurprisingly given the reduced sample size, these trends generally fall below the threshold for statistical significance (Table \ref{fluxtab}). The exceptions are: the low-mass sample for $\log{B_{\rm d}}$ vs.\ $\log{t}$; all sub-samples for $\log{B_{\rm d}}$ vs.\ $\tau_{\rm TAMS}$; and the high-mass sample for $\log{\Phi}$ vs.\ $\tau_{\rm TAMS}$. In all cases the low- and mid-mass samples have similar slopes, while the high-mass sample exhibits a steeper slope than either. For $\log{t}$ this might plausibly be ascribed to the much shorter main sequence lifetimes of more massive stars; this should not affect fractional age. 

Turning to radius, the slopes of fits to the individual mass bins relative to $\log{t}$ are between 0.15 and 0.3, and the slopes relative to $\tau_{\rm TAMS}$ range between 0.4 and 0.7. All are less steep than the corresponding slope relative to $B_{\rm d}$. This is further indication that the decline in $B_{\rm d}$ over time is too rapid to be explained as purely a function of flux conservation.

One cause for concern in the interpretation of these results is that the mid-mass sample -- which has the best sampling of the main sequence, and also the narrowest range in mass -- yields the {\em lowest} level of statistical significance of the three mass bins. This is probably because it is the smallest of the 3 samples (14 stars, as compared to 25 in the low-mass sample and 17 in the high-mass sample). The mid-mass bin is examined in more detail below in \S~\ref{subsec:flux_field_strength}.

If $\tau_{\rm TAMS}$ values from \cite{2011A&A...530A.115B} models are used instead, the slopes in Table \ref{fluxtab} are not changed outside the error bars. The more compressed main sequence means that the slopes with $\tau_{\rm TAMS}$ are a bit steeper, particularly in the case of the high-mass stars, but also leads to larger uncertainties and higher $p$ values; however, the significance of the majority of the slopes in Table \ref{fluxtab} remains unchanged.

We conclude that the inferred decrease in magnetic flux with time appears to be robust against random and systematic error (already accounted for in the shaded regions in Fig.\ \ref{bd_fract_age}), and against sub-division of the sample. This confirms the basic results of \cite{land2007,land2008}, extended to higher mass stars. There is also evidence that flux decreases more rapidly for more massive stars, as suggested by \cite{2016A&A...592A..84F}; this question will be considered in more detail in \S~\ref{subsec:flux_decay_with_mass}. 

\subsection{Rotational Evolution}\label{subsec:rot_evol}

   \begin{figure}
   \centering
   \includegraphics[trim = 0 25 0 25, width=8.5cm]{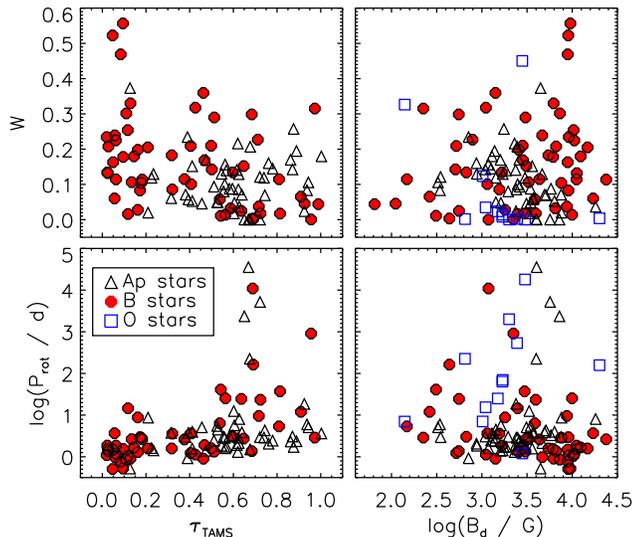}
      \caption[]{Rotation parameter $W$ ({\em top}) and logarithmic rotation period ({\rm bottom}) as functions of fractional main sequence age $\tau_{\rm TAMS}$ ({\em left}) and $\log{B_{\rm d}}$ ({\em right}). $\tau_{\rm TAMS}$ is not known for the magnetic O-type stars.}
         \label{prot_fract}
   \end{figure}

Magnetic braking is expected to lead to rapid spindown. Given this, we expect that the most rapidly rotating stars should be systematically younger than the slowest rotators. Fig.\ \ref{prot_fract} demonstrates that this basic pattern appears to hold. The left panels show the rotation parameter $W$ (which emphasizes rapid rotators) and $\log{P_{\rm rot}}$ (emphasizing slow rotators) as functions of $\tau_{\rm TAMS}$. While there is a spread of rotational properties at any given evolutionary stage, extremely slow rotators ($\log{P_{\rm rot}} > 3$) are all in the second half of the main sequence, while in the first half of the main sequence there are no stars with $\log{P_{\rm rot}}$ much greater than 1. Conversely, the most rapidly rotating stars ($W > 0.4$) are quite young, and the majority of the stars with fractional ages $\tau_{\rm TAMS} > 0.5$ also have $W < 0.2$.  The basic trend of dramatically slowing rotation with increasing $\tau_{\rm TAMS}$ is unchanged if values from \cite{2011A&A...530A.115B} models are used. As can be seen in the top left panel, stars at the ZAMS have rotation parameters $W$ spanning almost the entire range between 0 and 0.5; the presence of relatively slow rotators at the ZAMS could be consistent with extremely slow rotators having evolved from such stars (but see below).

Magnetic braking should also be more rapid with increasing $\dot{M}$ and increasing $B_{\rm d}$. In Paper I we showed that O-type stars do indeed have significantly longer rotational periods than magnetic B-type stars, as expected (P13), whereas magnetic A-type stars and B-type stars have similar distributions of $P_{\rm rot}$. The relation of $B_{\rm d}$ to rotation is shown in the right-hand panels of Fig.\ \ref{prot_fract}. We see that the most rapidly rotating stars have some of the strongest magnetic fields, while the most slowly rotating stars have surface fields of intermediate strength. While at first this might seem counterintuitive, this makes sense if one recalls the evidence that magnetic fields weaken with time (Fig.\ \ref{bd_fract_age}): since very strong magnetic fields are seen only in young stars, these will also tend to be rapid rotators. The stars that experience the strongest spindown will tend to be those with the initially strongest magnetic fields, but over the time that it takes for their rotation to become exceptionally slow, their magnetic fields will weaken. 

There are very few stars with very weak magnetic fields and either very rapid or very slow rotation. The former may well be an observational bias, since Zeeman signatures are inherently harder to detect in broad-lined stars. It may also be due to destabilization of fossil fields below a critical field strength in rapid rotators, as suggested by \cite{2007AA...475.1053A}. Such a bias does not affect slow rotators; the absence of extremely slowly rotating, very weakly magnetized stars is naturally explained by the dependence of the spindown timescale on $B_{\rm d}$. 

In Fig.\ \ref{prot_fract} we also compare these results to those for magnetic A-type stars (from \cite{2019MNRAS.483.2300S,2019MNRAS.483.3127S}, who provided masses, radii, and fractional ages) and magnetic O-type stars (from P13, with modifications as described in \S~\ref{subsec:bd}). Both A-type stars and O-type stars show rotational behaviour that is qualitatively similar to that of the B-type stars. While little can be said about the young, rapidly rotating A-type stars (since the \citeauthor{2019MNRAS.483.2300S} sample contains few very young stars), slowly rotating A-type stars are all old. The most slowly rotating A- and O-type stars are not the most strongly magnetic, but rather those with intermediate surface magnetic fields; conversely, the most rapid rotators are relatively strongly magnetized. 

A similar distribution of $B_{\rm d}$ and $P_{\rm rot}$ was noted by \cite{2011Ap.....54..231G}, who suggested it to be a consequence of dynamo activity in pre-stellar cores which peaks in efficiency at a medium cloud motion velocity; the increase in $P_{\rm rot}$ over time makes this hypothesis unlikely, since in this case slow rotation would necessarily be primordial.  

   \begin{figure}
   \centering
   \includegraphics[width=8.5cm]{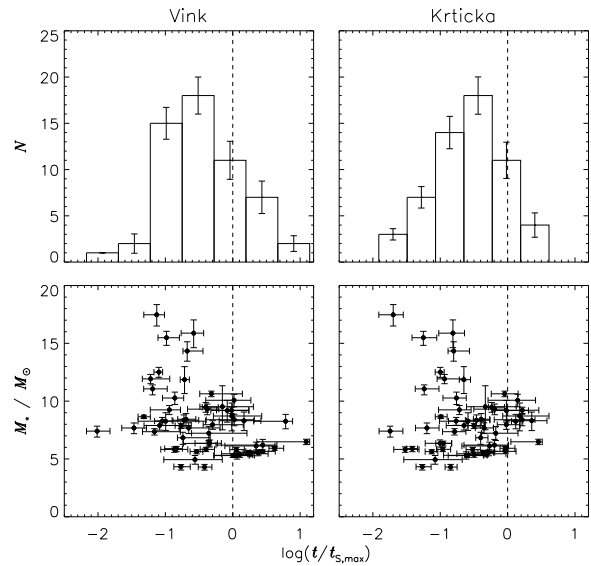}
      \caption[]{{\em Top}: histograms of the logarithmic ratio of age to $t_{\rm S,max}$ for Vink (left) and \protect\citeauthor{krticka2014} (right) mass-loss rates. {\em Bottom}: $\log{(t/t_{\rm S,max})}$ as a function of stellar mass.}
         \label{age_tsmax}
   \end{figure}

The qualitative picture in Fig.\ \ref{prot_fract} does not tell us whether $P_{\rm rot}$ is precisely as expected given the properties of the stars' magnetic fields and stellar winds. Fig.\ \ref{age_tsmax} shows the logarithmic ratio of the stellar age $t$ to the maximum spindown age $t_{\rm S,max}$ for mass-loss rates calculated using the \cite{vink2001} and \cite{krticka2014} prescriptions. Both mass-loss prescriptions result in differences of up to about 2 dex. The Vink mass-loss rates result in ratios that are approximately normally distributed about $\log{(t/t_{\rm S,max})} = 0$, with a standard deviation of 0.75 dex, while the \citeauthor{krticka2014} mass-loss rates result in a systematic shift to $\log{(t/t_{\rm S,max})} = -0.6 \pm 0.6$, i.e.\ the \citeauthor{krticka2014} mass-loss rates generally yield $t < t_{\rm S,max}$. In principle, this is acceptable, however in some cases these differences are very large. The bottom panels of Fig.\ \ref{age_tsmax} demonstrate that differences between $t$ and $t_{\rm S,max}$ are not random. In both cases, the most massive stars ($M_* > 11 M_\odot$) have $t \ll t_{\rm S,max}$. The Vink mass-loss rates also yield $t \gg t_{\rm S,max}$ for less massive stars ($M_* < 11 M_\odot$), while the \citeauthor{krticka2014} mass-loss rates largely resolve, or at least reduce, this discrepancy. Thus, \citeauthor{krticka2014} mass-loss rates yield spindown ages that are more consistent with the ages determined from isochrones for less massive stars. Neither prescription is able to resolve the large differences obtained for the most massive stars. One possible, obvious interpretation of the result that $t \ll t_{\rm S,max}$ for so many stars in the sample is that these stars may have arrived at the ZAMS already rotating slowly. This may seem plausible given that there are some stars at the ZAMS with $W$ already close to 0. However, in the case of HD\,46328, the most slowly rotating star in the sample, this scenario would require that an initial rotation parameter $\log{W_0} = -3.2$, corresponding to $P_{\rm rot} \sim 10^3$~d, about two dex higher than is observed for any young star. Alternately, this difference may be entirely illusory, as these $t_{\rm S,max}$ values assume a constant $\tau_{\rm J}$.

\subsection{Magnetospheric evolution: the life story of a typical magnetic B-type star}\label{subsec:magnetosphere_evol}

Having explored the evolution of the magnetic and rotational properties of magnetic early B-type stars, we now assemble these components to provide an outline of magnetospheric evolution. We have seen that stars with H$\alpha$ emission are both more rapidly rotating and have more strongly magnetically confined winds than stars without emission, consistent with an origin of their emission in CMs. Rotational periods increase with fractional age, an expected result of magnetic braking. Since magnetic braking should efficiently spin down the most rapidly rotating stars, stars exhibiting H$\alpha$ emission formed in CMs are expected to be young. 

   \begin{figure}
   \centering
   \includegraphics[width=8.5cm]{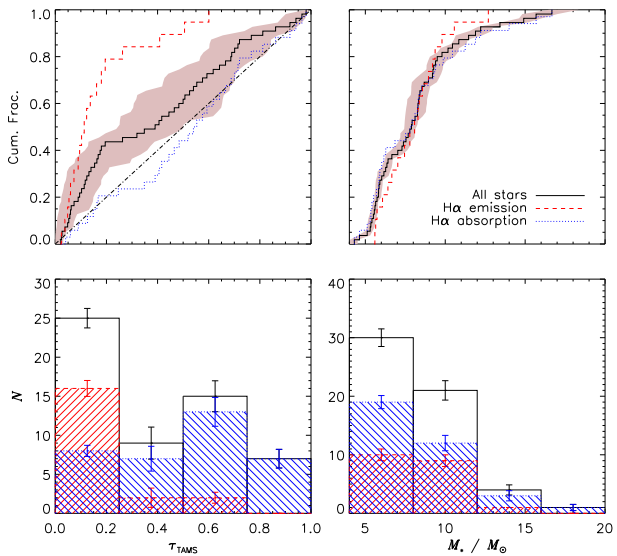}
      \caption[]{As. Fig.\ \ref{cluster_field_mass_fract}, this time with the sample divided into stars with H$\alpha$ in emission and absorption.}
         \label{emission_absorption_mass_fract}
   \end{figure}

Fig.\ \ref{emission_absorption_mass_fract} repeats the analysis of Fig.\ \ref{cluster_field_mass_fract}, this time dividing the sample into stars with (17 stars) and without (37 stars) H$\alpha$ emission. Stars with H$\alpha$ in absorption approximately follow an even distribution in $\tau_{\rm TAMS}$ (K-S significance of 0.46), and the distribution in mass between the two sub-samples is nearly identical (K-S significance of 0.71). However, the emission-line stars are systematically younger than absorption-line stars (two-sample K-S significance of $7\times 10^{-6}$). Indeed, while only about 1/3 of the overall sample displays H$\alpha$ emission, amongst the youngest cohort ($\tau_{\rm TAMS} < 0.25$), emission-line stars are a majority (15 in emission vs.\ 7 stars in absorption or 68\% in emission), while for stars older than $\tau_{\rm TAMS} = 0.25$ absorption-line stars are in the great majority (4 in emission vs.\ 27 stars in absorption or 87\% in absorption). None of the stars older than $\tau_{\rm TAMS}=0.75$ are in emission. It is worth noting that the most evolved H$\alpha$ emission star in the sample, HD 46328 ($\tau_{\rm TAMS} = 0.77 \pm 0.05$), is actually an extremely slow rotator ($P_{\rm rot} > 30~{\rm yr}$) with an emission morphology consistent with an origin in a dynamical magnetosphere \citep{2017MNRAS.471.2286S}, and its emission is thus unrelated to rotation. Removing HD\,46328 from the emission sample reduces the fraction of older stars with emission to 10\%.

   \begin{figure}
   \centering
   \includegraphics[width=\hsize]{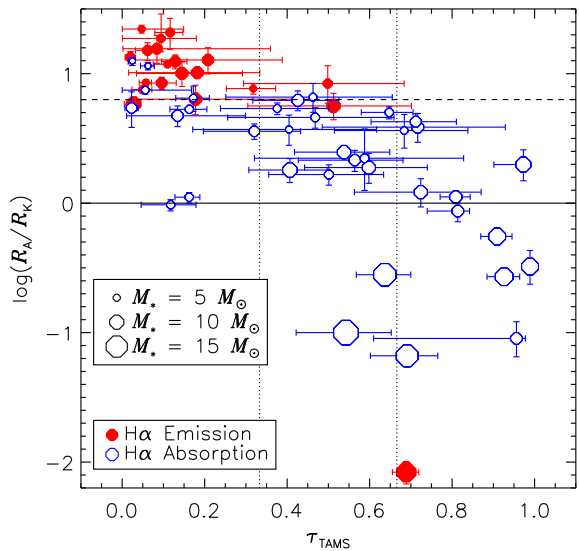}
      \caption[]{Logarithmic ratio of \ra~over \rk~as a function of fractional main sequence age. The solid line separates CMs (above) from DMs (below). The dashed line shows the threshold for H$\alpha$ emission. The dotted lines separate the three eras of a typical star's magnetospheric evolution (see text).}
         \label{rark_time}
   \end{figure}

P13 used the dimensionless parameter \rark~to quantify the size of a star's CM. Stars with \rark~$>0$ formally possess CMs; those with \rark~$>0.8$ tend to display H$\alpha$ emission. {\sc genec} models \citep{2008Ap&SS.316...43E} self-consistently including mass-loss quenching and magnetic spindown calculated by \cite{2019MNRAS.485.5843K} showed that magnetic stars should in general evolve from having CMs in their youth to DMs as they approach the TAMS. Fig.\ \ref{rark_time} tests this prediction by showing \rark~as a function of fractional age. 

The typical star begins its life on the main sequence with a strong magnetic field of about 8 kG (see Fig.\ \ref{bd_fract_age}). It is also a rapid rotator, with a rotation period on the order of 1 d (Fig.\ \ref{prot_fract}). The combination of strong magnetic wind confinement and rapid rotation provides sufficient centrifugal support to the corotating circumstellar plasma to form a large CM (\ra~$\sim 22~R_*$, \rk~$\sim 3~R_*$). All of the sample stars in the first third of the main sequence have CMs, and a majority of these have CMs detectable via H$\alpha$ emission. The exceptions either possess small CMs (due to relatively slow rotation or weak magnetic fields: HD\,36982, HD\,37058, HD\,47777, HD\,67621, HD\,105382, and HD\,125823), or are low in mass (and therefore have weak stellar winds: HD\,35298, HD\,36526, HD\,130807, and HD\,175362). 

As the star ages its surface magnetic field weakens. As a result, magnetic wind confinement also weakens, and \ra~moves closer to the star. By half-way through the main sequence, its surface magnetic field strength has declined to about 3 kG, and \ra~is about 10 $R_*$. The star simultaneously sheds angular momentum, and its rotational period is now on the order of a few days. Thus \rk~moves further from the star, to about 4 $R_*$. Referring to Fig.\ \ref{rark_time}, the star still has a CM, but the CM is now too small to be detectable in H$\alpha$. 

By the final third of the main sequence the star's surface magnetic field has decreased to about 1 kG, while its rotational period has increased to about 10 d. As \rk~$\sim 8~R_*$ is now greater than \ra~$\sim 5 R_*$, the star no longer has a CM, and if its dynamical magnetosphere is detectable at all in visible light it is only if the mass-loss rate is high enough to fill its dynamical magnetosphere (which is the case, so far uniquely among B stars, for HD\,46328).

The basic picture described here does not change qualitatively if fractional main sequence ages determined using \cite{2011A&A...530A.115B} models are used instead. The main difference is that very few stars are found in the final third of the main sequence; the overall trend of decreasing \rark~over time remains unchanged, and it remains true that the first third of the MS is composed entirely of stars with CMs and overwhelmingly of stars with H$\alpha$ emission.

\section{Discussion}\label{sec:discussion}

\subsection{Flux decay and mass}\label{subsec:flux_decay_with_mass}

   \begin{figure}
   \centering
   \includegraphics[trim=20 20 20 20, width=8.4cm]{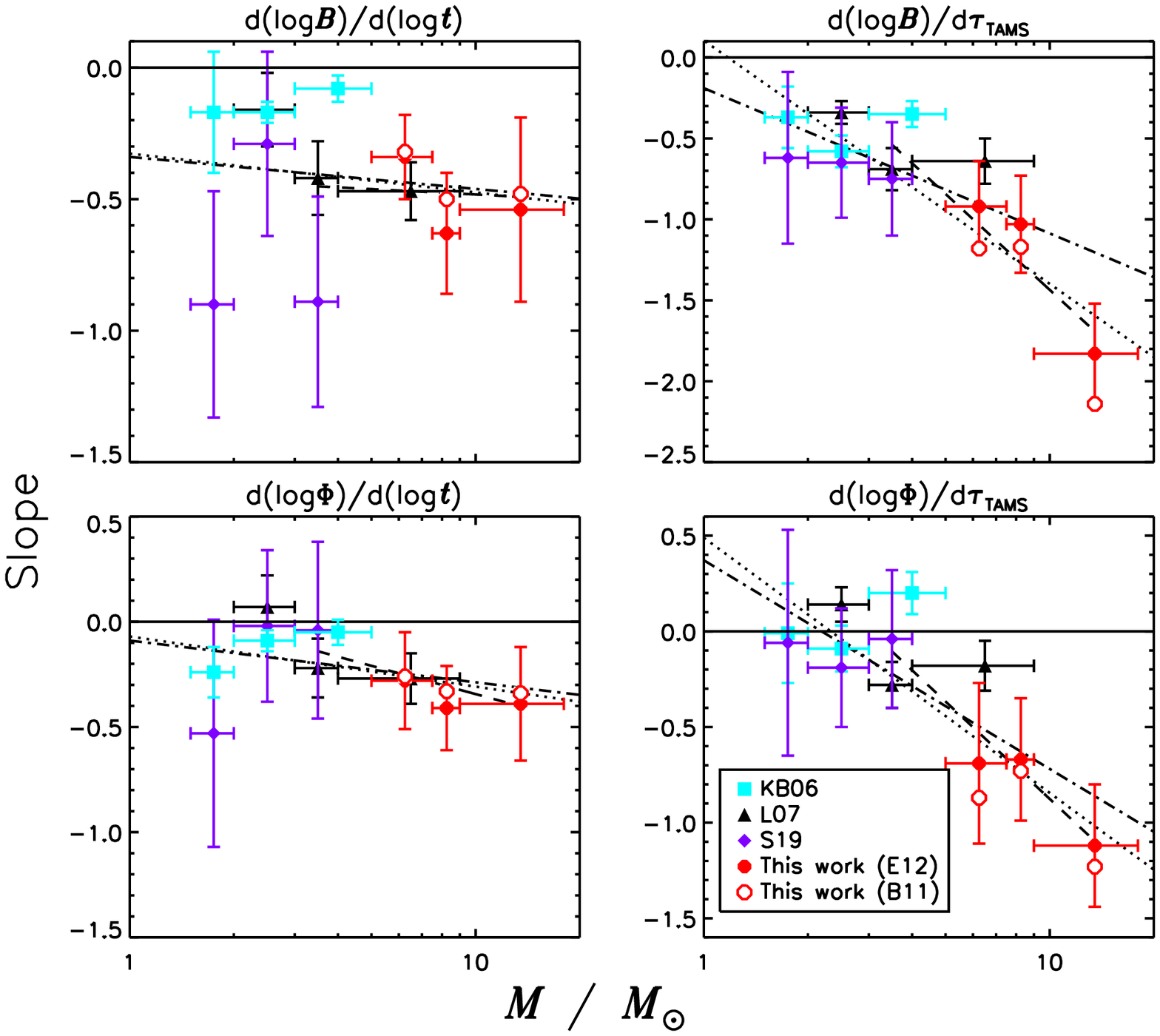}
      \caption[]{Linear regression slopes for $B_{\rm d}$ and $\Phi$ vs.\ $t$ and $\tau_{\rm TAMS}$ as functions of stellar mass from this and previous works (see text). Horizontal dashed lines indicate constant $B_{\rm d}$ or $\Phi$ with time. Diagonal dashed lines show linear fits to measurements above 4 \msun. The legend indicates the studies from which the data were taken: \protect\citet[KB06]{2006AA...450..763K}, \protect\citet[L07]{land2007}, \protect\citet[S19]{2019MNRAS.483.2300S}, and the current work for results obtained from \protect\citet[E12]{ekstrom2012} and \protect\citet[B11]{2011A&A...530A.115B} models. For clarity, error bars are omitted for the B11 slopes. }
         \label{bd_flux_slopes}
   \end{figure}

\cite{1987ApJS...64..219T} reported an increase in the surface magnetic field strength of Ap/Bp stars with mass. This is not supported by our results. \citet[KB06]{2006AA...450..763K} found no correlation with surface magnetic field strength and mass, but did find that magnetic flux increased with mass. \citet[L07]{land2007} also looked for trends in magnetic properties with mass. They found slopes consistent with no change in surface magnetic field strength, consistent with our results, but reported a possible slight increase in $\Phi$ with mass, with slopes for their full sample of $0.5 \pm 0.3$ for $\log{B}$ vs.\ $\log{M_*}$ and $1.44 \pm 0.26$ for $\log{\Phi}$ vs.\ $\log{M}$. While their results for smaller samples sub-divided by age were more ambiguous than our own, in all cases they found a slight increase in $\Phi$ with $M_*$.

The increase in $\Phi$ with $M_*$ raises the question of whether the apparent increase in the decay rate of $\Phi$ with increasing $M_*$ might be an artifact. Detrending $\log{\Phi}$ using the linear fit in Fig.\ \ref{bd_fract_age} does not change the slopes of the various sub-samples outside of the uncertainties; for the full sample, it actually results in a steeper and more statistically significant relationship vs.\ $\tau_{\rm TAMS}$ ($-0.84 \pm 0.13$, $p=10^{-5}$). Therefore the temporal decay in $\Phi$ cannot be due to the mass range of the sample. 

In addition to the direct evidence for flux decay, the number of magnetic stars declines with increasing fractional age (Fig.\ \ref{cluster_field_mass_fract}), an observation compatible with the finding of \cite{2016A&A...592A..84F} that the incidence fraction of magnetic vs.\ non-magnetic early-type stars declines over time. \citeauthor{2016A&A...592A..84F} also found that this evolutionary difference in incidence fraction is more pronounced for more massive stars, which they interpreted as evidence that magnetic flux decays more rapidly with increasing mass. 

Direct investigation of the evolution of fossil magnetic fields has been undertaken for several different samples. KB06 examined Ap/Bp stars, using ages determined from HRD positions and using the root-mean-square \bz~as a proxy to $B_{\rm d}$. L07 also used the r.m.s.\ \bz, but limited their study to stars in open clusters in order to maximize the precision of stellar ages for stars close to the ZAMS. S19 examined a complete volume-limited sample of Ap stars, obtaining precise measurements of $B_{\rm d}$ from high-resolution spectropolarimetry. Fig.\ \ref{bd_flux_slopes} shows, as functions of stellar mass, the slopes found in these studies and in the present work via linear regression of $\log{B}$ and $\log{\Phi}$ vs.\ $\log{t}$ and $\tau_{\rm TAMS}$. For the sake of completeness, the slopes achieved using \cite{2011A&A...530A.115B} models are also shown; as is clear, they do not change the results outside the uncertainties.

Below about 4~\msun, KB06, L07, and S19 all agree that $\Phi$ is constant in time. Flux appears to decay only above 4 \msun; furthermore, the rate of flux decay appears to increase with mass. The solid lines in Fig.\ \ref{bd_flux_slopes} show constant $B_{\rm d}$ and $\Phi$ with time; almost all studies are consistent with a declining $B_{\rm d}$, whereas only the studies of low-mass stars are consistent with constant $\Phi$. The dotted, dashed, and dot-dashed lines show three different linear fits to the slopes.  The dotted line shows the fit to all measurements. The dashed line shows a fit to all measurements of samples with masses above 4 \msun. The dot-dashed line fits all measurements except the highest-mass sub-sample, which might plausibly be considered an outlier. In all cases a negative slope is obtained i.e.\ the rate of change of $B_{\rm d}$ and $\Phi$ increases with $M_*$. In the cases of ${\rm d}\log{\Phi}/{\rm d}\log{t}$, the slopes are all very shallow, and could be consistent with a constant rate of flux decay with $M_*$; however, a constant rate cannot provide a good match to ${\rm d}\log{\Phi}/{\rm d}\log{\tau_{\rm TAMS}}$, even with the highest-mass bin excluded. 

The mass range of the lower mass sample in the present work overlaps with the upper mass range examined by L07. We obtain compatible results within uncertainty for $\log{t}$, but steeper slopes vs.\ $\tau_{\rm TAMS}$. This is despite only partial overlap in the samples themselves: 10 stars with masses below 9~\msun~appear in both the present sample and the L07 sample, as compared to 38 stars total in this sample and 25 stars total in the L07 sample with masses in this range. 

An important difference between L07 and the present study is that the former was based upon just a few \bz~measurements per star, many of which were non-detections. In order to avoid biasing their results, they added 200 G to the r.m.s. of all stars. Since $B_{\rm d}$ is at least 3.5$\times$ higher than \bz$_{\rm max}$, this is equivalent to assuming a minimum dipole strength of 700 G. The present study, which is based entirely on magnetic detections and magnetic dipole strengths, contains 12 stars with $B_{\rm d} \le 700$~G, all of which have $\tau_{\rm TAMS} \ge 0.4$. Thus, the L07 results are insensitive to the weakest magnetic fields in this population, which are predominantly found at more advanced ages. If the L07 slopes are recalculated without the 200 G correction, the lowest-mass sub-sample in the L07 study continues to be consistent with no flux decay, but the middle ($3 < M_* < 4$~\msun) and high ($M_* \ge 4$~\msun) samples yield steeper slopes: respectively about ${\rm d}\log{\Phi}/{\rm d}\log{t} \sim -0.5$ (consistent with both S19 and this work) and ${\rm d}\log{\Phi}/{\rm d}\tau_{\rm TAMS} \sim -0.7$ (consistent with the present work, but not with S19). 

Another key point of comparison is age bias. As demonstrated in Figs.\ \ref{cluster_field_mass_fract} and \ref{mass_age}, while the present sample is biased towards young stars, especially regarding cluster stars and stars with $M_* < 7.5$~\msun, the middle and high-mass sub-samples are relatively evenly distributed across the main sequence. In contrast, the S19 sample is heavily biased towards relatively old stars, while the L07 sample is biased towards young stars, particularly in the highest-mass bin for which the median fractional main-sequence age is about 0.1. Comparing the three samples to a flat distribution via the one-sample K-S test, the present sample yields the closest resemblance with a probability of 0.05, as compared to $7\times10^{-5}$ for the L07 sample and 0.006 for the S19 sample. Looking only at the overlapping mass ranges of the L07 sample and the present work increases the difference: $8\times10^{-6}$ for the L07 high-mass sub-sample, and 0.4 for our mid-mass sample. Our high-mass sub-sample yields a probability of 0.68 to belong to a flat distribution. We conclude that the present study provides a more even sampling of the main sequence than previous works, and does so particularly within the mass range in which flux decay becomes unambiguous. 

We can use the slopes given in Table \ref{fluxtab} to estimate the ZAMS flux $\Phi_0$ of the more evolved stars in the sample via $\log{\Phi_0} = \log{\Phi} - ({\rm d}\log{\Phi}/{\rm d}\tau_{\rm TAMS})\tau_{\rm TAMS}$. For HD\,46328, which has a very strong magnetic field ($B_{\rm d} = 1.2$~kG) for its age ($\tau_{\rm TAMS} = 0.69$), $\log{\Phi} = 4.9 \pm 0.2$, yielding $\log{\Phi_0} = 5.8 \pm 0.2$. Accounting for the increase of about a factor of 1.8 in $R_*$ from the ZAMS then leads to a surface dipole strength on the ZAMS of about 30 kG, similar to that of the most strongly magnetic stars in the sample. Conservation of flux for this star would imply $B_{\rm d} \sim 4$~kG on the ZAMS, which is still compatible with the range of ZAMS $B_{\rm d}$ but on the low end: it would be strange for one of the most strongly magnetic of the evolved massive stars to have one of the weakest ZAMS fields. The same analysis for HD\,52089, which is the most evolved and also the most weakly magnetic star in the sample, yields a ZAMS dipole strength of about 8 kG, consistent with the typical magnetic field strengths of the younger stars. By contrast, flux conservation would imply a ZAMS field of about 1 kG for HD\,52089; while not impossible, it is worth noting that the weakest field amongst stars with $\tau_{\rm TAMS} < 0.1$ is 1.8 kG (HD\,36982). Excluding HD\,36982 (which is technically a Herbig Be star, and is therefore likely to still be contracting as it approaches the ZAMS), the weakest field belongs to HD\,142990 (4.9 kG). 

Extending the fits shown in Fig.\ \ref{bd_flux_slopes} to higher masses enables us to estimate how rapid flux decay should be for the magnetic O-type stars. Over a mass interval of 30 to 60 \msun, we predict ${\rm d}\log{\Phi}/{\rm d}\tau_{\rm TAMS}$ to range from $-1.6$ to $-2.1$, as compared to $-1.1 \pm 0.3$ for the highest-mass sample of B stars. Extending the investigation of flux decay to the upper end of the main sequence is beyond the scope of this paper, as for the magnetic O-type stars the effects of magnetic fields on stellar evolution due to e.g.\ mass-loss quenching are expected to be highly significant, and their evolutionary status cannot be accurately investigated without evolutionary models taking these effects into account. However, it is worth noting that this extrapolation predicts that if a magnetic O-type star starts at the ZAMS with a field strength equal to the mean ZAMS value for the B-type stars (8 kG), then by $\tau_{\rm TAMS} = 0.5$ its surface magnetic field strength would have decreased to 700 G. Such a rapid decline would explain the lack of magnetic O-type stars in the second half of the MS noted by \cite{2016A&A...592A..84F}. 

\subsection{Flux decay and field strength}\label{subsec:flux_field_strength}

In their review of the phenomena associated with fossil magnetic fields, \cite{2017RSOS....460271B} speculate on 3 possible mechanisms to explain flux decay: i) Ohmic decay; ii) thermal diffusion and magnetic buoyancy; and iii) meridional circulation. i) seems unlikely, as the Ohmic decay timescale increases slightly in more massive stars, while the flux of such stars instead appears to decay more quickly. In the case of iii), flux decay should be more rapid for more rapidly rotating stars. This scenario is inherently difficult to evaluate since, as seen in \S~\ref{subsec:rot_evol}, magnetic braking rapidly removes angular momentum: since both hypotheses predict that evolved stars with slow rotation should have relatively strong magnetic fields, it is not obvious how to disentangle them. In the following we focus upon scenario ii). 

   \begin{figure}
   \centering
   \includegraphics[trim=20 20 20 20, width=8.4cm]{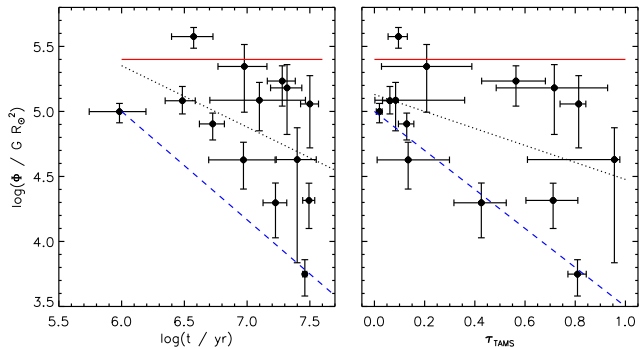}
      \caption[]{Magnetic flux vs.\ time for stars with $7.5~M_\odot~< M_* < 9~M_\odot$. The dotted black line shows the least-squares linear fit; the solid red and dashed blue lines indicate a possible magnetic field strength-dependent flux decay scenario discussed in the text.}
         \label{midmassflux}
   \end{figure}

According to \cite{2017RSOS....460271B}, the characteristic decay timescale expected for scenario ii) goes as $B^{-2}$, and should therefore be negligible for stars with weak magnetic fields. In this case, the lower range of magnetic flux should be approximately constant, while the upper range falls. Fig.\ \ref{midmassflux} shows the evolution of $\Phi$ for the mid-mass sample, which fully samples the main sequence and is narrow enough in mass that differences in stellar structure should be negligible. We see the opposite of the pattern predicted from scenario ii). The upper limit of $\Phi$ appears to be approximately constant (red line in Fig.\ \ref{midmassflux}), dropping by about 0.5 dex from early to late stages. Conversely, the lower limit seems to decrease more rapidly, by about 1 dex (dashed blue line). 

A possible explanation may be that strong magnetic fields may inhibit whatever mechanism is responsible for flux decay. \cite{2013MNRAS.433.2497S} demonstrated that inhibition of convection within sub-surface Fe and He opacity bumps is a viable explanation for the lack of macroturbulent broadening in the spectral lines of NGC 1624-2. \cite{2015ApJ...808L..31G} have further shown that macroturbulence very likely originates due to high-order pulsations excited by turbulent convection within sub-surface convective zones. Paper I presented evidence that the distribution of macroturbulent broadening of the present sample is systematically much lower than that of non-magnetic early B-type stars (although a clear dependence on $B$ was difficult to discern)\footnote{The conclusion that magnetic stars have systematically lower macroturbulent velocities than non-magnetic stars is actually somewhat strengthened with the discovery by \cite{2019MNRAS.482.3950S} that the magnetic field of HD\,37061 belongs to a previously unknown tertiary, which has a much lower $v_{\rm mac}$ than the primary previously believed to host the system's magnetic field, which in turn had the highest $v_{\rm mac}$ in the sample as presented in Paper I.}. Sub-surface convection zones increase in size, strength, and proximity to the surface with increasing mass, and are proposed to support convective dynamos \citep{cant2009,cb2011}. If envelope convection plays a role in dissipating magnetic flux, these properties would be a natural explanation for the apparent acceleration in the flux decay rate with mass.

Another candidate mechanism may be interactions between the extremely strong ($\sim$~MG) dynamo fields that are expected within the convective cores of massive stars \citep{2016ApJ...829...92A} and the fossil fields within the radiative zone. The remnants of core dynamo fields have been inferred within the radiative cores of red giants descended from main-sequence A stars \citep{2015Sci...350..423F,2016Natur.529..364S}. As with sub-surface convection zones, the convective cores of massive stars increase in mass fraction with stellar mass. Core dynamo fields are predicted to be intensified to super-equipartition strength by interaction with fossil fields in radiative envelopes \citep{2009ApJ...705.1000F}; what effect, if any, core dynamos have upon the strength of the radiative zone fossil field is unexplored (although \cite{2009ApJ...705.1000F} noted that the interaction should change the $\beta$ angle of the fossil field). Some speculations on this topic were made by \cite{2017RSOS....460271B}.

We emphasize that Fig.\ \ref{midmassflux} is highly speculative, and a significantly larger sample could reveal an entirely different picture. This is especially true if there are stars close to the ZAMS with fields much weaker or much stronger than the mean ZAMS field strength, since in this case the overall change in flux could be consistent with a constant rate of flux decay regardless of surface field strength. 

\subsection{Do complex magnetic fields decay more rapidly than dipoles?}

   \begin{figure}
   \centering
   \includegraphics[width=8.5cm]{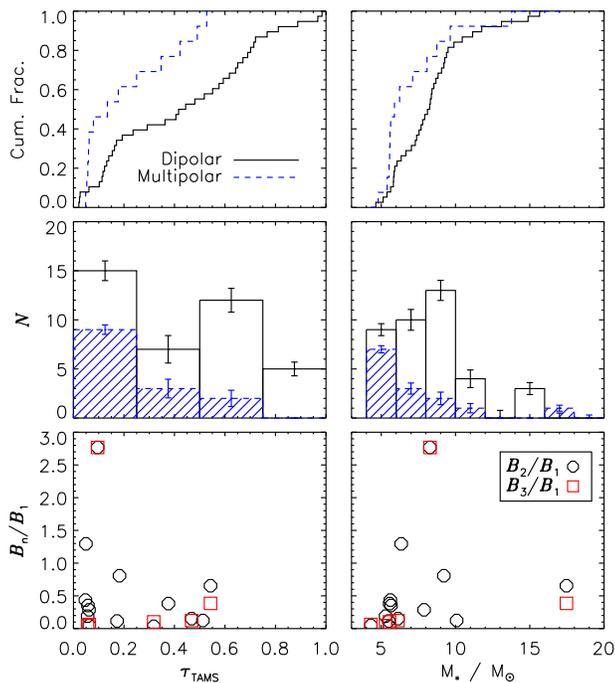}
      \caption[]{{\em Left}: Cumulative distributions (top) and histograms (middle) of fractional main sequence age $\tau_{\rm TAMS}$ for stars with and without detectable contributions to their \bz~curves from multipolar surface magnetic fields. The bottom panel shows the ratio of 2$^{nd}$ and 3$^{rd}$ harmonic amplitudes to the first-harmomic amplitude, as a function of $\tau_{\rm TAMS}$. {\em Right}: as the left, but for $M_*$.}
         \label{bz_fitpars_ttams}
   \end{figure}

\cite{2008MNRAS.386.1947B} demonstrated that, as with magnetic dipoles, non-axisymmetric magnetic configurations can be stable over evolutionary timescales within radiative zones; it was also indicated that the characteristic evolutionary timescales of axisymmetric and non-axisymmetric field configurations may differ, with magnetic dipoles appearing first, and more complex surface fields appearing later. 

In Paper I several stars were identified with detectable contributions to their \bz~curves from higher-order (quadrupolar or octupolar) magnetic field components, i.e.\ with non-zero amplitudes $B_2$ and/or $B_3$ for the second and third harmonics. Anharmonic \bz~curves are a well-known consequence of surface chemical abundance inhomogeneities, but should affect primarily \bz~measurements obtained from metallic lines (indeed, in Paper I this effect was quantified and investigated in detail). H line \bz~measurements, which were used for all stars for which anharmonic \bz~curves were reported in Paper I, should not be affected by the presence of chemical spots. Thus, non-zero values of $B_2$ and $B_3$ measured from H line \bz~curves serve as semi-quantitative measures of the degree of departure of the large-scale field from a dipole. While modelling complex surface magnetic fields in detail is outside the scope of this paper, it is still possible to ask whether there are any trends in the incidence of such fields with age. Fig.\ \ref{bz_fitpars_ttams} shows cumulative distributions and histograms of $\tau_{\rm TAMS}$ for stars with and without detected higher-order harmonic terms. The distribution shows some evidence for stars with complex magnetic fields being systematically younger than stars with dipolar fields: the median fractional age of stars with complex magnetic fields is about 0.1, while that of stars with dipolar fields is about 0.5. Furthermore, there are no complex magnetic fields above $\tau_{\rm TAMS} = 0.6$. However, the two-sample K-S test probability for $\tau_{\rm TAMS}$ is only 0.1. Comparing the histograms, the difference in distributions may be due simply to the smaller number of stars with complex magnetic fields. Given the generally weaker magnetic fields of the oldest stars, it is also possible that complex magnetic fields at such ages exist, but simply remain undetected. These results are unchanged using absolute age. 

The bottom panels of Fig.\ \ref{bz_fitpars_ttams} show the ratios of $B_2$ and $B_3$ to the dipolar amplitude $B_1$ as functions of $\tau_{\rm TAMS}$. This gives an idea of the relative importance of higher-order terms in the \bz~curve. Both of the stars with multipole terms with higher amplitudes than their dipolar terms (HD 36485 and HD 37776) are very young ($\tau_{\rm TAMS} < 0.1$, $t < 10$~Myr). HD\,149438, which is known to have a complex surface magnetic field, has ratios of $B_2/B_1 = 0.65$ and $B_3/B_1=0.39$, i.e.\ the dipolar term is still the strongest according to the crude criteria used here; its fractional age is about 0.5. The majority of other stars with multipolar terms in their \bz~curve have ratios around 20\%, and these appear at all ages. There is thus an apparently rapid decline, by a factor of several, in the relative contributions of first and higher-order terms to \bz~curves. 

\cite{2019A&A...621A..47K} noted that extremely complex magnetic fields have so far been seen only in the most massive stars that have so far been mapped using ZDI. However, there are no stars that have been mapped between about 4 and 7 \msun. The right panels of Fig.\ \ref{bz_fitpars_ttams} show the distributions with mass. There is actually a slightly more significant difference in the sample distributions with $M_*$ (K-S probability of 0.04) than with $\tau_{\rm TAMS}$, with higher harmonics being more common close to the bottom of the mass range. However, complex fields are seen across the full mass range. However, the relative strength of higher harmonics increases towards higher masses. 

One possible objection to this analysis is that, as seen in Fig.\ \ref{bd_fract_age}, the surface magnetic field strength declines over time, making more complex magnetic fields intrinsically more difficult to detect in older stars. However, the Stokes $V$ signatures of quadrupoles are not much lower in amplitude than those of dipoles. Furthermore, as shown in Paper I (Fig.\ 16), numerous stars with \bz~curves compatible with dipoles have been observed at levels of magnetic precision comparable to, and in some cases exceeding, the relative precision of stars in which the signatures of more complex fields have been detected. Furthermore, since we are looking at the relative contributions of higher-order terms, and older stars have weaker magnetic fields, if anything the results for older stars should be biased to larger relative contributions since these would be easier to detect amongst stars with intrinsically weaker magnetic signatures. That is, if the relative contribution of small-scale magnetic structure to \bz~is constant with time, amongst older stars such contributions should be detectable only when they are very large; this is the opposite of what we see in Fig.\ \ref{bz_fitpars_ttams}.

\subsection{Expanding the sample of H$\alpha$-bright CM stars}

   \begin{figure}
   \centering
   \includegraphics[width=8.5cm]{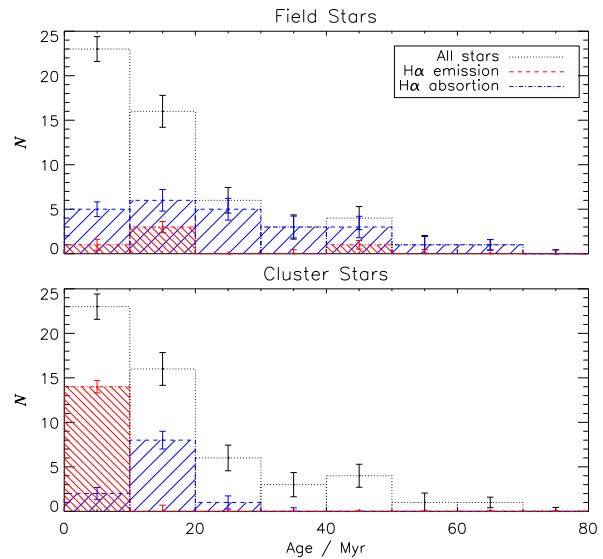}
      \caption[]{Histograms of stellar ages, with field stars and cluster stars divided into stars with and without H$\alpha$ emission.}
         \label{field_cluster_emission_age_hist}
   \end{figure}

The predominance at young ages of rapidly rotating, strongly magnetized stars displaying H$\alpha$ emission suggests that future searches for such objects should focus on young stellar clusters. Indeed, as demonstrated in Fig.\ \ref{field_cluster_emission_age_hist}, almost all of the known emission-line stars are found in young clusters. Of the five field stars with H$\alpha$ emission, one is HD 46328 (which is predicted to have a DM). Of the remaining four, two (HD 64740 and HD 176582) have very weak emission (Shultz et al., in prep.; \citealt{bohl2011}). HD 182180 is an extremely rapid rotator with amongst the strongest emission lines ever detected \citep{rivi2013}; it is listed in the {\em Catalogue of Runaway Stars} by \cite{2011MNRAS.410..190T}, suggesting it was ejected from its birthplace. The kinematics of the final H$\alpha$-bright field star, HD 345439, have not been studied in detail.

One possible location for such an investigation may be $h+\chi$ Per (NGC\,869 and 884), two clusters of approximately 13-14 Myr ages \citep{2010ApJS..186..191C}. The clusters' stellar populations extend up to about B0 in spectral type, and the spectroscopic properties have already been characterized \citep{2010ApJS..186..191C,2012AJ....144..158M}. The brightest stars in the cluster are just at the limit currently attainable with high-resolution spectroscopy ($V \sim 10$); as such, a limited magnetic survey of the top of the HRD is feasible, but has not yet been carried out. Existing spectroscopic data should be searched for signs of chemical peculiarity and CM-pattern emission. The age of the clusters suggests that any magnetic B-type stars with CMs may well have already spun down; failure to detect CM emission would validate the hypothesis that such emission is unique to very young magnetic stars. 

Another tempting location is the 30 Dor star-forming region in the Large Magellanic Cloud, which consists of multiple young open clusters on the order of 10 Myr in age. 30 Dor contains over 400 B-type stars with spectral types B5 and earlier, of which many show H$\alpha$ emission \citep{2015A&A...574A..13E}. If 10\% of these stars are magnetic, and 50\% of the magnetic B-type stars show H$\alpha$ emission, there could be an additional 20 stars with detectable CMs waiting to be discovered in 30 Dor alone. Such a survey, if succesful, would more than double the number of such stars known. This may be a somewhat conservative estimate: as Fig.\ \ref{field_cluster_emission_age_hist} demonstrates, if clusters younger than 10 Myr are considered, 75\% of the stars display H$\alpha$ emission, which would increase the expected number of such stars in 30 Dor to 30. Interestingly, \cite{2015A&A...574A..13E} noted that 30 stars show H$\alpha$ emission without accompanying Fe {\sc ii} emission; Fe {\sc ii} is frequently in emission in classical Be stars, but has never been seen in emission in CM host stars. These objects should be examined more closely, looking for signs of chemical peculiarity and comparing their emission morphologies to the characteristics expected of CMs. 

This work has concentrated on magnetic early B-type stars, because it is in this range that CMs are most often detectable in emission. There is one case of a relatively cool magnetic Bp star, 36\,Lyn, which does not show emission but is known to show enhanced H$\alpha$ absorption due to eclipsing by its magnetospheric plasma \citep{2006A&A...458..569W,2006AA...458..581S}. 36\,Lyn's rotational and stellar parameters were provided by \cite{2006A&A...458..569W}, from which we can determine \rk$=4.4~R_*$. Using \cite{vink2001} mass-loss rates determined from the stellar parameters, and adopting $B_{\rm d} \sim 3$~kG from the ZDI map presented by \cite{2018MNRAS.473.3367O}, we find \ra~$=31~R_*$ and $\log{(R_{\rm A}/R_{\rm K})} = 0.84$. 36\,Lyn is therefore in the same magnetospheric regime as the emission-line stars examined here, strongly suggesting that its eclipses are a consequence of a CM. This suggests that a similar phenomenon might be seen in other cool Bp stars. It is worth noting that eclipses are only expected for a relatively narrow range of $i_{\rm rot}$ and $\beta$ \citep{town2008}, and 36\,Lyn's ORM geometry \citep[$i \sim 60^\circ$, $\beta \sim 80^\circ$][]{2006A&A...458..569W,2018MNRAS.473.3367O} is in the ideal range. Furthermore, they are detectable only during a brief segment of the rotational phase curve, when the plasma is directly in front of the star; therefore excellent coverage of the rotational phase curve with high-resolution spectroscopy is essential. Emission, by contrast, is detectable almost independently of geometry or rotational phase: we should therefore expect stars showing only enhanced absorption due to eclipses to be relatively rare. It is curious to note, however, that HD\,175362 (for which a very large ESPaDOnS dataset with excellent phase coverage is available) has similar rotational properties and ORM angles as 36\,Lyn, but a much stronger magnetic field, yet shows no sign of enhanced absorption in the core of its H$\alpha$ line.

\subsection{Limitations and future steps}

There are several fundamental limitations to this analysis. 

First, the evolutionary models used to determine the ages of the field stars do not include the effects of magnetic fields, which are expected to include inhibition of mixing and, possibly, enforcement of internal solid-body rotation. Therefore these results should also be checked when reliable stellar evolutionary tracks and isochrones accounting for magnetic effects are developed \citep[e.g.][]{2018CoSka..48..124K}. 

Second, the main sequence is not comprehensively sampled at the low-mass ($M_* < 7.5 M_\odot$) end of our sample; future surveys should aim to fill this gap. 

Third, the evaluation of observational biases in this sample is hampered by the absence of a thorough analysis of existing spectropolarimetric surveys of B-type stars in general. It would be useful to know how the upper limits on the surface magnetic fields of stars without detected Zeeman signatures change as a function of spectral type, rotation, and age. 

Fourth, younger stars tend to be more rapidly rotating. Rapid rotators have broader spectral lines, making weaker magnetic fields harder to detect. The apparent absence of very weak magnetic fields amongst the youngest stars may therefore be a consequence of observational bias. The dot-dashed lines in the middle panels of Fig.\ \ref{bd_fract_age} show the median dipole sensitivity of the MiMeS survey, as evaluated from the median \bz~error bar \citep{2016MNRAS.456....2W}. This is about 1 dex lower than the dipole strength of the average star close to the ZAMS, suggesting that if young stars with very weak magnetic fields exist, they should have been detected. The 75\% and 95\% dipole sensitivities of the MiMeS Survey are 1 and 3 kG \citep{2016MNRAS.456....2W}, significantly below the inferred median value of about 8 kG close to the ZAMS (although the 95\% value overlaps with the lower tail of the ZAMS distribution). The MiMeS survey sensitivity is a function of \vsini~\citep[][]{2016MNRAS.456....2W,2017MNRAS.465.2432G}, which should decrease over time for non-magnetic as well as magnetic stars: therefore the sensitivity of the MiMeS survey in principle decreases for younger stars. However, given that only a very small fraction of the MiMeS survey targets have dipole sensitivities in the range of several kG, whereas the majority of the survey targets are likely to be more rapidly rotating than the magnetic B-type stars \citep{2018MNRAS.475.5144S}, it seems plausible that the absence of very young stars with weak magnetic fields is real. A rigorous answer to this question will require a precise evaluation of the properties of the MiMeS B-star survey as a function of stellar age.

Finally, the values of $\Phi$ used here are something of an estimate, as they are based on the assumption that surface magnetic fields are appropriately described by a dipole. While this is likely reasonable for the majority of stars, and approximately correct even for the 20\% of the sample showing some evidence of topological complexity, it is certainly not correct for some stars, e.g.\ HD 37776 which has a highly complex surface magnetic field with a maximum strength of 30 kG \citep{koch2011}. Some stars with \bz~curves consistent with a dipole show signs in Stokes $V$ diagnostics of significant contributions from higher-order components \citep[e.g.][]{2018MNRAS.478L..39S}. A study utilizing ZDI maps for a large sample of magnetic stars is required to obtain a precise determination of the change in magnetic flux over time. Such a study would also be able to more accurately explore the question of whether magnetic complexity itself evolves, e.g.\ if complex magnetic fields are less stable than dipoles. A related issue deserving of attention is why some stars possess such complex magnetic fields in the first place. HD\,149438 and HD\,61556 have been suggested as potential merger products \citep{2016MNRAS.457.2355S}, and both stars possess complex magnetic fields \citep{2006MNRAS.370..629D,2015MNRAS.449.3945S}, especially HD\,149438 which has one of the most complex magnetic fields known amongst early-type stars \citep{2016A&A...586A..30K}; might the same be true of HD\,37776? 

Having explored rotational and magnetic evolution on the main sequence, the next logical step is to put these results in the context of the pre- and post-main sequence evolution of magnetic massive stars. \cite{2013MNRAS.429.1001A,2013MNRAS.429.1027A} have investigated the rotational properties of the magnetic Herbig Ae/Be stars, finding them to be more slowly rotating than non-magnetic Herbig stars; \cite{2015Ap.....58...29G} presented evidence that surface magnetic fields are weaker on the pre-main sequence. These results have been limited by the heavy extinction of most such objects; a survey using the new IR spectropolarimeter SPIRou may be able to significantly enlarge the size of this sample. On the other end of the main sequence, \cite{2017MNRAS.471.1926N} and \cite{2018MNRAS.475.1521M} have detected weak magnetic fields in several hot supergiants, a population which remains largely unexplored. 

Within the rotation-magnetic confinement diagram (Fig.\ \ref{ipod}), there is an overlap between the H$\alpha$ emission and absorption regions, and the stars with and without emission might be distinguishible by mass-loss rate rather than by CM size alone. If so, this may be evidence for a steady-state magnetospheric leakage mechanism \citep{petit2013,2018MNRAS.474.3090O}. A more conclusive test of this hypothesis can be provided via a detailed spectroscopic analysis of the emission properties of these stars, in the context of the stellar and magnetospheric parameters determined here.

\section{Conclusions}\label{sec:conclusion}

We utilized the rotational, magnetic, and atmospheric measurements presented in Papers I and II, together with additional constraints such as binary mass ratios, independent constraints on rotational inclination angles, and cluster ages, to obtain self-consistent stellar, rotational, magnetic, and magnetospheric properties for the magnetic early B-type stars. These were in turn used to perform the first detailed examination of the magnetic and rotational evolution of this population of magnetic stars. The fundamental conclusions of this work are as follows: 

\begin{enumerate}
\item There is a systematic offset between the distribution of rotational axis inclination angles $i_{\rm rot}$ and the expected random distribution, which also affects some previous, comparable studies. This appears to be an artifact of a bias towards sharp-lined stars. The magnetic obliquity angle $\beta$ is statistically consistent with a random distribution.
\item The distribution of the dipolar magnetic field strengths $B_{\rm d}$ is approximately log-normal, but with a cutoff at high values. Ap/Bp stars and O-type stars have almost identical $B_{\rm d}$ distributions, and there is no systematic change in $B_{\rm d}$ with mass. Magnetic flux $\Phi$, however, increases strongly with mass. 
\item Both $B_{\rm d}$ and $\Phi$ decrease with time; the decrease in $\Phi$ is especially pronounced for the most massive stars, which have the highest $\Phi$ of the full sample close to the ZAMS but the lowest $\Phi$ close to the TAMS. This suggests that flux decay occurs, and that the rate is mass dependent.
\item There is a tentative indication that stars with initially weaker surface magnetic fields may experience more rapid flux decay than stars with initially stronger magnetic fields; if confirmed, this may point to the mechanism responsible for flux decay.

\item Magnetic topologies appear to simplify over time; this needs to be confirmed using ZDI. 

\item Rotation slows over time, which is qualitatively consistent with magnetospheric braking. Indeed, magnetic braking appears to be more efficient than predicted, leading to much longer rotational periods than can currently be explained. 

\item We have revisited the rotation-magnetic confinement diagram introduced by P13. There is now very little ambiguity in the placement of H$\alpha$-bright stars. Without exception, all stars displaying H$\alpha$ emission consistent with an origin in a centrifugal magnetosphere (CM) have very strong magnetic fields (hence large Alfv\'en radii) and are rapid rotators (hence small Kepler radii)

\item H$\alpha$-bright CM stars share a third property: they are all very young. Indeed, H$\alpha$ emission is found amongst a majority (2/3) of the young stars, despite being seen in only about 1/3 of the total sample. This youth is consistent with rapid rotation.

\end{enumerate}

These conclusions can be assembled into an outline of the biography of the typical magnetic early B-type star. It starts its main sequence life with a very strong magnetic field ($\sim 8$ kG), and very rapid rotation; the resulting large CM is detectable in H$\alpha$. As it ages, its surface magnetic field weakens due to the twin effects of its growing radius, which reduces the surface flux density, and gradual decay of its total magnetic flux. At the same time it loses angular momentum via its magnetized wind. By middle-age the CM remains but its H$\alpha$ emission is gone. As it approaches the TAMS the CM disappears entirely, with its magnetic field dropping to several hundred G and its rotational period increasing to tens of days. 

The youth of H$\alpha$-bright CM stars immediately suggests that any future survey attempting to find more such stars should focus on young stellar clusters, where the overwhelming majority of these stars are located. 

\section*{Acknowledgements}

This work is based on observations obtained at the Canada-France-Hawaii Telescope (CFHT) which is operated by the National Research Council of Canada, the Institut National des Sciences de l'Univers (INSU) of the Centre National de la Recherche Scientifique (CNRS) of France, and the University of Hawaii; at the La Silla Observatory, ESO Chile with the MPA 2.2 m telescope; and at the Observatoire du Pic du Midi (France), operated by the INSU. This work has made use of the VALD database, operated at Uppsala University, the Institute of Astronomy RAS in Moscow, and the University of Vienna. This work has made use of data from the European Space Agency (ESA) mission Gaia (https://www.cosmos.esa.int/gaia), processed by the Gaia Data Processing and Analysis Consortium (DPAC, https://www.cosmos.esa.int/web/gaia/dpac/consortium). Funding for the DPAC has been provided by national institutions, in particular the institutions participating in the Gaia Multilateral Agreement. This research has made use of the WEBDA database, operated at the Department of Theoretical Physics and Astrophysics of the Masaryk University. MES acknowledges the financial support provided by the European Southern Observatory studentship program in Santiago, Chile; the Natural Sciences and Engineering Research Council (NSERC) Postdoctoral Fellowship program; and the Annie Jump Cannon Fellowship, supported by the University of Delaware and endowed by the Mount Cuba Astronomical Observatory. GAW acknowledges support from an NSERC Discovery Grant. VP acknowledges support from the National Science Foundation under Grant No.\ 1747658. AuD acknowledges support from NASA through Chandra Award number TM7-18001X issued by the Chandra X-ray Observatory Center, which is operated by the Smithsonian Astrophysical Observatory for and on behalf of NASA under contract NAS8- 03060. OK acknowledges financial support from the Knut and Alice Wallenberg Foundation, the Swedish Research Council, and the Swedish National Space Board. The MiMeS and BinaMIcS collaborations acknowledge financial support from the Programme National de Physique Stellaire (PNPS) of INSU/CNRS. We acknowledge the Canadian Astronomy Data Centre (CADC). 

\bibliography{bib_dat.bib}{}

%\newpage

\begin{center}

\onecolumn
\thispagestyle{plain} % empty
\vspace*{\fill}
\textbf{\Huge Online Material}
\vspace*{\fill}
\end{center}
%\mbox{}

\pagebreak
\clearpage
%\widetext
%

%

\twocolumn

\renewcommand{\thetable}{\Alph{section}\arabic{table}}
\appendix
\section{Monte Carlo sampling of the HRD}\label{monte_carlo}

   \begin{figure}
   \centering
   \includegraphics[width=8.5cm, trim={0 3cm 0 0}]{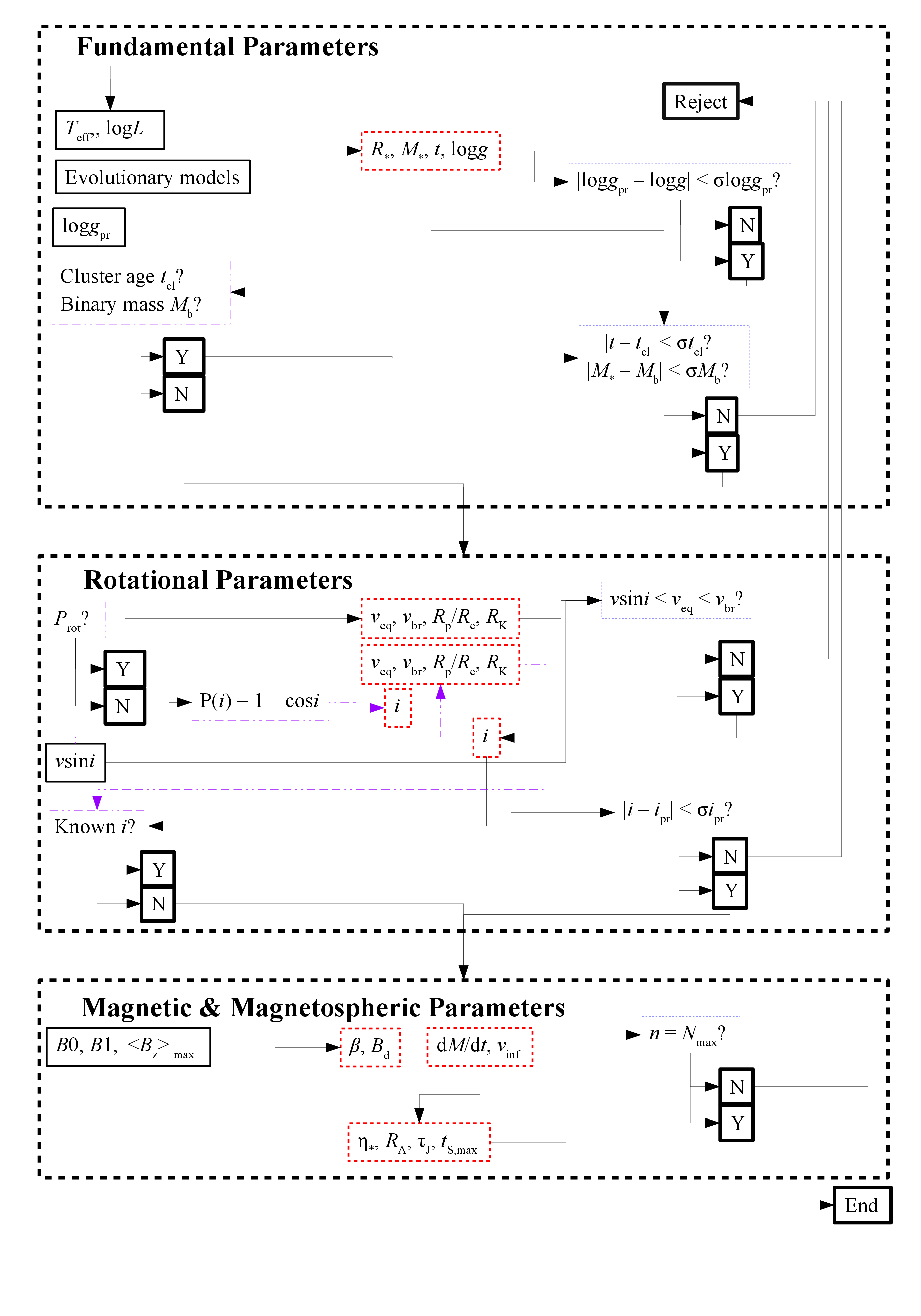} %&
      \caption[]{Flowchart illustrating the Monte Carlo algorithm used to determine sample parameters. Solid black outlines indicate necessary priors; dot-dashed purple outlines, optional priors; dashed red outlines, posteriors; dotted blue outlines, tests; thick black outlines, decision points. The dot-dashed purple arrows in the rotational parameters box illustrate the alternative pathway used to determine surface rotational parameters if the rotational period is unknown.}
         \label{flowchart}
   \end{figure}

\begin{figure}
   \centering
   \includegraphics[width=8.5cm]{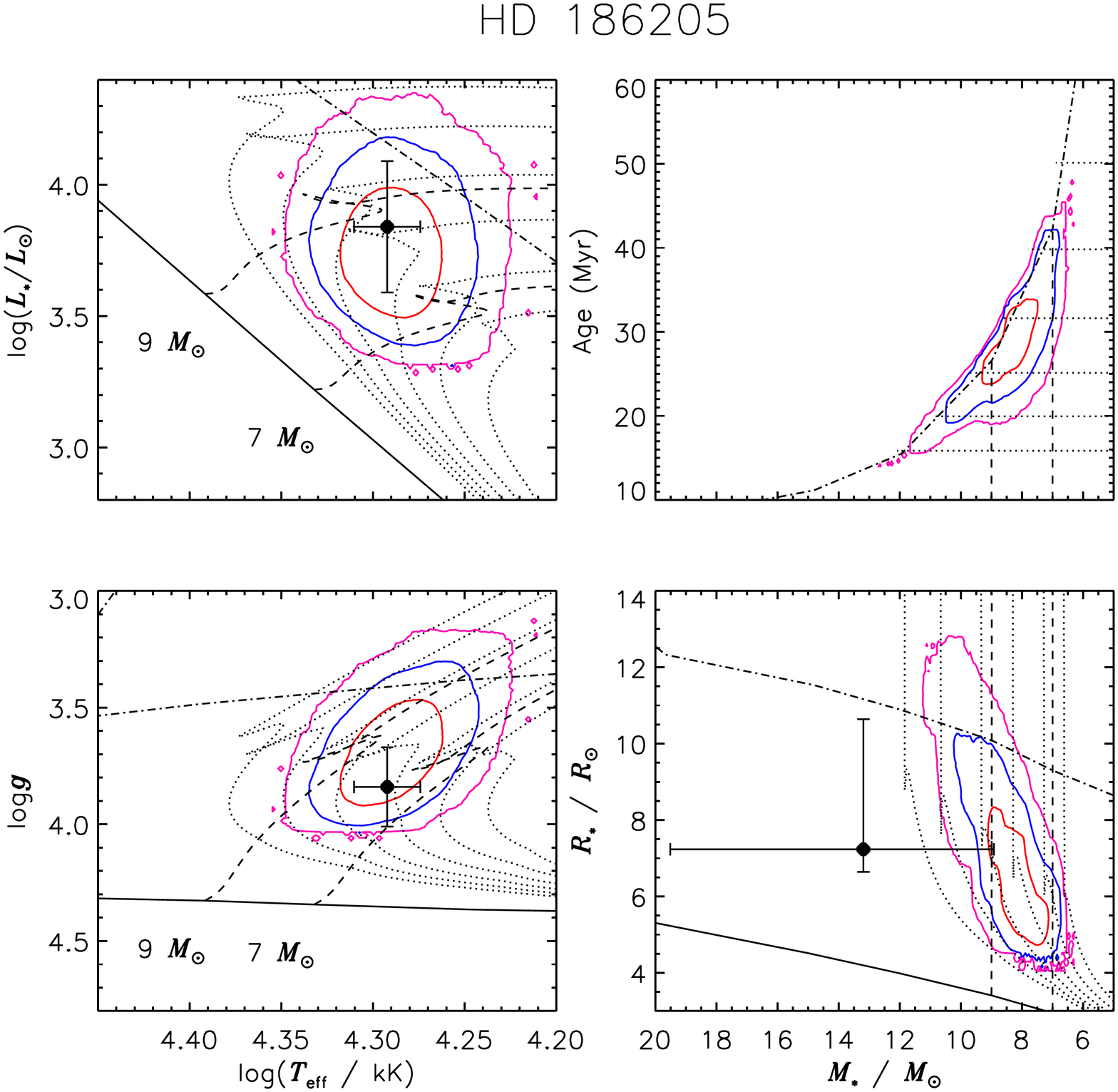} %&
      \caption[]{Monte Carlo stellar parameter determination for HD 186205. The panels show test point density contours (red, blue, and purple are 1, 2, and 3$\sigma$ contours, respectively), in the HRD (top left), the \teff-$\log{g}$ plane (bottom left), the $M_*$-$R_*$ plane (bottom right) and the $M_*$-age plane (top right). The ZAMS and TAMS are indicated by solid and dot-dashed lines; dashed lines show evolutionary tracks, and dotted lines show isochrones from $\log{(t/{\rm Myr})} = 7.2$ to 7.7 in increments of 0.1. Filled black circles indicate the spectroscopic parameters. Note the tilt in the contours in the \teff-$\log{g}$ plane due to the correlation function (Paper II). Requiring consistency between $\log{L}$ and $\log{g}$ leads to improved constraints on $M_*$ and $R_*$. The spectroscopic mass is much higher than the mass obtained from the HRD.}
         \label{HD186205_physrotmag_mcpars}
   \end{figure}

\begin{figure}
   \centering
   \includegraphics[width=8.5cm]{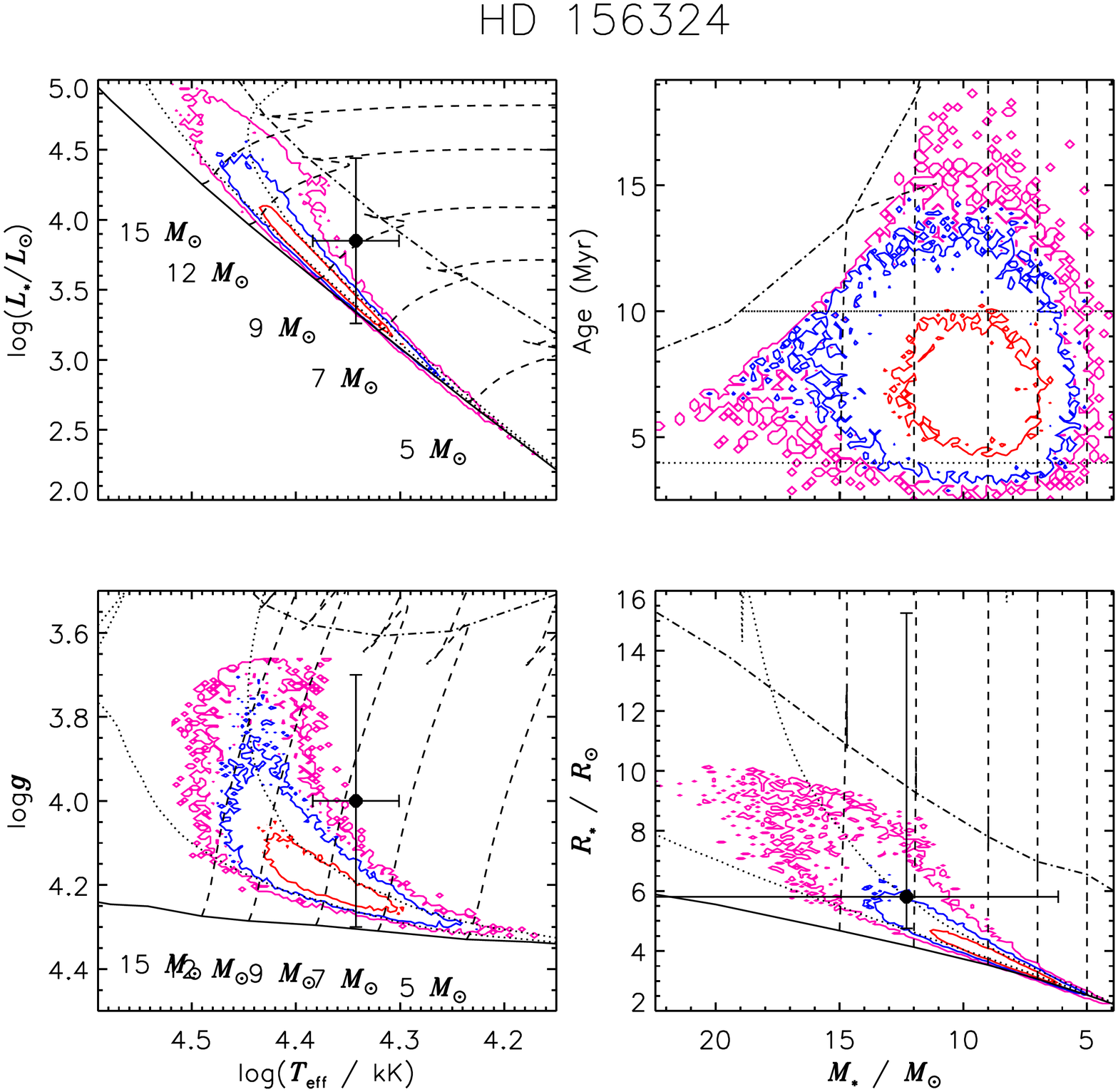} 
      \caption[]{As Fig.\ \ref{HD186205_physrotmag_mcpars}, for the cluster star HD 156324. In this case the star's parameters have been limited to be consistent with the star's cluster age, the upper and lower bounds of which are indicated by the two isochrones (dotted lines). This results in a much more precise determination of the mass and radius of the star than would be possible using its luminosity and surface gravity, both of which are highly uncertain.}
         \label{HD156324_physrotmag_mcpars}
   \end{figure}

\begin{figure}
   \centering
   \includegraphics[width=8.5cm]{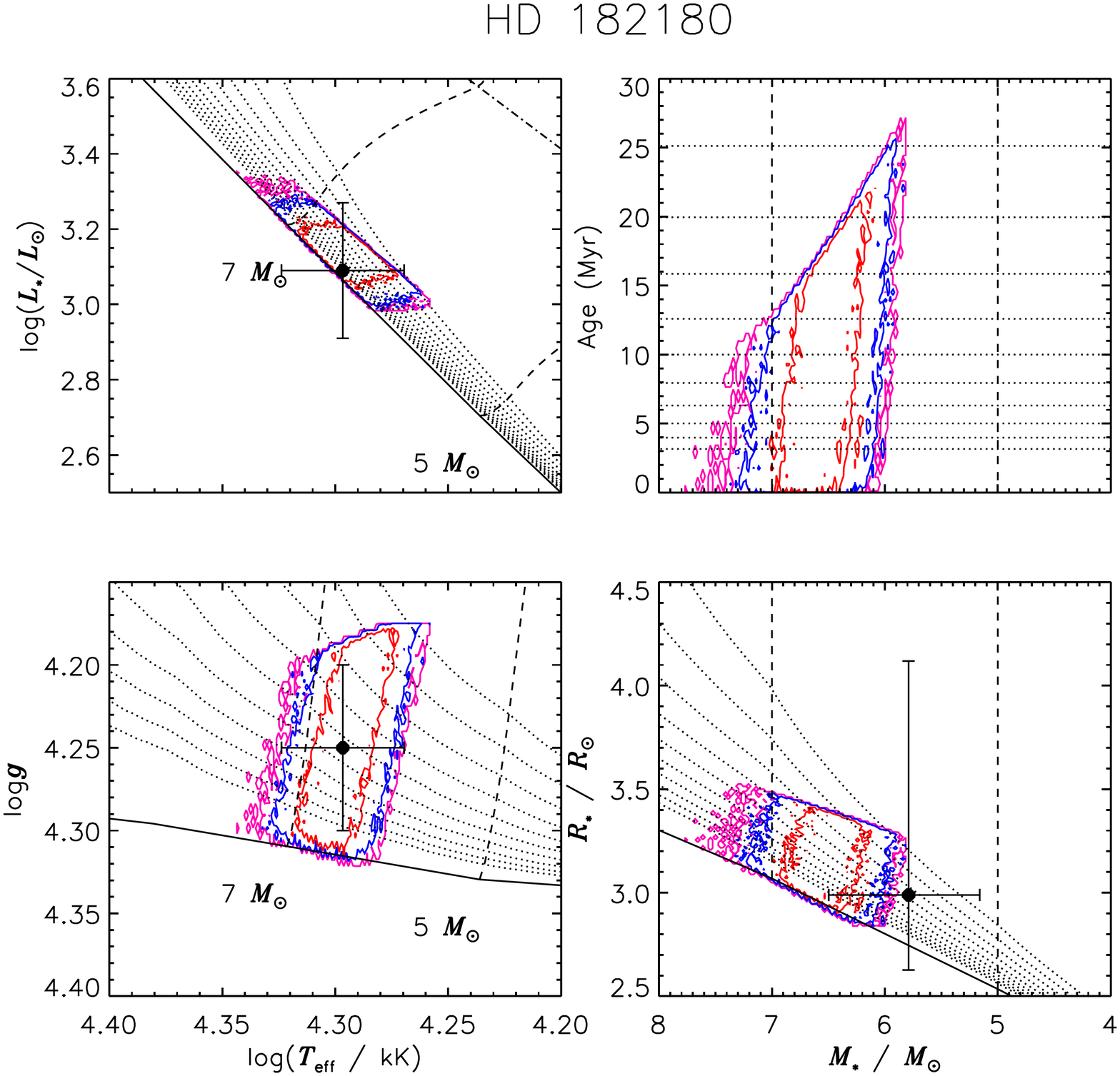} \\
      \caption[]{As Fig.\ \ref{HD186205_physrotmag_mcpars}, for the rapid rotator HD 182180. Isochrones are from $\log{(t /{\rm yr})} = 6.5$ to 7.4 in incremenets of 0.1. Rotational constraints lead to sharp cutoffs in the density contours at large radii, and a slightly higher mass than is obtained spectroscopically.}
         \label{HD182180_physrotmag_mcpars}
   \end{figure}

A flow-chart illustrating the relationship of priors to posteriors, and the tests for physical plausibility and consistency, is shown in Fig.\ \ref{flowchart}. Examples of the output are shown in Figs.\ \ref{HD186205_physrotmag_mcpars}-\ref{HD182180_physrotmag_mcpars}.

The algorithm proceeds as follows: 

\noindent 1. The HRD is sampled by drawing a test point from \teff~and $\log{L}$ distributions. 

\noindent 2. The stellar radius is obtained directly from \teff~and $\log{L}$ as $R_*/R_\odot = \sqrt{(L_*/L_\odot)/(T_{\rm eff}/T_\odot)^4}$, where $T_\odot = 5780$~K. The mass $M_*$ and age $t$ are obtained via linear interpolation between the evolutionary tracks and isochrones from stellar evolutionary models. Test points without either an evolutionary track or an isochrone in each quadrant of the HRD are discarded; in practice, this amounts to requiring that the test point lies above the ZAMS, an assumption compatible with the restriction of the sample to include main sequence stars only. 

\noindent 3. {\em Test 1}: The surface gravity is calculated from $R_*$ and $M_*$, and is compared to a target value drawn from a Gaussian distribution based upon the measured value of $\log{g}$, tilted according to the \teff-$\log{g}$ correlation function (Paper II). Test points with $\log{g}$ differing from the target value by greater than 1$\sigma$ are discarded. 

\noindent 4. {\em Test 2}: If the cluster age $t_{\rm cl}$ is available, $t$ is compared to a target value drawn from a Gaussian distribution based upon $t_{\rm cl}$. Test points yielding an age differing by more than 1$\sigma$ from the target value are discarded. 

\noindent 5. {\em Test 3}: If the binary mass is available, the same test is applied to $M_*$ as to $\log{g}$ and $t$.

\noindent 6a. If $P_{\rm rot}$ is known, the equatorial rotational velocity $v_{\rm eq}$ is calculated from $R_*$ and $P_{\rm rot}$. In this step the rotational oblateness $R_{\rm p}/R_{\rm eq}$, breakup velocity $v_{\rm br}$, and Kepler corotation radius \rk~are calculated from the mass, radius, and angular velocity \citep{1928asco.book.....J,town2005c,ud2008}. 

\noindent 6b. If $P_{\rm rot}$ is not known, a random value of $i_{\rm rot}$ is assigned, and $v_{\rm eq}$, $v_{\rm br}$, $R_{\rm p}/R_{\rm eq}$, and \rk~are calculated using $i_{\rm rot}$ and \vsini. 

\noindent 7. {\em Test 4}: Test points yielding $v\sin{i} < v_{\rm eq} < v_{\rm br}$ are rejected.

\noindent 8. The rotational inclination $i_{\rm rot}$ is calculated from \vsini~and $v_{\rm eq}$.

\noindent 9. {\em Test 5}: If a prior constraint on $i_{\rm rot}$ is available, a similar test is applied as in Steps 2, 3, and 5. 

\noindent 10. Obliquities $\beta$ and dipole magnetic field strengths $B_{\rm d}$ are determined using $i_{\rm rot}$ and the sinusoidal fitting parameters to \bz, to solve Preston's equations \citep{preston1967}. Calculating $B_{\rm d}$ also requires the limb darkening coefficient, which is obtained via linear interpolation within the tables calculated by \cite{2016MNRAS.456.1294R}. 

\noindent 11. Mass-loss rates \mdot~and wind terminal velocities \vinf~are determined using both the recipes determined by \cite{vink2001} and \cite{krticka2014}, and from these Alfv\'en radii \ra, magnetic braking timescales $\tau_{\rm J}$, and spindown ages $t_{\rm S,max}$ are calculated \citep{ud2002,ud2008,ud2009,petit2013}. 

\noindent 12. If the total number of accepted test points reaches a pre-determined number (we selected 30,000, so as to achieve a reasonably smooth sampling of parameter space), the algorithm terminates. The final values of all parameters are determined from the peaks of the posterior probability density functions, with asymmetric uncertainties arising naturally from the standard deviations above and below the peak values. 

Fig.\ \ref{HD186205_physrotmag_mcpars} illustrates the basic method for HD 186205. In this case due to the star's distance the surface gravity is measured with greater precision than the luminosity, and the final distribution reflects the better constraints on $\log{g}$. The evolutionary models intersecting the star's surface parameters in the HRD and the \teff-$\log{g}$ plane are not perfectly overlapping: the $\log{L}$ would suggest $M_* \sim 9 M_\odot$, while $\log{g}$ is more compatible with $7 M_\odot$. The algorithm finds a compromise solution between the two at about 8 $M_\odot$. Forcing agreement leads to relatively precise constraints on stellar parameters as compared to those available via the photometric radius and spectroscopic mass. In addition to a reduced uncertainty in $M_*$, the spectroscopic mass is almost twice as high as the mass determined via the MC algorithm; thus, the MC algorithm avoids a systematic substantial systematic error. 

The effect of the cluster age constraint is shown for the cluster star HD 156324 in the middle panels of Fig.\ \ref{HD156324_physrotmag_mcpars}. For this star, the uncertainties in both $\log{L}$ and $\log{g}$ are significant, due both to distance, and to its status as an SB3, which makes its surface parameters difficult to determine. Requiring that the parameters be consistent with the age of the cluster leads to much tighter constraints on $R_*$ and $M_*$ than would be possible via photometric and spectroscopic parameters alone. Note that at least in the case of HD 156324, imposing consistency with the cluster age yields a rotational inclination identical to the orbital inclination, as must be the case for this tidally locked star \citep{2018MNRAS.475..839S}; by contrast, the value determined directly from the photometric luminosity yields a much smaller inclination.

Fig.\ \ref{HD182180_physrotmag_mcpars} shows the effect of rotational constraints for the rapidly rotating star HD 182180 ($P_{\rm rot} \sim 0.5$~d, \citealt{rivi2013}). The requirement that the star be rotating at or below its critical velocity rules out radii above about 3.5 $R_*$, which are formally consistent with its \teff~and $\log{L}$. 

\section{Additional constraints}\label{appendix:optional_priors}

\begin{table}
\caption[]{Cluster ages. Where two references are provided, the top reference is for membership, the bottom reference for age.}
\label{clustertab}
\resizebox{8.5 cm}{!}{
\begin{tabular}{l | l r l }
\hline\hline
HD No. & Cluster & $\log{(t/Myr)}$ & References \\
\hline
 23478   & IC348 & $6.78\pm0.10$ & {\cite{2003A&A...402..587S}} \\
 & & &{\cite{2013MNRAS.434..806B}} \\
 35298   & Ori OB1a & $7.00\pm0.10$ & {\cite{land2007}} \\
 35502   & Ori OB1a & $7.00\pm0.10$ & {\cite{land2007}} \\
 36485   & Ori OB1b & $6.55\pm0.15$ & {\cite{land2007}} \\
 36526   & Ori OB1b & $6.55\pm0.15$ & {\cite{land2007}} \\
 36982   & Ori OB1d & $6.00\pm0.20$ & {\cite{land2007}} \\
 37017   & NGC 1981 & $6.69\pm0.09$ & {\cite{2010MNRAS.407.1875M}} \\
 37058   & Ori OB1c & $6.66\pm0.20$ & {\cite{land2007}} \\
 37061   & ONC/M43 & $5.75\pm0.25$ & {\cite{2008MNRAS.386..261M}} \\
 & & &{\cite{2018ApJ...859..166F}} \\
 37479   & Ori OB1b & $6.55\pm0.15$ & {\cite{land2007}} \\
 37776   & Ori OB1b & $6.55\pm0.15$ & {\cite{land2007}} \\
 47777   & NGC 2264 & $6.82\pm0.32$ & {\cite{2014A&A...562A.143F}} \\
 66765   & Vel OB2 & $7.30\pm0.20$ & {\cite{1999AJ....117..354D}} \\
 67621   & Vel OB2 & $7.30\pm0.20$ & {\cite{1999AJ....117..354D}} \\
105382   & Lower Cen Crux & $7.10\pm0.10$ & {\cite{2007AJ....133.1092W}} \\
 & & &{\cite{1995ApJ...454..151M}} \\
121743   & Upper Cen Lup & $7.16\pm0.10$ & {\cite{2007AJ....133.1092W}} \\
 & & &{\cite{1995ApJ...454..151M}} \\
125823   & Upper Cen Lup & $7.20\pm0.10$ & {\cite{land2007}} \\
130807   & Upper Cen Lup & $7.16\pm0.10$ & {\cite{2007AJ....133.1092W}} \\
 & & &{\cite{1995ApJ...454..151M}} \\
%142184   & Upper Sco & $7.04\pm0.07$ & {\cite{2005AJ....129..856H}} \\
%         &          &               & {\cite{2012ApJ...746..154P}} \\
142990   & Upper Sco & $6.70\pm0.10$ & {\cite{land2007}} \\
149277   & NGC 6178 & $7.15\pm0.22$ & {\cite{land2007}} \\
142438   & Upper Sco & $7.04\pm0.07$ & {\cite{1989A&A...216...44D}} \\
         &          &               & {\cite{2012ApJ...746..154P}} \\
156324   & Sco OB4 & $6.82\pm0.20$ & {\cite{2005AA...438.1163K}} \\
156424   & Sco OB4 & $6.82\pm0.20$ & {\cite{2005AA...438.1163K}} \\
164492   & Trifid & $5.75\pm0.25$ & {\cite{2008AA...489..157L}} \\
ALS 3694 & NGC 6193 & $6.69\pm0.20$ & {\cite{land2007}} \\
CPD $-57^\circ$ 3509   & NGC 3293 & $7.02\pm0.10$ & {\cite{1994A&AT....4..153L}} \\
CPD $-62^\circ$ 2124   & IC 2944 & $6.50\pm0.25$ & {\cite{2014MNRAS.443..411B}} \\
\hline\hline
\end{tabular}
}
\end{table}

In many cases, information is available that can be used to further constrain stellar properties. For about half of the sample, their membership in a star cluster provides the age via the main sequence turnoff. Several stars are in close binary systems, and the mass ratios of the components can be used to constrain their masses. Finally, in some cases the rotational inclination has been constrained via emission properties, Doppler Imaging (DI), or asteroseismology. The additional constraints utilized for individual stars are indicated in Table \ref{physrottab}. 

\subsection{Cluster ages}\label{subsubsec:cluster_ages}

As discussed by \cite{bagn2006}, ages inferred from evolutionary tracks and the positions of individual stars on the HRD are fairly imprecise for stars close to the Zero-Age Main Sequence (ZAMS). The evolution of main sequence stars across the HRD is initially relatively slow, speeding up in later evolutionary stages. Conversely, cluster ages, which are principally determined by the location of the main-sequence turnoff, are much more precise for younger stars. This has motivated a careful search for magnetic stars that are members of stellar clusters \citep{bagn2006,land2007,land2008}. In the present sample, 28 stars have been identified as probable cluster members. These stars are listed with their cluster memberships and ages, along with the relevant citations, in Table \ref{clustertab}. 

Cluster age constraints were not adopted for HD\,96446 or HD\,149438. HD 96446 was suggested to be the central star of the Feinstein 1 association by \cite{2014A&A...564A..42P}, however the Gaia distance to this star, $540 \pm 20$ pc, is less than half the distance to the cluster of $1200^{+400}_{-200}$ pc given by \citeauthor{2014A&A...564A..42P}. We therefore consider it unlikely to be a member. HD 149438 ($\tau$ Sco) is a member of the Upper Sco OB association \citep{1989A&A...216...44D}, however its surface parameters are impossible to reconcile with the 11 Myr age given by \cite{2012ApJ...746..154P}, leading to the suggestion that the star is a blue straggler \citep{2016MNRAS.457.2355S}. It was therefore treated as a field star for the purposes of parameter determination. 

\subsection{Close binaries} 

There are several close binaries in our sample for which orbital solutions are available in the literature. In these cases, stellar masses were constrained using a Monte Carlo algorithm similar to the one described in Appendix \ref{monte_carlo}, but adapted for close binary systems. A detailed description of this variant is given by \cite{pablo_epslup} and \cite{2019MNRAS.482.3950S}. Briefly, the algorithm uses the distance, the observed $V$ magnitude, the \teff~and $\log{g}$ of the individual components, and the mass ratio $M_1/M_2$ obtained via the semi-amplitudes of the radial velocity curves. Test points are generated via \teff~and $\log{g}$, and retained if they are consistent with $M_1/M_2$ and $V$. Since close binaries are expected to be coeval, test points for the individual stellar components are also required to be within $\Delta\log{(t/{\rm yr})}=0.1$ of the same age. Derived rotational, magnetic, and magnetospheric parameters are then obtained as above. 

The systems for which this method was utilized are: HD 36485 \citep[$M_1/M_2=2.6\pm0.4$;][]{leone2010}; HD 37017 \citep[$M_1/M_2=2.0\pm0.1$;][]{1998AA...337..183B}; HD 37061 \citep[$M_1/M_2=2.3\pm0.4$;][]{2019MNRAS.482.3950S}; HD 122451 \citep[$M_1/M_2=1.14\pm0.03$;][]{2005MNRAS.356.1362D,2006AA...455..259A,2016A&A...588A..55P}; HD 136504 \citep[$M_1/M_2=1.19\pm0.02$;][]{uytterhoeven2005,pablo_epslup}; HD 149277 \citep[$M_1/M_2=1.11\pm0.05$;][]{2016PhDT.......390S,2018MNRAS.481L..30G}; and HD 156324 \citep[$M_1/M2=0.38\pm0.09$;][]{2018MNRAS.475..839S}. 

In the cases of HD\,37061, HD\,122451, and HD\,136504, the masses of individual components were also constrained interferometrically, with the masses of the magnetic components as $7.0\pm1.7$~\msun~for HD 37061 C \citep{2019MNRAS.482.3950S}; $10.6\pm0.2$~\msun~for HD 122451 Ab \citep{2016A&A...588A..55P}; and $8^{+4}_{-1}$~\msun~and $7^{+3}_1$~\msun~for HD 136504 Aa and Ab \citep{pablo_epslup}.

\subsection{Rotational inclinations}\label{subsubsec:inc}

If $i_{\rm rot}$ is already known, this can be used to constrain the allowable values of $R_*$. Constraints from emission properties are available for HD 37479 \citep[$i>80^\circ$;][]{2015MNRAS.451.2015O} and HD 176582 \cite[$i>85^\circ$;][]{bohl2011}. Asteroseismic constraints are available for HD 44743 \citep[$55.3\pm1.5^\circ$;][]{2015AA...574A..20F}. Constraints from DI are available for HD 125823 \citep[$i=70\pm5^\circ$][]{2010AA...520A..44B}. Asteroseismic and DI constraints are also available for several other stars \citep[HD 37776, HD 43317, HD 149438, HD 184927;][]{koch2010,2012A&A...542A..55P,2006MNRAS.370..629D,2015MNRAS.447.1418Y}; however, in these cases identical results were found directly from the stellar and rotational properties. 

For HD 136504Aab $P_{\rm rot}$ is not known, but the evidence strongly favours spin-orbit alignment, thus we fixed $i_{\rm rot}$ to the orbital inclination \citep[$i_{\rm orb}=20\pm2^\circ$,][]{pablo_epslup}. 

$P_{\rm rot}$ is also unknown for HD\,58260 due to the lack of variation in \bz~(see Paper I). \cite{2018A&A...618L...2J} claimed to detect a slight variation in \bz~consistent with very slow rotation, however $P_{\rm rot}$ still cannot be determined for this star. \bz~is formally compatible with small $\beta$, small $i_{\rm rot}$, or an extremely long rotation period. The same is true of \vsini, which is effectively 0. However, in the case of small $\beta$, large $i_{\rm rot}$, and negligible \vsini~the Stokes $V$ signature should effectively disappear due to flux cancellation. The same is true for small $i_{\rm rot}$ and large $\beta$. The hypothesis that is most consistent with very small variation in \bz~and the strong, positive Stokes $V$ signature is that both $i_{\rm rot}$ and $\beta$ are small; therefore in this case $i_{\rm rot}$ was fixed to be below $10^\circ$. If a random distribution is instead adopted as per Eqn.\ 1, the most likely value and uncertainties of $\beta$ remain the same; the main difference is an extended low-probability tail towards very high $B_{\rm d}$, with the most likely value of $B_{\rm d}$ being unchanged. 

\section{Tabulated Parameters}\label{parameter_tables}

\begin{table*}
\caption[]{Stellar and rotational parameters derived from \protect\cite{ekstrom2012} models. Labels in superscripts next to the stellar identifier in the first column indicate the type of model used ({\em R}otating or {\em N}on-rotating), as well as additional prior constraints ({\em a}, cluster age; {\em i}, inclination; {\em m}, mass) or special conditions ({\em r}, assumed random inclination; {\em u}, unknown period). The columns are equatorial stellar radius $R_*$, stellar mass $M_*$, logarithmic age, fractional main-sequence age $\tau_{\rm TAMS}$, the rotational oblateness $R_{\rm p}/R_{\rm e}$, the equatorial rotational velocity $v_{\rm eq}$, the rotation parameter $W$, and the Kepler co-rotation radius $R_{\rm K}$.}
\label{physrottab}
\begin{tabular}{l | r r r r r r r r}
\hline\hline
HD No. & $R_*$       & $M_*$       & log(Age)   & $\tau_{\rm TAMS}$ & $R_{\rm p}/R_{\rm e}$ & $v_{\rm eq}$ & $\log{W}$ & $R_{\rm K}$      \\
       & ($R_\odot$) & ($M_\odot$) & (yr)      &                   &                       & (\kms)       &           & $(R_*)$          \\
\hline
  3360$^{R}$ & $6.2\pm0.3$ & $8.6\pm0.1$ & $7.46\pm0.02$ & $0.81\pm0.04$ & 1 & $58\pm2$ & $0.94\pm0.03$ & $4.2\pm0.2$   \\  [3pt]
 23478$^{R, a}$ &$3.2\pm0.2$ & $6.8\pm0.5$ & $6.8\pm0.1$ & $0.12\pm0.03$ & $0.971\pm0.003$ & $159\pm9$ & $0.59\pm0.02$ & $2.48\pm0.09$   \\  [3pt]
 25558$^{R}$ & $2.6^{+0.6}_{-0.5}$ & $4.9\pm0.4$ & $7.8\pm0.3$ & $0.4\pm0.2$ & $0.98^{+0.00}_{-0.02}$ & $107^{+27}_{-18}$ & $0.68\pm0.09$ & $2.5\pm0.4$   \\  [3pt]
 35298$^{R, a}$ &$2.42\pm0.09$ & $4.3\pm0.2$ & $7.0\pm0.1$ & $0.06\pm0.02$ & 1 & $66\pm2$ & $0.94\pm0.01$ & $4.23\pm0.08$   \\  [3pt]
 35502$^{R, a}$ &$2.96\pm0.10$ & $5.8\pm0.2$ & $7.0\pm0.1$ & $0.11\pm0.03$ & $0.960\pm0.003$ & $180\pm5$ & $0.52\pm0.01$ & $2.22\pm0.05$   \\  [3pt]
 36485$^{R, am}$ &$2.97\pm0.08$ & $6.3\pm0.2$ & $6.6\pm0.1$ & $0.05\pm0.02$ & 1 & $102\pm2$ & $0.79\pm0.01$ & $3.35\pm0.06$   \\  [3pt]
 36526$^{R, a}$ &$2.35^{+0.09}_{-0.12}$ & $4.3^{+0.2}_{-0.3}$ & $6.68^{+0.09}_{-0.13}$ & $0.025^{+0.006}_{-0.010}$ & 1 & $77^{+2}_{-3}$ & $0.88^{+0.01}_{-0.01}$ & $3.84^{+0.06}_{-0.08}$   \\  [3pt]
 36982$^{R, a}$ &$3.2\pm0.1$ & $7.4\pm0.5$ & $6.1\pm0.2$ & $0.02\pm0.02$ & 1 & $88\pm3$ & $0.87\pm0.01$ & $3.82\pm0.06$   \\  [3pt]
 37017$^{R, am}$ &$3.6\pm0.2$ & $8.4\pm0.5$ & $6.73\pm0.10$ & $0.13\pm0.03$ & $0.952\pm0.004$ & $212\pm9$ & $0.48\pm0.02$ & $2.09\pm0.06$   \\  [3pt]
 37058$^{N, a}$ &$2.8\pm0.1$ & $5.8\pm0.2$ & $7.0\pm0.2$ & $0.12\pm0.07$ & 1 & $9.8\pm0.5$ & $1.80\pm0.02$ & $15.6\pm0.6$   \\  [3pt]
 37061$^{R, am}$ &$3.3\pm0.1$ & $7.7\pm0.5$ & $6.0\pm0.2$ & $0.02\pm0.01$ & $0.974\pm0.001$ & $156\pm4$ & $0.63\pm0.01$ & $2.62\pm0.04$   \\  [3pt]
 37479$^{R, ai}$ &$3.39^{+0.04}_{-0.06}$ & $7.9^{+0.2}_{-0.3}$ & $6.5^{+0.1}_{-0.1}$ & $0.06^{+0.02}_{-0.03}$ & $0.9759^{+0.0006}_{-0.0008}$ & $147^{+1}_{-2}$ & $0.649^{+0.006}_{-0.008}$ & $2.69^{+0.02}_{-0.03}$   \\  [3pt]
 37776$^{R, a}$ &$3.5\pm0.1$ & $8.3\pm0.3$ & $6.6\pm0.2$ & $0.10\pm0.04$ & $0.985\pm0.001$ & $118\pm4$ & $0.75\pm0.02$ & $3.13\pm0.08$   \\  [3pt]
 43317$^{R}$ & $3.1\pm0.4$ & $5.5\pm0.2$ & $7.7\pm0.2$ & $0.5\pm0.2$ & $0.95^{+0.02}_{-0.03}$ & $184\pm28$ & $0.44\pm0.10$ & $1.9\pm0.3$   \\  [3pt]
 44743$^{N, iu}$ &$8.4\pm0.5$ & $12.5\pm0.4$ & $7.10\pm0.03$ & $0.93\pm0.04$ & 1 & $24\pm2$ & $1.34\pm0.05$ & $7.8\pm0.6$   \\  [3pt]
 46328$^{N, r}$ &$8.0\pm0.7$ & $14.3\pm0.8$ & $6.97\pm0.04$ & $0.69\pm0.03$ & 1 & $0.037\pm0.003$ & $4.19\pm0.05$ & $623\pm44$   \\  [3pt]
 47777$^{R, a}$ &$3.6\pm0.3$ & $7.9\pm0.4$ & $7.0\pm0.3$ & $0.1^{+0.2}_{-0.1}$ & 1 & $68\pm6$ & $0.97\pm0.05$ & $4.4\pm0.3$   \\  [3pt]
 52089$^{N, ru}$ &$10.1\pm0.5$ & $11.9\pm0.4$ & $7.19\pm0.02$ & $0.9889^{+0.0008}_{-0.0084}$ & 1 & $21^{+6}_{-2}$ & $1.36^{+0.12}_{-0.06}$ & $7^{+1}_{-1}$   \\  [3pt]
 55522$^{R}$ & $4.2\pm0.3$ & $5.9\pm0.2$ & $7.75\pm0.05$ & $0.65\pm0.06$ & 1 & $74\pm6$ & $0.82\pm0.04$ & $3.5\pm0.2$   \\  [3pt]
 58260$^{N, iu}$ &$3.3\pm0.7$ & $6.2\pm0.5$ & $7.5\pm0.2$ & $0.6\pm0.3$ & 1 & $12^{+17}_{-10}$ & $1.5\pm0.3$ & $7\pm3$   \\  [3pt]
 61556$^{R}$ & $3.3\pm0.6$ & $6.1\pm0.3$ & $7.6\pm0.2$ & $0.5\pm0.2$ & 1 & $95\pm15$ & $0.8\pm0.1$ & $3.2\pm0.5$   \\  [3pt]
 63425$^{N}$ & $6.9\pm0.6$ & $15.5\pm0.6$ & $6.86\pm0.06$ & $0.69\pm0.08$ & 1 & $2.1\pm0.2$ & $2.48\pm0.05$ & $44\pm3$   \\  [3pt]
 64740$^{R}$ & $4.5\pm0.7$ & $10.1\pm0.5$ & $7.1\pm0.2$ & $0.5\pm0.2$ & $0.97^{+0.01}_{-0.02}$ & $178\pm29$ & $0.54\pm0.10$ & $2.1\pm0.3$   \\  [3pt]
 66522$^{N, r}$ &$4.4\pm1.2$ & $8.3\pm0.8$ & $7.4\pm0.2$ & $0.96^{+0.02}_{-0.35}$ & 1 & $0.24\pm0.07$ & $3.3\pm0.2$ & $145\pm37$   \\  [3pt]
 66665$^{N}$ & $7.5\pm1.2$ & $15.9\pm1.2$ & $6.92\pm0.05$ & $0.64\pm0.07$ & 1 & $15\pm2$ & $1.61\pm0.08$ & $10\pm1$   \\  [3pt]
 66765$^{R, a}$ &$3.9\pm0.7$ & $7.2\pm0.6$ & $7.5\pm0.2$ & $0.5\pm0.2$ & $0.98^{+0.01}_{-0.02}$ & $123\pm23$ & $0.7\pm0.1$ & $2.8\pm0.4$   \\  [3pt]
 67621$^{R, a}$ &$3.7\pm0.3$ & $7.4\pm0.3$ & $7.3\pm0.1$ & $0.3\pm0.1$ & 1 & $53\pm4$ & $1.07\pm0.05$ & $5.1\pm0.4$   \\  [3pt]
 96446$^{N}$ & $4.9\pm0.7$ & $9.3\pm0.5$ & $7.23\pm0.10$ & $0.7\pm0.2$ & 1 & $10\pm1$ & $1.75\pm0.09$ & $13\pm1$   \\  [3pt]
105382$^{R, a}$ &$3.0\pm0.1$ & $5.8\pm0.2$ & $7.2\pm0.1$ & $0.16\pm0.04$ & $0.982\pm0.002$ & $118\pm4$ & $0.70\pm0.02$ & $2.93\pm0.08$   \\  [3pt]
121743$^{R, a}$ &$4.2\pm0.4$ & $8.3\pm0.5$ & $7.23\pm0.09$ & $0.4\pm0.1$ & $0.95\pm0.01$ & $192\pm20$ & $0.50\pm0.06$ & $2.0\pm0.2$   \\  [3pt]
122451$^{R, m}$ &$8.4\pm1.6$ & $10.6\pm0.2$ & $7.40^{+0.03}_{-0.07}$ & $0.97^{+0.01}_{-0.07}$ & $0.95\pm0.03$ & $154\pm31$ & $0.5\pm0.1$ & $2.1\pm0.4$   \\  [3pt]
125823$^{N, ai}$ &$2.97^{+0.07}_{-0.10}$ & $5.9^{+0.2}_{-0.3}$ & $7.14^{+0.07}_{-0.10}$ & $0.16^{+0.03}_{-0.03}$ & 1 & $17.0^{+0.4}_{-0.6}$ & $1.56^{+0.01}_{-0.01}$ & $10.9^{+0.2}_{-0.2}$   \\  [3pt]
127381$^{R}$ & $4.7\pm0.5$ & $9.2\pm0.4$ & $7.3\pm0.1$ & $0.6\pm0.1$ & 1 & $79\pm8$ & $0.87\pm0.07$ & $3.8\pm0.4$   \\  [3pt]
130807$^{R, a}$ &$2.88\pm0.09$ & $5.4\pm0.2$ & $7.2\pm0.1$ & $0.17\pm0.04$ & 1 & $49\pm1$ & $1.08\pm0.02$ & $5.3\pm0.1$   \\  [3pt]
136504Aa$^{R, imau}$ &$5.1\pm0.7$ & $7.7\pm0.5$ & $7.49\pm0.05$ & $0.7\pm0.1$ & $0.976\pm0.005$ & $116\pm11$ & $0.64\pm0.05$ & $2.7\pm0.2$   \\  [3pt]
136504Ab$^{R, imau}$ &$3.7\pm0.5$ & $6.4\pm0.3$ & $7.59\pm0.08$ & $0.5\pm0.1$ & 1 & $76\pm7$ & $0.85\pm0.05$ & $3.6\pm0.3$   \\  [3pt]
\hline\hline
\end{tabular}
\end{table*}
 
\begin{table*}
\contcaption{}
\label{physrottab:continued}
\begin{tabular}{l | r r r r r r r r}
\hline
\hline
HD No. & $R_*$       & $M_*$       & log(Age)   & $\tau_{\rm TAMS}$ & $R_{\rm p}/R_{\rm e}$ & $v_{\rm eq}$ & $\log{W}$ & $R_{\rm K}$      \\
       & ($R_\odot$) & ($M_\odot$) & (yr)      &                   &                       & (\kms)       &           & $(R_*)$          \\
\hline
142184$^{R}$ & $2.8\pm0.1$ & $5.7\pm0.1$ & $7.2\pm0.3$ & $0.05^{+0.10}_{-0.05}$ & $0.89^{+0.01}_{-0.02}$ & $312\pm21$ & $0.28^{+0.05}_{-0.03}$ & $1.5^{+0.1}_{-0.1}$   \\  [3pt]
142990$^{R, a}$ &$2.79\pm0.06$ & $5.6\pm0.2$ & $6.7\pm0.1$ & $0.06\pm0.01$ & $0.973\pm0.001$ & $147\pm3$ & $0.621\pm0.009$ & $2.58\pm0.04$   \\  [3pt]
149277$^{N, am}$ &$4.2\pm0.4$ & $8.0\pm0.4$ & $7.3\pm0.1$ & $0.6\pm0.1$ & 1 & $8.3\pm0.9$ & $1.85\pm0.06$ & $16\pm1$   \\  [3pt]
149438$^{N}$ & $6.4\pm0.6$ & $17.5\pm0.9$ & $6.73\pm0.10$ & $0.5\pm0.1$ & 1 & $7.8\pm0.7$ & $1.96\pm0.06$ & $19\pm1$   \\  [3pt]
156324$^{R, am}$ &$3.9\pm0.6$ & $8.7\pm1.2$ & $7.0\pm0.2$ & $0.2\pm0.2$ & $0.98^{+0.01}_{-0.01}$ & $134\pm20$ & $0.69\pm0.08$ & $2.9\pm0.3$   \\  [3pt]
156424$^{R, a}$ &$4.2\pm1.0$ & $9.5\pm1.9$ & $7.0\pm0.2$ & $0.2^{+0.3}_{-0.2}$ & 1 & $74\pm17$ & $1.0\pm0.1$ & $4.2\pm0.7$   \\  [3pt]
163472$^{R}$ & $4.6\pm0.2$ & $10.3\pm0.5$ & $7.0\pm0.1$ & $0.41\pm0.10$ & 1 & $63\pm3$ & $1.00\pm0.04$ & $4.6\pm0.2$   \\  [3pt]
164492$^{R, a}$ &$4.02^{+0.09}_{-0.13}$ & $11.1^{+0.3}_{-0.5}$ & $5.9^{+0.1}_{-0.2}$ & $0.03^{+0.02}_{-0.02}$ & $0.978^{+0.001}_{-0.001}$ & $151^{+3}_{-4}$ & $0.683^{+0.008}_{-0.012}$ & $2.83^{+0.05}_{-0.07}$   \\  [3pt]
175362$^{R}$ & $2.7\pm0.2$ & $5.3\pm0.2$ & $7.2\pm0.3$ & $0.06^{+0.11}_{-0.06}$ & 1 & $36\pm2$ & $1.22^{+0.04}_{-0.03}$ & $6.5\pm0.4$   \\  [3pt]
176582$^{R, i}$ &$3.21\pm0.06$ & $5.6\pm0.3$ & $7.5\pm0.1$ & $0.32\pm0.06$ & $0.984\pm0.001$ & $104\pm2$ & $0.74\pm0.02$ & $3.09\pm0.08$   \\  [3pt]
182180$^{R}$ & $3.2\pm0.1$ & $6.5\pm0.2$ & $7.1^{+0.3}_{-0.4}$ & $0.09\pm0.09$ & $0.89^{+0.02}_{-0.03}$ & $336\pm26$ & $0.25\pm0.05$ & $1.45\pm0.09$   \\  [3pt]
184927$^{N}$ & $4.2\pm0.8$ & $8.4\pm0.5$ & $7.3\pm0.1$ & $0.7\pm0.2$ & 1 & $22\pm4$ & $1.4\pm0.1$ & $8\pm1$   \\  [3pt]
186205$^{N}$ & $6.2\pm1.1$ & $8.3\pm0.6$ & $7.50\pm0.07$ & $0.81^{+0.03}_{-0.07}$ & 1 & $8\pm1$ & $1.8\pm0.1$ & $14\pm2$   \\  [3pt]
189775$^{R}$ & $3.2\pm0.3$ & $5.6\pm0.2$ & $7.7\pm0.1$ & $0.4\pm0.1$ & 1 & $62\pm5$ & $0.94\pm0.05$ & $4.2\pm0.3$   \\  [3pt]
205021$^{N}$ & $8.3\pm0.6$ & $11.9\pm1.2$ & $7.19\pm0.09$ & $0.91\pm0.04$ & 1 & $35\pm2$ & $1.19\pm0.04$ & $6.1\pm0.4$   \\  [3pt]
208057$^{R}$ & $3.8\pm0.5$ & $5.3\pm0.3$ & $7.8\pm0.1$ & $0.7\pm0.2$ & $0.97\pm0.02$ & $141\pm23$ & $0.5\pm0.1$ & $2.2\pm0.4$   \\  [3pt]
345439$^{R}$ & $3.7\pm0.5$ & $8.3\pm0.8$ & $7.1\pm0.4$ & $0.08^{+0.28}_{-0.08}$ & $0.93^{+0.02}_{-0.04}$ & $276\pm37$ & $0.39\pm0.09$ & $1.6\pm0.2$   \\  [3pt]
ALS 3694$^{R, a}$ &$3.9\pm0.4$ & $9.2\pm0.9$ & $6.8\pm0.2$ & $0.1^{+0.1}_{-0.1}$ & $0.984\pm0.004$ & $121\pm14$ & $0.75\pm0.06$ & $3.0\pm0.2$   \\  [3pt]
CPD $-62^\circ 2124$$^{R, a}$ &$3.9\pm0.4$ & $9.2\pm0.3$ & $7.1\pm0.3$ & $0.2\pm0.2$ & 1 & $75\pm8$ & $0.94\pm0.06$ & $4.2\pm0.4$   \\  [3pt]
CPD $-57^\circ 3509$$^{N, a}$ &$4.5^{+0.6}_{-0.4}$ & $9.5\pm0.4$ & $7.20\pm0.06$ & $0.5\pm0.1$ & 1 & $36^{+4}_{-3}$ & $1.24\pm0.06$ & $6.6\pm0.6$   \\  [3pt]
\hline\hline
\end{tabular}
\end{table*}

\begin{table*}
\caption[]{ORM parameters. The second column gives the Preston $r$ parameter. Inclinations $i_{\rm rot}$ are not given for stars for which this could not be determined due to unknown or very long $P_{\rm rot}$. Stars with \bz~curves indicative of multipolar surface magnetic fields are indicated with a superscript $m$.}
\label{ormtab}
\resizebox{18.5 cm}{!}{
\begin{tabular}{l | r r r r l | r r r r}
\hline\hline
HD No. & $r$ & $i_{\rm rot}$ & $\beta$    & $B_{\rm d}$ & HD No. & $r$ & $i_{\rm rot}$ & $\beta$    & $B_{\rm d}$ \\
       &     & ($^\circ$)    & ($^\circ$) & (kG)        &        &     & ($^\circ$)    & ($^\circ$) & (kG)        \\
\hline
  3360   & $-0.5\pm0.1$ & $18.8\pm1.2$  & $83\pm2$  & $0.15\pm0.03$  &  105382   & $-0.0\pm0.1$ & $37\pm2$  & $52\pm7$  & $2.6\pm0.1$  \\ [2.5pt]
 23478   & $0.75\pm0.08$ & $56\pm5$  & $4\pm2$  & $10\pm2$  &  121743   & $0.6\pm0.1$ & $21\pm2$  & $31\pm11$  & $1.1\pm0.3$  \\ [2.5pt]
 25558   & $0.0\pm0.2$ & $16\pm3$  & $76\pm5$  & $0.5\pm0.3$  &  122451   & $0.4\pm0.2$ & $28\pm6$  & $37\pm14$  & $0.23\pm0.04$  \\ [2.5pt]
35298$^m$   & $-0.79\pm0.02$ & $64\pm4$  & $78\pm2$  & $11.2\pm1.0$  &  125823   & $-0.87\pm0.04$ & $76\pm3$  & $76\pm6$  & $1.8\pm0.2$  \\ [2.5pt]
 35502   & $-0.15\pm0.03$ & $25.6\pm1.1$  & $70.4\pm1.3$  & $7.3\pm0.5$  &  127381   & $-0.8\pm0.1$ & $47\pm7$  & $82\pm5$  & $0.7\pm0.2$  \\ [2.5pt]
36485$^m$   & $0.95\pm0.02$ & $18.6\pm0.7$  & $4\pm2$  & $8.9\pm0.2$  &  130807$^m$   & $-0.25\pm0.04$ & $77\pm6$  & $22\pm9$  & $9^{+5}_{-2}$   \\ [2.5pt]
 36526   & $-0.25\pm0.03$ & $46\pm2$  & $57\pm2$  & $11.2\pm0.5$  &  136504Aa   & $0.52\pm0.07$ & $20\pm2$  & $<30.7$ & $0.79\pm0.08$  \\ [2.5pt]
 36982   & $0.20\pm0.09$ & $71\pm6$  & $10\pm5$  & $1.7\pm1.1$  &  136504Ab   & $0.53\pm0.08$ & $20\pm2$  & $<21.4$ & $0.5\pm0.2$  \\ [2.5pt]
 37017   & $-0.10\pm0.04$ & $39\pm2$  & $57\pm3$  & $6.2\pm0.9$  &  142184$^m$   & $0.48\pm0.04$ & $64\pm6$  & $9\pm3$  & $9\pm2$  \\ [2.5pt]
 37058   & $-0.60\pm0.04$ & $74\pm7$  & $55\pm13$  & $2.5\pm0.5$  &  142990$^m$   & $-0.87\pm0.08$ & $55\pm2$  & $84\pm3$  & $4.7\pm0.4$  \\ [2.5pt]
 37061   & $-0.16\pm0.08$ & $40\pm2$  & $59\pm4$  & $9.2\pm1.0$  &  149277   & $-0.70\pm0.04$ & $56\pm12$  & $78\pm9$  & $9.9\pm1.2$  \\ [2.5pt]
37479$^m$   & $-0.56\pm0.04$ & $77\pm4$  & $38\pm9$  & $10^{+2}_{-1}$   &  149438$^m$   & $-0.68\pm0.03$ & $57\pm10$  & $75\pm7$  & $0.31\pm0.02$  \\ [2.5pt]
37776$^m$   & $-0.3\pm0.2$ & $57\pm4$  & $47\pm9$  & $6.1\pm0.7$  &  156324   & $-0.13\pm0.07$ & $21\pm4$  & $75\pm4$  & $14\pm3$  \\ [2.5pt]
 43317   & $-0.7\pm0.1$ & $32\pm5$  & $84\pm3$  & $1.4\pm0.3$  &  156424   & $0.3\pm0.1$ & $5.1\pm1.2$  & $81\pm3$  & $7\pm3$  \\ [2.5pt]
 44743   & $0.0\pm0.4$ &  -- & $13\pm8$  & $0.11\pm0.02$  &  163472   & $-0.47\pm0.08$ & $67\pm8$  & $47\pm14$  & $1.1\pm0.5$  \\ [2.5pt]
 46328   & $-0.44\pm0.04$ &  -- & $71\pm22$  & $1.2^{+0.6}_{-0.1}$   &  164492   & $-0.17\pm0.04$ & $62\pm5$  & $35\pm7$  & $6.6\pm0.8$  \\ [2.5pt]
 47777   & $-0.8\pm0.1$ & $55\pm7$  & $83\pm5$  & $3.3\pm0.7$  &  175362$^m$   & $-0.654\pm0.008$ & $61\pm7$  & $69\pm6$  & $17.0^{+0.6}_{-0.4}$   \\ [2.5pt]
 52089   & $-0.3\pm0.5$ &  -- & $7^{+31}_{-7}$  & $0.07\pm0.05$  &  176582$^m$   & $-0.99\pm0.03$ & $84\pm2$  & $89.3^{+0.6}_{-1.4}$  & $5.4\pm0.2$  \\ [2.5pt]
 55522   & $-0.98\pm0.10$ & $61\pm8$  & $>88$ & $3.1\pm0.4$  &  182180   & $-0.82\pm0.06$ & $53^{+9}_{-5}$  & $82\pm4$  & $9.5\pm0.6$  \\ [2.5pt]
 58260   & $0.87\pm0.02$ & $<10$  & $<5.6$ & $6.5\pm0.2$  &  184927   & $0.06\pm0.02$ & $19\pm3$  & $68\pm4$  & $8.7\pm1.5$  \\ [2.5pt]
61556$^m$   & $-0.10\pm0.03$ & $35\pm7$  & $59\pm6$  & $2.8\pm0.3$  &  186205   & $0.71\pm0.04$ & $45\pm12$  & $7\pm4$  & $3.0^{+1.0}_{-0.5}$   \\ [2.5pt]
 63425   & $0.31\pm0.03$ & $55\pm17$  & $17\pm10$  & $0.44^{+0.18}_{-0.07}$   &  189775$^m$   & $-0.23\pm0.04$ & $59\pm9$  & $43\pm11$  & $4.3\pm0.7$  \\ [2.5pt]
64740$^m$   & $-0.47\pm0.04$ & $42\pm8$  & $72\pm5$  & $3.0\pm0.5$  &  205021   & $-0.94\pm0.04$ & $69\pm9$  & $87\pm3$  & $0.26\pm0.03$  \\ [2.5pt]
 66522   & $-0.94\pm0.04$ &  -- & $89.4^{+0.4}_{-4.6}$  & $2.2\pm0.6$  &  208057   & $-1.0\pm0.1$ & $42\pm9$  & $>89$ & $0.6\pm0.2$  \\ [2.5pt]
 66665   & $-0.34\pm0.03$ & $28\pm5$  & $76\pm3$  & $0.56\pm0.10$  &  345439   & $-0.26\pm0.08$ & $59\pm10$  & $46\pm13$  & $8.9\pm1.1$  \\ [2.5pt]
 66765   & $-0.53\pm0.04$ & $44\pm9$  & $73\pm5$  & $2.8\pm0.5$  &  ALS\,3694 & $0.7\pm0.2$ & $22\pm2$  & $26\pm11$  & $12\pm3$  \\ [2.5pt]
 67621   & $-0.22\pm0.02$ & $22\pm2$  & $75\pm2$  & $1.6\pm0.2$  &  CPD\,$-62^\circ 2124^m$ & $0.64\pm0.04$ & $26\pm3$  & $25\pm4$  & $23\pm1$  \\ [2.5pt]
 96446   & $0.41\pm0.05$ & $15\pm4$  & $56\pm7$  & $3.9\pm0.8$  &  CPD\,$-57^\circ 3509$ & $-0.73\pm0.09$ & $66\pm11$  & $77\pm10$  & $3.9\pm0.5$  \\ [2.5pt]
\hline\hline
\end{tabular}
}
\end{table*}

\begin{table*}
\caption[]{Stellar wind, magnetospheric, and magnetic braking parameters. The second column indicates H$\alpha$ emission status; references are given by P13, except for: HD 25558 \citep{2014MNRAS.438.3535S}; HD 23478 \citep{2015MNRAS.451.1928S}; HD 35502 \citep{2016MNRAS.460.1811S}; HD 44743, HD 52089 \citep{2015AA...574A..20F}; HD 130807 \citep{alecian2011}; HD 164492C \protect\citep{2017MNRAS.465.2517W}; and ALS 3694 \citep{2016ASPC..506..305S}; or the present work for HD 37058, HD 189775, and HD 36526. The subsequent columns give: the logarithmic mass-loss rate $\log{\dot{M}}$; the wind terminal velocity \vinf; the logarithmic wind magnetic confinement parameter $\log{\eta_*}$; the Alfv\'en radius \ra; the logarithmic ratio of \ra~to \rk; the logarithmic spindown timescale $\tau_{\rm J}$; the logarithmic maximum spindown age $t_{\rm S,max}$; and the logarithmic ratio of the age $t$ to $t_{\rm S,max}$.}
\label{magnetosphere_table}
\begin{tabular}{l | l r r r r r r r r}
HD No. & H$\alpha$ & $\log{\dot{M}}$           & \vinf  & $\log{\eta_*}$ & \ra     & \rark     & $\log{\tau_{\rm J}}$ & $\log{t_{\rm S,max}}$ & $\log{(t/t_{\rm S,max})}$ \\
       &           & ($M_\odot {\rm yr}^{-1}$) & (\kms) &                & ($R_*$) &           & (yr)          & (yr)           &                           \\
\hline
  3360   &  No  & $-9.90\pm0.08$ & $1882\pm42$ & $2.6\pm0.2$ & $4.8\pm0.5$ & $0.05\pm0.05$ & $8.4\pm0.1$ & $8.8\pm0.1$ & $-1.32\pm0.09$ \\ [2.5pt]
 23478   &  Yes & $-11.1^{+0.7}_{-0.3}$ & $2351^{+56}_{-70}$ & $6.8\pm0.6$ & $30^{+14}_{-8}$ & $1.3\pm0.1$ & $7.3^{+0.2}_{-0.3}$ & $7.5^{+0.2}_{-0.3}$ & $-0.7^{+0.4}_{-0.1}$ \\ [2.5pt]
 25558   &  No  & $-10.5\pm0.3$ & $1048\pm66$ & $4.1\pm0.4$ & $9\pm2$ & $0.6\pm0.1$ & $8.0\pm0.3$ & $8.3\pm0.4$ & $-0.6\pm0.4$ \\ [2.5pt]
 35298   &  No  & $-10.9\pm0.2$ & $1075\pm13$ & $6.7\pm0.2$ & $48\pm3$ & $1.06\pm0.03$ & $7.1\pm0.1$ & $7.5\pm0.1$ & $-0.4\pm0.1$ \\ [2.5pt]
 35502   &  Yes & $-10.0\pm0.1$ & $1129\pm12$ & $5.64\pm0.10$ & $26\pm1$ & $1.07\pm0.02$ & $6.9\pm0.1$ & $6.95\pm0.09$ & $0.1^{+0.1}_{-0.1}$ \\ [2.5pt]
 36485   &  Yes & $-9.8\pm0.1$ & $1163\pm13$ & $5.6\pm0.1$ & $25\pm1$ & $0.88\pm0.02$ & $6.7\pm0.1$ & $6.99\pm0.06$ & $-0.4^{+0.1}_{-0.4}$ \\ [2.5pt]
 36526   &  No  & $-11.0^{+0.2}_{-0.3}$ & $1086\pm19$ & $6.7^{+0.1}_{-0.2}$ & $47^{+3}_{-4}$ & $1.10^{+0.02}_{-0.04}$ & $7.1\pm0.1$ & $7.4^{+0.1}_{-0.1}$ & $-0.8^{+0.1}_{-0.1}$ \\ [2.5pt]
 36982   &  No  & $-10.8\pm0.3$ & $2439\pm36$ & $5.2\pm0.6$ & $15\pm6$ & $0.7\pm0.1$ & $7.9\pm0.1$ & $8.2\pm0.1$ & $-2.0^{+0.2}_{-0.2}$ \\ [2.5pt]
 37017   &  Yes & $-10.3\pm0.3$ & $2419\pm31$ & $5.6\pm0.2$ & $24\pm3$ & $1.09\pm0.05$ & $7.3\pm0.2$ & $7.4\pm0.2$ & $-0.7\pm0.1$ \\ [2.5pt]
 37058   &  No  & $-10.0\pm0.1$ & $1149\pm24$ & $4.7\pm0.2$ & $15\pm1$ & $-0.01\pm0.04$ & $7.3\pm0.1$ & $7.9\pm0.1$ & $-0.9\pm0.2$ \\ [2.5pt]
 37061   &  Yes & $-10.6\pm0.2$ & $2455\pm34$ & $6.2\pm0.2$ & $36\pm4$ & $1.13\pm0.05$ & $7.3\pm0.1$ & $7.5\pm0.1$ & $-1.5\pm0.2$ \\ [2.5pt]
 37479   &  Yes & $-10.6^{+0.1}_{-0.1}$ & $2437\pm31$ & $6.5^{+0.2}_{-0.3}$ & $36\pm9$ & $1.18^{+0.06}_{-0.08}$ & $7.1\pm0.1$ & $7.33^{+0.06}_{-0.09}$ & $-0.8^{+0.2}_{-0.1}$ \\ [2.5pt]
 37776   &  Yes & $-10.4\pm0.2$ & $2448\pm35$ & $5.6\pm0.2$ & $24\pm3$ & $0.93\pm0.05$ & $7.4\pm0.1$ & $7.59\pm0.09$ & $-1.0^{+0.2}_{-0.1}$ \\ [2.5pt]
 43317   &  No  & $-10.0\pm0.1$ & $991\pm84$ & $4.4\pm0.2$ & $12\pm1$ & $0.8\pm0.1$ & $7.5\pm0.2$ & $7.6\pm0.3$ & $0.3^{+0.3}_{-0.4}$ \\ [2.5pt]
 44743   &  No  & $-8.4\pm0.1$ & $1885\pm62$ & $1.1\pm0.2$ & $2.2\pm0.2$ & $-0.57\pm0.04$ & $7.7\pm0.1$ & $8.22\pm0.09$ & $-1.10^{+0.06}_{-0.07}$ \\ [2.5pt]
 46328   &  Yes & $-8.0\pm0.2$ & $2116\pm79$ & $2.8\pm0.3$ & $5.3\pm1.0$ & $-2.08\pm0.09$ & $6.7\pm0.2$ & $7.6\pm0.2$ & $-0.7^{+0.2}_{-0.1}$ \\ [2.5pt]
 47777   &  No  & $-10.4\pm0.2$ & $2396\pm111$ & $5.3\pm0.3$ & $20\pm3$ & $0.68\pm0.08$ & $7.7\pm0.2$ & $8.0\pm0.2$ & $-1.1^{+0.3}_{-0.2}$ \\ [2.5pt]
 52089   &  No  & $-8.61\pm0.08$ & $1721\pm34$ & $1.2\pm0.5$ & $2.2\pm0.6$ & $-0.5\pm0.1$ & $7.9\pm0.2$ & $8.4\pm0.2$ & $-1.2^{+0.3}_{-0.1}$ \\ [2.5pt]
 55522   &  No  & $-9.6\pm0.1$ & $962\pm36$ & $4.9\pm0.1$ & $16\pm1$ & $0.70\pm0.05$ & $6.9\pm0.1$ & $7.2\pm0.2$ & $0.6^{+0.1}_{-0.1}$ \\ [2.5pt]
 58260   &  No  & $-9.5\pm0.4$ & $1030\pm127$ & $5.3\pm0.2$ & $21\pm3$ & $0.3\pm0.2$ & $6.6\pm0.3$ & $7.1\pm0.3$ & $0.4^{+0.2}_{-0.7}$ \\ [2.5pt]
 61556   &  No  & $-9.7\pm0.3$ & $1027\pm83$ & $4.7\pm0.2$ & $15\pm1$ & $0.66\pm0.08$ & $7.0\pm0.2$ & $7.3\pm0.3$ & $0.3^{+0.3}_{-0.5}$ \\ [2.5pt]
 63425   &  No  & $-7.9\pm0.1$ & $2329\pm121$ & $1.6\pm0.2$ & $2.8\pm0.4$ & $-1.18\pm0.08$ & $7.1\pm0.1$ & $7.8\pm0.1$ & $-1.0^{+0.2}_{-0.1}$ \\ [2.5pt]
 64740   &  Yes & $-9.4\pm0.2$ & $2302\pm169$ & $4.4\pm0.2$ & $12\pm1$ & $0.8\pm0.1$ & $7.2\pm0.2$ & $7.3\pm0.3$ & $0.0^{+0.2}_{-0.4}$ \\ [2.5pt]
 66522   &  No  & $-10.3\pm0.9$ & $913^{+1222}_{-326}$ & $5.0\pm0.6$ & $10^{+8}_{-3}$ & $-1.0\pm0.1$ & $7.5\pm0.7$ & $8.4\pm0.7$ & $-1.0^{+0.6}_{-0.4}$ \\ [2.5pt]
 66665   &  No  & $-7.8\pm0.2$ & $2215\pm137$ & $1.9\pm0.2$ & $3.2\pm0.3$ & $-0.55\pm0.09$ & $6.9\pm0.2$ & $7.4\pm0.2$ & $-0.6^{+0.1}_{-0.2}$ \\ [2.5pt]
 66765   &  Yes & $-9.4\pm1.0$ & $1033^{+1087}_{-264}$ & $4.4^{+0.9}_{-0.2}$ & $13^{+8}_{-1}$ & $0.9\pm0.1$ & $7.7^{+0.5}_{-0.8}$ & $7.9^{+0.5}_{-0.9}$ & $-0.4^{+0.6}_{-0.3}$ \\ [2.5pt]
 67621   &  No  & $-10.7\pm0.2$ & $2248\pm88$ & $5.0\pm0.2$ & $18\pm1$ & $0.56\pm0.06$ & $8.0\pm0.2$ & $8.4\pm0.2$ & $-1.2^{+0.4}_{-0.2}$ \\ [2.5pt]
 96446   &  No  & $-9.7\pm0.2$ & $2209\pm163$ & $5.0\pm0.3$ & $17\pm2$ & $0.1\pm0.1$ & $7.0\pm0.2$ & $7.6\pm0.2$ & $-0.4^{+0.2}_{-0.3}$ \\ [2.5pt]
105382   &  No  & $-9.9\pm0.1$ & $1115\pm14$ & $4.71\pm0.08$ & $15.3\pm0.7$ & $0.72\pm0.02$ & $7.3\pm0.1$ & $7.50\pm0.08$ & $-0.4^{+0.1}_{-0.1}$ \\ [2.5pt]
121743   &  No  & $-10.2\pm0.3$ & $2240\pm80$ & $4.3\pm0.3$ & $12\pm2$ & $0.79\pm0.08$ & $7.9\pm0.2$ & $8.0\pm0.2$ & $-0.7^{+0.2}_{-0.2}$ \\ [2.5pt]
122451   &  No  & $-8.96\pm0.08$ & $1768\pm171$ & $2.3\pm0.2$ & $4.1\pm0.6$ & $0.3\pm0.1$ & $7.7\pm0.1$ & $7.7\pm0.2$ & $-0.3^{+0.5}_{-0.2}$ \\ [2.5pt]
125823   &  No  & $-9.9^{+0.1}_{-0.1}$ & $1127\pm17$ & $4.4^{+0.1}_{-0.2}$ & $12\pm1$ & $0.05^{+0.04}_{-0.04}$ & $7.4\pm0.1$ & $7.97^{+0.06}_{-0.09}$ & $-0.8^{+0.1}_{-0.2}$ \\ [2.5pt]
127381   &  No  & $-9.7\pm0.1$ & $2191\pm147$ & $3.4\pm0.3$ & $7\pm1$ & $0.3\pm0.1$ & $7.8\pm0.2$ & $8.1\pm0.2$ & $-0.9^{+0.2}_{-0.3}$ \\ [2.5pt]
130807   &  No  & $-10.2\pm0.1$ & $1100\pm18$ & $6.1\pm0.4$ & $32\pm10$ & $0.8\pm0.1$ & $6.8\pm0.1$ & $7.2\pm0.1$ & $0.1^{+0.2}_{-0.1}$ \\ [2.5pt]
136504Aa   &  No  & $-10.2\pm0.3$ & $2004\pm144$ & $4.2\pm0.2$ & $11\pm1$ & $0.63\pm0.07$ & $8.0\pm0.2$ & $8.1\pm0.2$ & $-0.7^{+0.5}_{-0.1}$ \\ [2.5pt]
136504Ab   &  No  & $-9.4\pm0.2$ & $1004\pm52$ & $3.1\pm0.3$ & $6.2\pm1.0$ & $0.22\pm0.08$ & $7.6\pm0.2$ & $7.9\pm0.2$ & $-0.3\pm0.2$ \\ [2.5pt]
142184   &  Yes & $-10.10\pm0.09$ & $1135\pm26$ & $6.0\pm0.2$ & $31\pm3$ & $1.34\pm0.03$ & $6.7\pm0.1$ & $6.48\pm0.04$ & $0.4^{+0.2}_{-0.2}$ \\ [2.5pt]
142990   &  Yes & $-10.1\pm0.1$ & $1142\pm9$ & $5.3\pm0.1$ & $21\pm1$ & $0.93\pm0.02$ & $7.1\pm0.1$ & $7.25\pm0.08$ & $-0.54\pm0.09$ \\ [2.5pt]
149277   &  No  & $-10.2\pm0.3$ & $2209\pm124$ & $6.3\pm0.2$ & $38\pm5$ & $0.33\pm0.08$ & $7.0\pm0.2$ & $7.6\pm0.2$ & $-0.3^{+0.5}_{-0.1}$ \\ [2.5pt]
149438   &  No  & $-7.7\pm0.2$ & $2540\pm123$ & $0.9\pm0.1$ & $2.0\pm0.1$ & $-1.00\pm0.05$ & $7.2\pm0.1$ & $7.9\pm0.1$ & $-1.1^{+0.1}_{-0.2}$ \\ [2.5pt]
156324   &  Yes & $-9.9\pm0.5$ & $2373\pm121$ & $6.2\pm0.3$ & $30\pm8$ & $1.1\pm0.1$ & $6.6\pm0.4$ & $6.8\pm0.4$ & $-0.0^{+0.4}_{-0.3}$ \\ [2.5pt]
156424   &  Yes & $-9.7\pm0.7$ & $2379\pm160$ & $5.3\pm0.5$ & $21\pm6$ & $0.8\pm0.1$ & $6.8\pm0.5$ & $7.1\pm0.5$ & $-0.2^{+0.5}_{-0.3}$ \\ [2.5pt]
\hline
\hline
\end{tabular}
\end{table*}
 
\begin{table*}
\contcaption{}
\label{magnetosphere_table:continued}
\begin{tabular}{l | l r r r r r r r r}
\hline
\hline
HD No. & H$\alpha$ & $\log{\dot{M}}$           & \vinf  & $\log{\eta_*}$ & \ra     & \rark     & $\log{\tau_{\rm J}}$ & $\log{t_{\rm S,max}}$ & $\log{(t/t_{\rm S,max})}$ \\
       &           & ($M_\odot {\rm yr}^{-1}$) & (\kms) &                & ($R_*$) &           & (Myr)          & (Myr)           & \\
\hline
163472   &  No  & $-9.3\pm0.2$ & $2385\pm96$ & $3.5\pm0.4$ & $7\pm1$ & $0.26\pm0.09$ & $7.5\pm0.2$ & $7.8\pm0.2$ & $-0.9^{+0.1}_{-0.2}$ \\ [2.5pt]
164492   &  Yes & $-9.2^{+0.2}_{-0.2}$ & $2627\pm35$ & $4.8^{+0.2}_{-0.2}$ & $15^{+1}_{-2}$ & $0.77^{+0.04}_{-0.06}$ & $6.8\pm0.1$ & $6.99^{+0.06}_{-0.08}$ & $-1.2^{+0.2}_{-0.1}$ \\ [2.5pt]
175362   &  No  & $-10.3\pm0.1$ & $1117\pm30$ & $6.67\pm0.08$ & $46\pm2$ & $0.87\pm0.03$ & $6.6\pm0.1$ & $7.09\pm0.08$ & $0.0\pm0.3$ \\ [2.5pt]
176582   &  Yes & $-10.1\pm0.1$ & $1038\pm34$ & $5.6\pm0.1$ & $24\pm1$ & $0.89\pm0.04$ & $6.9\pm0.1$ & $7.17\pm0.04$ & $0.4^{+0.0}_{-0.1}$ \\ [2.5pt]
182180   &  Yes & $-9.7^{+0.2}_{-1.2}$ & $1124^{+1140}_{-54}$ & $5.6^{+0.9}_{-0.1}$ & $24^{+18}_{-1}$ & $1.3^{+0.2}_{-0.1}$ & $6.6^{+0.7}_{-0.4}$ & $6.2^{+0.7}_{-0.3}$ & $1.1^{+0.1}_{-0.7}$ \\ [2.5pt]
184927   &  No  & $-10.1\pm0.3$ & $2194\pm175$ & $6.1\pm0.3$ & $30\pm5$ & $0.6\pm0.1$ & $7.0\pm0.2$ & $7.4\pm0.3$ & $0.0^{+0.3}_{-0.4}$ \\ [2.5pt]
186205   &  No  & $-8.5\pm0.5$ & $870\pm87$ & $4.1\pm0.6$ & $10^{+4}_{-2}$ & $-0.06\pm0.08$ & $6.1\pm0.3$ & $6.7\pm0.3$ & $0.8^{+0.1}_{-0.5}$ \\ [2.5pt]
189775   &  No  & $-10.0\pm0.1$ & $1023\pm49$ & $5.3\pm0.2$ & $21\pm2$ & $0.73\pm0.04$ & $6.9\pm0.1$ & $7.28\pm0.08$ & $0.2^{+0.2}_{-0.1}$ \\ [2.5pt]
205021   &  No  & $-8.4\pm0.2$ & $1889\pm77$ & $1.9\pm0.2$ & $3.3\pm0.4$ & $-0.26\pm0.06$ & $7.4\pm0.1$ & $7.9\pm0.1$ & $-0.72\pm0.07$ \\ [2.5pt]
208057   &  No  & $-9.9\pm0.2$ & $923\pm81$ & $3.7\pm0.3$ & $8\pm1$ & $0.6\pm0.1$ & $7.7\pm0.2$ & $7.8\pm0.3$ & $0.1^{+0.3}_{-0.3}$ \\ [2.5pt]
345439   &  Yes & $-9.8\pm0.4$ & $2463\pm141$ & $5.5\pm0.5$ & $21\pm7$ & $1.2\pm0.1$ & $6.8\pm0.2$ & $6.6\pm0.2$ & $0.2^{+0.2}_{-0.4}$ \\ [2.5pt]
ALS 3694 &  Yes & $-9.8\pm0.4$ & $2447\pm93$ & $5.8\pm0.4$ & $27\pm6$ & $1.0\pm0.1$ & $6.7\pm0.2$ & $6.9\pm0.3$ & $-0.1^{+0.3}_{-0.2}$ \\ [2.5pt]
CPD\,$-62^\circ 2124$ &  Yes & $-9.9\pm0.2$ & $2431\pm94$ & $6.4\pm0.1$ & $41\pm2$ & $1.01\pm0.03$ & $6.5\pm0.1$ & $6.9\pm0.1$ & $-0.0\pm0.2$ \\ [2.5pt]
CPD\,$-57^\circ 3509$ &  No  & $-9.6\pm0.2$ & $2285\pm98$ & $4.7\pm0.1$ & $15\pm1$ & $0.39\pm0.04$ & $7.1\pm0.1$ & $7.5\pm0.2$ & $-0.4^{+0.2}_{-0.1}$ \\ [2.5pt]
\hline\hline
\end{tabular}
\end{table*}

\end{document}